\documentclass{emulateapj}

\begin{document}

\shortauthors{Luhman et al.}
\shorttitle{Disk Population of Taurus}

\title{The Disk Population of the Taurus Star-Forming Region\altaffilmark{1}}

\author{
K. L. Luhman\altaffilmark{2,3},
P. R. Allen\altaffilmark{2},
C. Espaillat\altaffilmark{4,5},
L. Hartmann\altaffilmark{4},
and N. Calvet\altaffilmark{4}
}

\altaffiltext{1}
{Based on observations performed with the {\it Spitzer Space Telescope}.}

\altaffiltext{2}{Department of Astronomy and Astrophysics, The Pennsylvania
State University, University Park, PA 16802; kluhman@astro.psu.edu.}

\altaffiltext{3}{Center for Exoplanets and Habitable Worlds, The Pennsylvania
State University, University Park, PA 16802.}

\altaffiltext{4}{Department of Astronomy, The University of Michigan,
Ann Arbor, MI 48109.}

\altaffiltext{5}{Current address: 
Harvard-Smithsonian Center for Astrophysics, Cambridge, MA 02138.}

\begin{abstract}

We have analyzed nearly all images of the Taurus star-forming region
at 3.6, 4.5, 5.8, 8.0, and 
24~\micron\ that were obtained during the cryogenic mission of the 
{\it Spitzer Space Telescope} (46~deg$^{2}$) and have measured photometry 
for all known members of the region that are within these data, corresponding 
to 348 sources, or 99\% of the known stellar population. 
By combining these measurements with previous observations with
the {\it Spitzer} Infrared Spectrograph and other facilities,
we have classified the members of Taurus according to whether they show 
evidence of circumstellar disks and envelopes (classes~I, II, and III).
Through these classifications, we find that the disk fraction in Taurus,
N(II)/N(II+III),
is $\sim75$\% for solar-mass stars and declines to $\sim45$\% for low-mass
stars and brown dwarfs (0.01--0.3~$M_\odot$). This dependence on stellar mass 
is similar to that measured for Chamaeleon~I, although the disk fraction
in Taurus is slightly higher overall, probably because of its younger age 
(1~Myr vs. 2--3~Myr).
In comparison, the disk fraction for solar-mass stars is much lower 
($\sim20$\%) in IC~348 and $\sigma$~Ori, which are denser than Taurus
and Chamaeleon~I and are roughly coeval with the latter. 
These data indicate that disk lifetimes for solar-mass stars are
longer in star-forming regions that have lower stellar densities. 
Through an analysis of multiple epochs of {\it Spitzer} photometry
that are available for $\sim200$ Taurus members, we find that stars with disks
exhibit significantly greater mid-infrared variability than diskless stars,
which agrees with the results of similar variability measurements for
a smaller sample of stars in Chamaeleon~I. The variability fraction for
stars with disks is higher in Taurus than in Chamaeleon~I, indicating that
the IR variability of disks decreases with age. Finally, 
we have used our data in Taurus to refine the observational criteria for 
primordial, evolved, and transitional disks.
The ratio of the number of evolved and transitional disks to the number
of primordial disks in Taurus is 15/98 for spectral types of K5--M5,
indicating a timescale of $0.15\times\tau_{\rm primordial}\sim0.45$~Myr for the
clearing of the inner regions of optically thick disks. 
After applying the same criteria to older clusters and associations
(2--10~Myr) that have been observed with {\it Spitzer},
we find that the proportions of evolved and transitional disks in those
populations are consistent with the measurements in Taurus when their star
formation histories are properly taken into account.

{\bf ERRATUM:}
In Table~7, we inadvertently omitted the spectral type bins in which class~II
sources were placed in Table~8 based on their bolometric luminosities
(applies only to stars that lack spectroscopic classifications).
The bins were K6--M3.5 for FT~Tau, DK~Tau~B, and IRAS 04370+2559,
M3.5--M6 for IRAS 04200+2759, IT~Tau~B, and ITG~1, and
M6--M8 for IRAS 04325+2402~C.
In addition, the values of $K_s-[3.6]$ in Table~13 and Figure~26 for
spectral types of M4--M9 are incorrect.
We present corrected versions of Table~13 and Figure~26.

\end{abstract}


\section{Introduction}
\label{sec:intro}

Much of our knowledge of the processes of star and planet formation has
been derived from observations of circumstellar disks around newborn stars. 
The Taurus complex of dark clouds is one of the principle sites for studies
of disks. 
Taurus is among the nearest star-forming regions to the Sun ($d=140$~pc) 
and contains more than 300 young stars and brown dwarfs \citep{ken08}.
Because of the low stellar density and the absence of photoionizing stars, 
most members of Taurus are not embedded within bright nebulae,
and thus are not subject to high levels of background emission 
that would hinder infrared (IR) observations of disks. 
The low density of Taurus also permitted resolved photometry of
individual stars with the low-resolution imaging that was offered by
the first major far-IR telescope, the
{\it Infrared Astronomical Satellite} \citep[IRAS,][]{bei86,ken90,ken94}.
By combining 12--100~\micron\ photometry from IRAS with 1--10~\micron\ data
from ground-based telescopes, \citet{kh95} performed the most
comprehensive census of disks in a star-forming region at that time.  
The disk-bearing stars identified in Taurus have served as targets
for a multitude of detailed observations of circumstellar disks 
\citep[][references therein]{ken08}.

The census of the stellar population of Taurus has expanded significantly since 
the disk study of \citet{kh95}, particularly at masses below 0.5~$M_\odot$. 
The sensitivities of IR telescopes also have improved dramatically, most 
notably with the deployment of the {\it Spitzer Space Telescope} \citep{wer04}, 
which is capable of detecting members of Taurus with masses below 
0.01~$M_\odot$.
The Taurus stellar population has been imaged extensively by {\it Spitzer} 
through pointed observations of individual stars as well as wide-field maps.
These images have been used to measure photometry for known members of Taurus
\citep{har05,luh06tau2,gui07} and search for new disk-bearing members 
\citep{luh06tau2,luh09fu,luh09tau1,reb10}.

With the recent completion of the cryogenic mission of {\it Spitzer},
we wish to present a study of disks in Taurus that makes use of all
{\it Spitzer} images of the region, which cover a total area
of 46~deg$^{2}$ and encompass 99\% of the known stellar population.
We begin by analyzing all {\it Spitzer} images of Taurus at 
3.6--24~\micron\ that are publicly available and measuring photometry for all 
known members that are detected in these data (\S~\ref{sec:images}).
We use these measurements in conjunction with previous observations
by the {\it Spitzer} Infrared Spectrograph and other telescopes to classify
the spectral energy distributions of all members of Taurus (\S~\ref{sec:sed}).
Through these classifications, we investigate how the prevalence of disks
depends on stellar mass (\S~\ref{sec:frac}) and location (\S~\ref{sec:spatial}).
We then characterize the mid-IR variability of disks in Taurus 
with the multi-epoch photometry from {\it Spitzer} (\S~\ref{sec:var}).
Finally, we use our mid-IR data in Taurus to refine the observational criteria
for the advanced stages of disk evolution and we apply these criteria to 
{\it Spitzer} measurements in nearby clusters and associations with ages of
2--10~Myr (\S~\ref{sec:trans}).

\section{Infrared Images}
\label{sec:images}

\subsection{Observations}

For our census of the disk population in Taurus, we use
images at 3.6, 4.5, 5.8, and 8.0~\micron\ obtained with
{\it Spitzer}'s Infrared Array Camera \citep[IRAC;][]{faz04} and
images at 24~\micron\ obtained with the Multiband Imaging Photometer for
{\it Spitzer} \citep[MIPS;][]{rie04}.
The fields of view are $5\farcm2\times5\farcm2$ and
$5\farcm4\times5\farcm4$ for IRAC and the 24~\micron\ channel of MIPS,
respectively. The cameras produced images with FWHM$=1\farcs6$--$1\farcs9$ 
from 3.6 to 8.0~\micron\ and FWHM=$5\farcs9$ at 24~\micron.

\begin{figure}
\epsscale{1.2}
\plotone{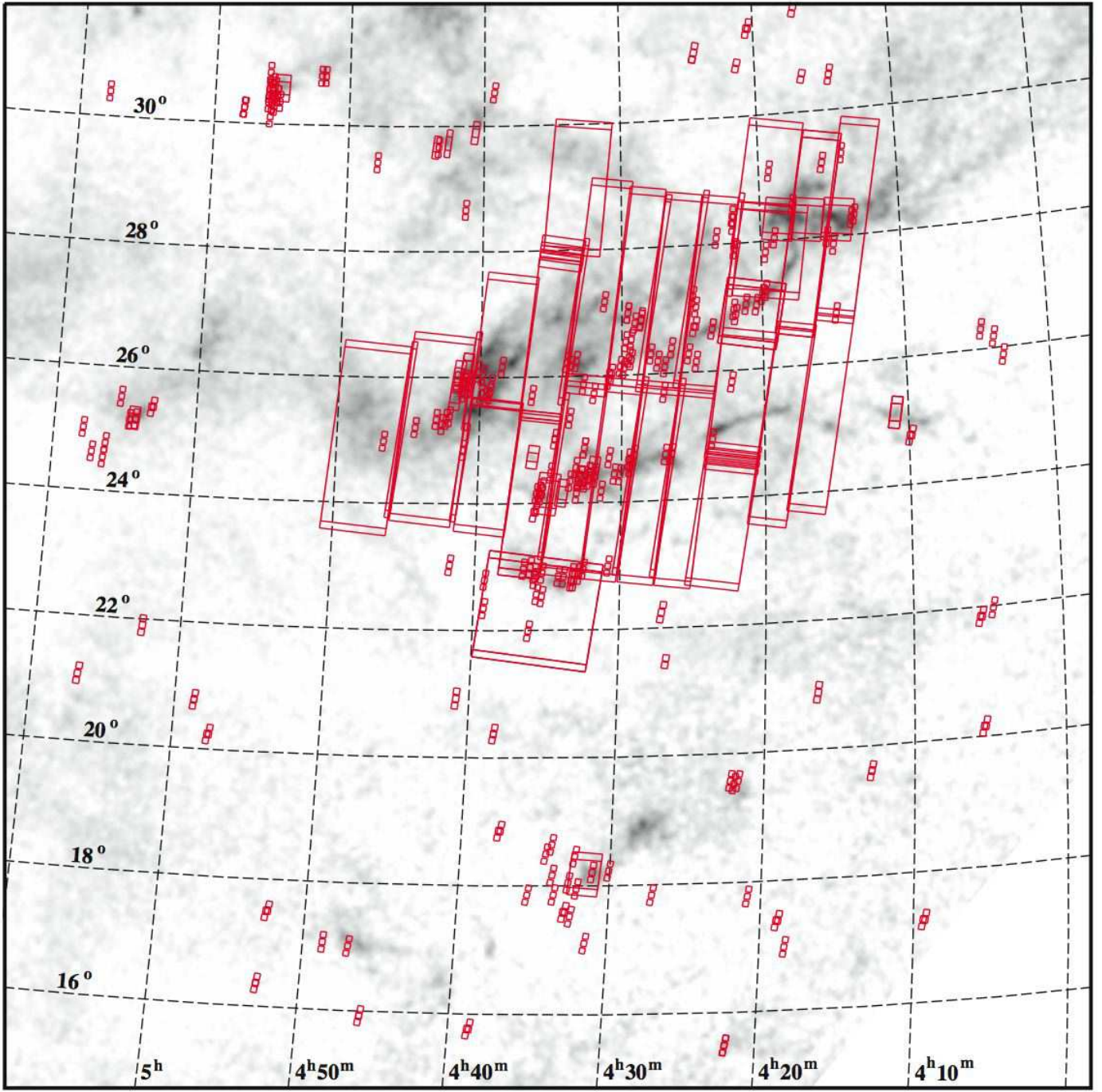}
\caption{
Fields in the Taurus star-forming region that have been imaged at 
3.6--8.0~\micron\ by IRAC on the {\it Spitzer Space Telescope} 
(Table~\ref{tab:iraclog}).
The dark clouds in Taurus are displayed with a map of extinction
\citep[{\it grayscale},][]{dob05}.
}
\label{fig:map1}
\end{figure}

\begin{figure}
\epsscale{1.2}
\plotone{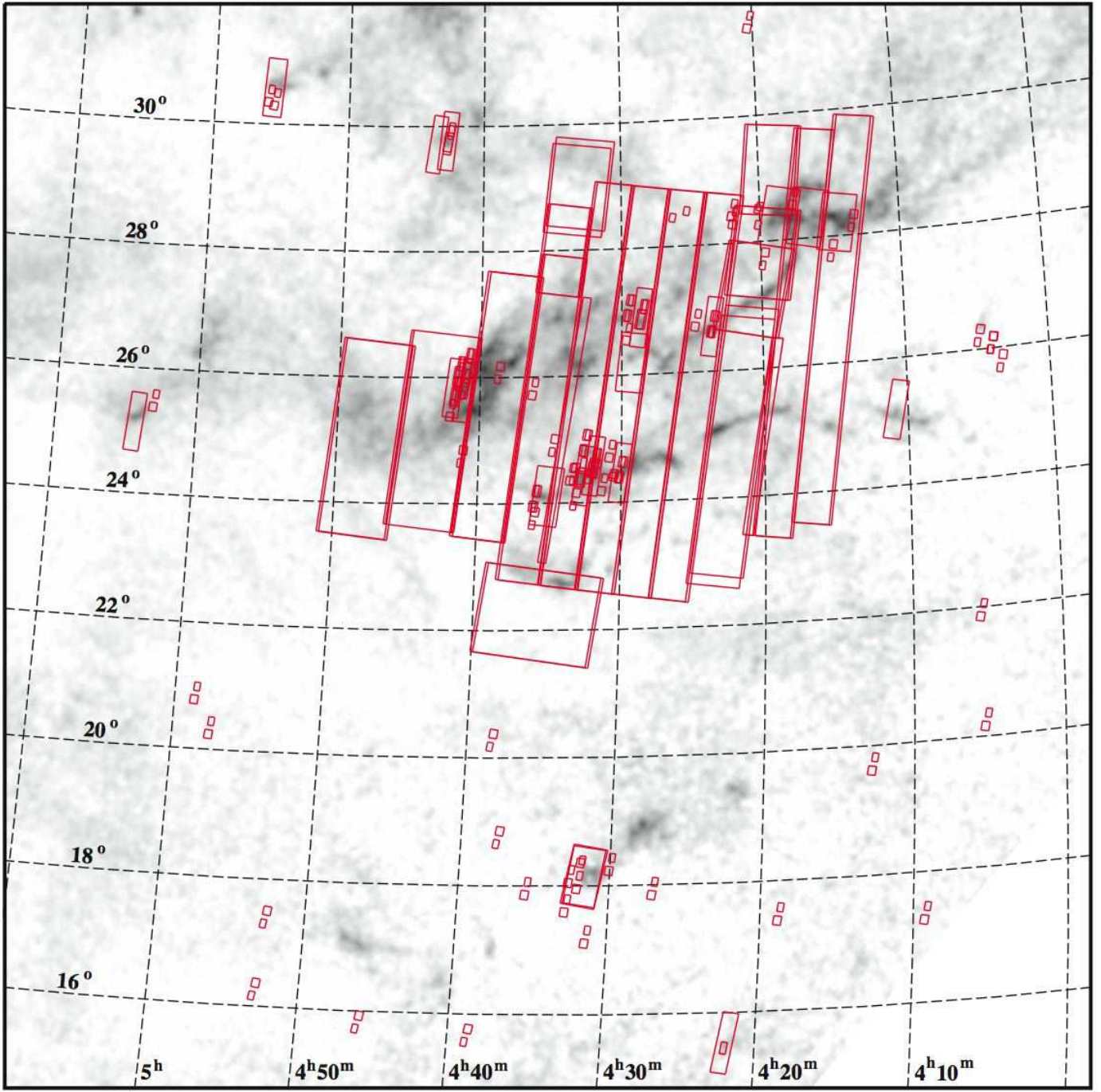}
\caption{
Fields in the Taurus star-forming region that have been imaged at 
24~\micron\ with MIPS on the {\it Spitzer Space Telescope} 
(Table~\ref{tab:mipslog}).
}
\label{fig:map2}
\end{figure}

We consider all IRAC and MIPS 24~\micron\ observations that have been performed
in the vicinity of the Taurus clouds with the exception of some of
the data with program identifications (PIDs) of 50477 and 50584, which are not 
yet available to the public. The images from the latter programs encompass 
only a small number of known members of Taurus, as discussed later in this 
section. The available data were obtained through Guaranteed Time 
Observations for PID=6, 36, 37 (G. Fazio), 53 (G. Rieke), 94 (C. Lawrence), 
30540 (G. Fazio, J. Houck), and 40302 (J. Houck), 
Director's Discretionary Time for PID=462 (L. Rebull), 
Legacy programs for PID=139, 173 (N. Evans), and 30816 (D. Padgett),
and General Observer programs for PID=3584 (D. Padgett), 20302 (P. Andr\'e), 
20386 (P. Myers), 20762 (J. Swift), 30384 (T. Bourke), 40844 (C. McCabe),
and 50584 (D. Padgett). 
The IRAC and MIPS observations were performed through 180 and 137 Astronomical 
Observation Requests (AORs), respectively. 
The characteristics of the resulting images are summarized in 
Tables~\ref{tab:iraclog} and \ref{tab:mipslog}.
All AORs in Taurus with publicly available data are included in these 
observing logs, even those that do 
not contain known members and thus do not contribute photometry to our disk 
census. The boundaries of the IRAC and MIPS images are indicated in maps 
of the Taurus dark clouds in Figures~\ref{fig:map1} and \ref{fig:map2}, 
respectively. Each camera observed a total area of 46~deg$^{2}$.
The fields covered by the two cameras closely overlap, as demonstrated in 
Figures~\ref{fig:map1} and \ref{fig:map2}.

We adopt the census of the stellar population in Taurus that was
described by \citet{luh09tau1}, which is an updated version of the one
presented by \citet{ken08}. 
This census consists of 352 and 341 sources that are resolvable by 
IRAC and MIPS 24~\micron\ images, respectively, 
346 and 299 of which are within the available IRAC and MIPS images. 
Two of the members in the MIPS images were not observed by IRAC.
Thus, a total of 348 resolved members appear within images from either
IRAC or MIPS, corresponding to 99\% of the known stellar population.
All known members of Taurus are included in the tabulations of photometry
in the next section. The stars that are outside of all available images
for a given camera are indicated in those tables.
The IRAC and MIPS images that are not yet available to the public 
encompass three and four known members, respectively. 
One and three of these stars are outside of the currently available 
IRAC and MIPS images, which consist of XEST~06-006 for IRAC and 
XEST~06-006, 2MASS J04163048+3037053, and 2MASS J04162725+2053091 for MIPS.

\subsection{Data Reduction}
\label{sec:reduction}

Initial processing of the IRAC and MIPS images was performed by the 
Spitzer Science Center (SSC) pipeline.
The pipeline versions were S14.0.0, S14.4.0, S15.3.0, S16.1.0, and
S18.7.0 for IRAC and S16.0.1, S16.1.0, S16.1.1, and S17.2.0 for MIPS.
The IRAC images produced by the SSC pipeline were combined into mosaics
using R. Gutermuth's WCSmosaic IDL package \citep{gut08}.
We selected a plate scale of $0\farcs86$~pixel$^{-1}$ for these final images.
For MIPS, we used the mosaics produced by the SSC pipeline, which
had a plate scale of $2\farcs45$~pixel$^{-1}$.

We measured photometry for the known members of Taurus with methods similar
to those employed in our previous {\it Spitzer} studies of star-forming
regions \citep{luh08cha1,luh08sig}. 
We used the IRAF task STARFIND to measure the positions of the Taurus members
in the IRAC and MIPS images and we performed aperture photometry at those 
locations using the IRAF task PHOT.
For MIPS sources, the radii of the apertures and inner and outer boundaries 
of the sky annuli were 3, 3, and 4 pixels, respectively.
For a given IRAC source, we selected one of the apertures listed in
Table~\ref{tab:apcor}. The 4 pixel apertures were the default choices.
We visually inspected the IRAC images of all stars to determine if smaller
apertures were needed because of close proximity to other sources. 
Based on this inspection, we adopted aperture radii of 3 pixels and
sky annuli extending from 3 to 4 pixels for
the IRAC measurements of JH~112 and 2MASS J04324938+2253082. 
For sources that were very close to other stars, we attempted to remove
the light from the latter by subtracting a scaled IRAC point spread function 
\citep[PSF,][]{mar06}. We performed this step prior to the measurement of 
photometry for UZ~Tau~Ba+Bb, DK~Tau~B, IT~Tau~B, HV~Tau~C, MHO~1, MHO~2, 
Haro~6-37~A and B, V710~Tau~A and B, J1-4872~A and B,
IRAS~04191+1523~B, HD~28867~A and B, and FU~Tau~B \citep{luh09fu}.
We measured photometry from each PSF-subtracted image with aperture radii
and inner and outer sky annuli of either 2, 14, and 15 pixels or
3, 3, and 4 pixels, respectively. We selected the aperture from these two
options that minimized contamination by the residuals of the neighboring
PSF-subtracted star. 

Many of the protostars in Taurus are surrounded by extended emission
in the IRAC images. We measured photometry for these sources using both
the default 4-pixel apertures as well as larger apertures that
encompass most of the scattered light.
L1521F-IRS and IRAS~04166+2706 exhibit only extended emission and no
point sources at 3.6--5.8 and 3.6~\micron, respectively. 
Therefore, we did not apply the 4-pixel apertures to these images
and measured photometry from only the large apertures.
Because IRAS~04325+2402~C is faint compared to its surrounding emission,
we estimated its IRAC photometry through PSF fitting.

To calibrate the photometry, we adopted zero point magnitudes 
($ZP$) of 19.670, 18.921, 16.855, 17.394, and 15.119 in the 3.6, 4.5, 5.8, 8.0, 
and 24~\micron\ bands, where $M=-2.5 \log (DN/sec) + ZP$ \citep{rea05,eng07}.
For each AOR that contained stars that were isolated, bright, and unsaturated,
we used these stars to measure aperture corrections between our adopted 
apertures and the larger ones employed by \citet{rea05} and \citet{eng07}.
For the MIPS data, we combined our aperture corrections with the one
estimated by \citet{eng07} between their aperture and an infinite one.
These aperture corrections were then applied to our photometry for all
other sources in that AOR.  If an AOR did not contain stars that were suitable 
for estimating aperture corrections, we adopted the averages of the corrections
measured among all of the AORs. The average aperture corrections
for IRAC are given in 
Table~\ref{tab:apcor}. The spread in these measurements was less than 
$\pm0.01$~mag for a given band and aperture. The corrections to an infinite
aperture for MIPS ranged between 0.78--0.84~mag with an average value 
of 0.80~mag. 

We present the IRAC and MIPS photometry for the known members of Taurus 
in Tables~\ref{tab:irac}, \ref{tab:ext}, and \ref{tab:mips}.
If an object is saturated or is not detected in all MIPS images of that
position, it appears only once in Table~\ref{tab:mips} and the observing
dates are omitted. Our quoted photometric errors
include the Poisson errors in the source and background emission
and the 2\% and 4\% uncertainties in the calibrations of IRAC and MIPS,
respectively \citep{rea05,eng07}.
The errors do not include an additional error of $\pm0.05$~mag due to
location-dependent variations in the IRAC calibration.

\section{Classifications of Spectral Energy Distributions}
\label{sec:sed}

\subsection{Known Class 0 and Class I Sources}
\label{sec:proto}

Spectral energy distributions (SEDs) of young stars are sensitive to
the presence of circumstellar dust, and thus can be used to constrain the
evolutionary stages of the members of a young stellar population.
These stages consist of a protostar surrounded by an accretion 
disk and an infalling envelope (classes 0 and I), a star with a disk but no 
envelope (class~II), and a star that is no longer surrounded by primordial dust
\citep[class~III,][]{lw84,lada87,and93,gre94}. We wish to classify the 
members of Taurus with SEDs constructed from our {\it Spitzer} photometry.
However, while class~III stars are easily distinguished from less evolved
systems, some class~I and class~II sources can exhibit similar
SEDs in the {\it Spitzer} bands \citep{har05}. 
The definitive identification of class~I sources requires other observations
that better constrain the presence of an envelope, such as mid-IR spectroscopy,
far-IR and millimeter photometry, and high-resolution images.
Therefore, before classifying the {\it Spitzer} SEDs in Taurus,
we assign classes~0 and I to all members of Taurus that exhibit evidence of
envelopes in previous data. These stars consist of all targets 
from \citet{wat04} and \citet{fur08}\footnote{
The IRS data from \citet{fur08} that were attributed to IRAS~04166+2706
actually apply to IRAS~04166+2708. This misidentification of IRAS~04166+2706
with the counterpart of IRAS~04166+2708 also appears in \citet{luh06tau1}, 
\citet{luh06tau2}, and \citet{ken08}, and probably originated in the former two
papers.},
IRAS~04108+2803~A \citep{zas09},
L1551NE \citep{mor95},
IRAS~04191+1523~A \citep{tam91,mor92},
IRAS~04191+1523~B \citep{mot98a,duc04,dun06},
IRAM~04191+1522 \citep{and99,dun06},
L1521F-IRS \citep{bou06},
IRAS~04111+2800G \citep{pru92}, and
IRAS~04166+2706 \citep{ken90,bon96,mot01}.
Haro~6-5B also is a class~I source according to unpublished data
from the {\it Spitzer} Infrared Spectrograph
(IRS, D. Watson, private communication).
Among these protostars, IRAS~04368+2557 and IRAM~04191+1522 are the clearest
examples of class~0 sources based on their low bolometric temperatures 
\citep[$T_{\rm bol}<70$~K,][]{chen95,and99,mot01}.
We adopt class~I designations for the remaining stars.
A few of these sources appear to be in transition between classes~0 and I 
\citep[L1551NE,][]{chen95} or between classes~I and II 
\citep[IRAS~04154+2823,][]{fur08}. The latter stage has been referred to as 
flat-spectrum or class~I-II in some studies, but we denote them as class~I for
the purposes of this work since they show evidence of envelopes (albeit
tenuous ones).

The stars described in this section that exhibit evidence of envelopes in 
previous observations comprise all class~0 and I sources in 
Table~\ref{tab:alpha}. 
Sources that have class~I SEDs but lack supporting observations 
that are sensitive to the presence of envelopes are listed as ``class I?'' 
(\S~\ref{sec:slopes}).

\subsection{Infrared Color-Color \& Color-Magnitude Diagrams}
\label{sec:color}

To provide an initial demonstration of how stars with disks are 
identified with {\it Spitzer} photometry, we can construct color-color 
and color-magnitude diagrams from our {\it Spitzer} data for Taurus.
These diagrams are frequently applied to young stellar
populations because they effectively discriminate between stars with and 
without disks and yet they do not require knowledge of the individual stellar 
properties, such as spectral types and extinctions \citep{all04,meg04}.

If multiple measurements are available in a given band for a star, we
adopt the mean of those measurements weighted by the inverse square
of their flux errors in the color-color and color-magnitude diagrams.
Some Taurus members lack photometry in one or more bands and thus
are absent from a given diagram because they 
are unresolved from brighter sources, outside the field of view
of the {\it Spitzer} images, saturated, detected only as extended emission,
or below the detection limit (only applies to the 24~\micron\ data). 
If a binary is resolved by IRAC but not by MIPS, we plot the total 
system magnitudes when data from both cameras are used in a diagram. 

\begin{figure}
\epsscale{1}
\plotone{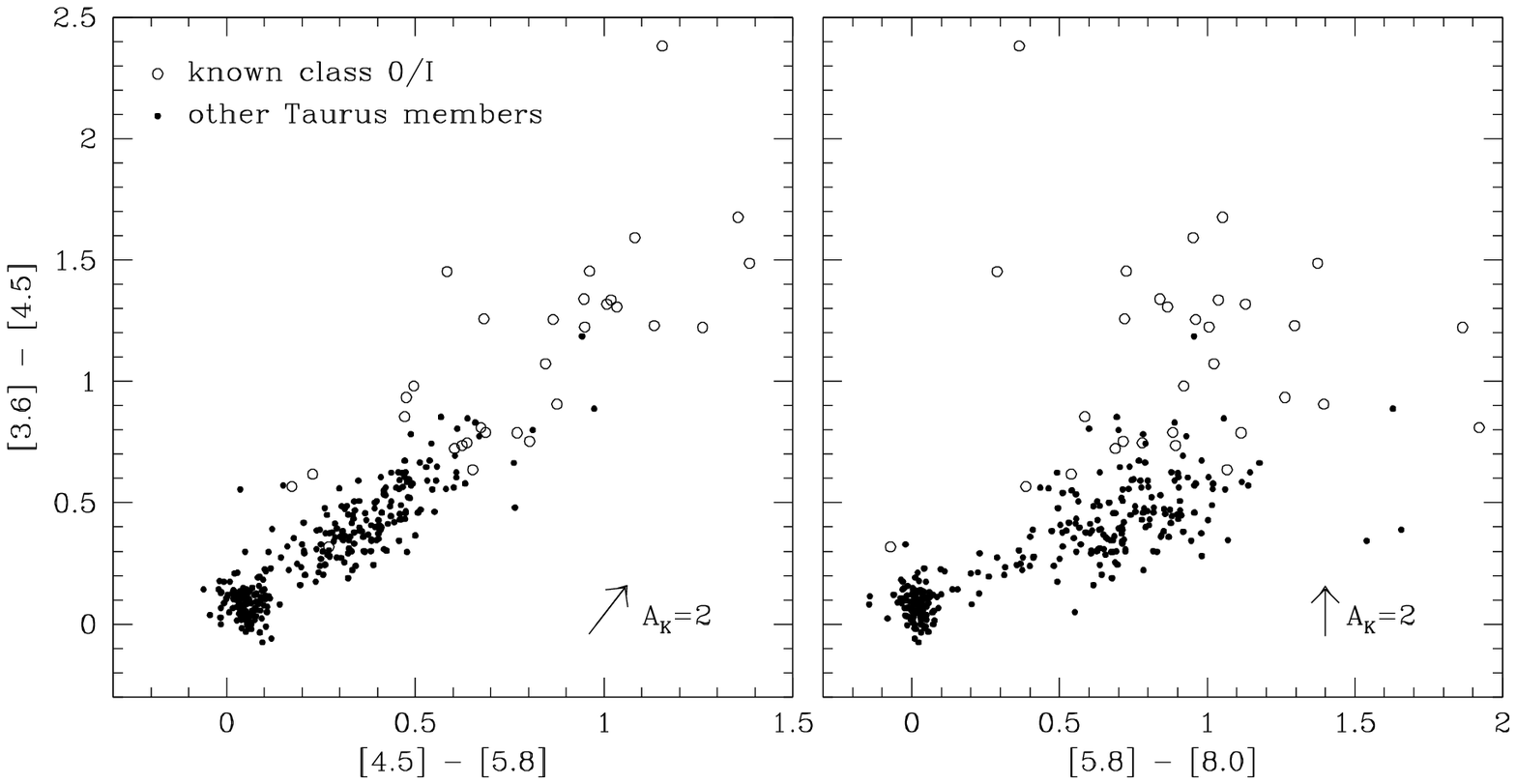}
\caption{
{\it Spitzer} IRAC color-color diagrams for members of the Taurus star-forming 
region. 
Stars that have been previously identified as class~0 and class~I sources 
through mid-IR spectroscopy and other diagnostics are indicated
({\it open circles}, \S~\ref{sec:proto}).
The reddening vectors are based on the extinction law from \citet{fla07}.
}
\label{fig:1234}
\end{figure}

In Figure~\ref{fig:1234}, we present two IRAC color-color diagrams
for the members of Taurus that have been measured in all of the IRAC bands.
Some of the stars reside in a tight group near the origin while others 
form a broad distribution of redder sources, which correspond to stellar 
photospheres and stars with disks, respectively.
As expected, the known class~0/I sources exhibit the reddest colors since
they are surrounded by the both circumstellar disks and envelope, although
a few protostars have relatively blue colors because they appear
nearly edge-on from our point of view (e.g., IRAS~04302+2247).

\begin{figure}
\epsscale{1}
\plotone{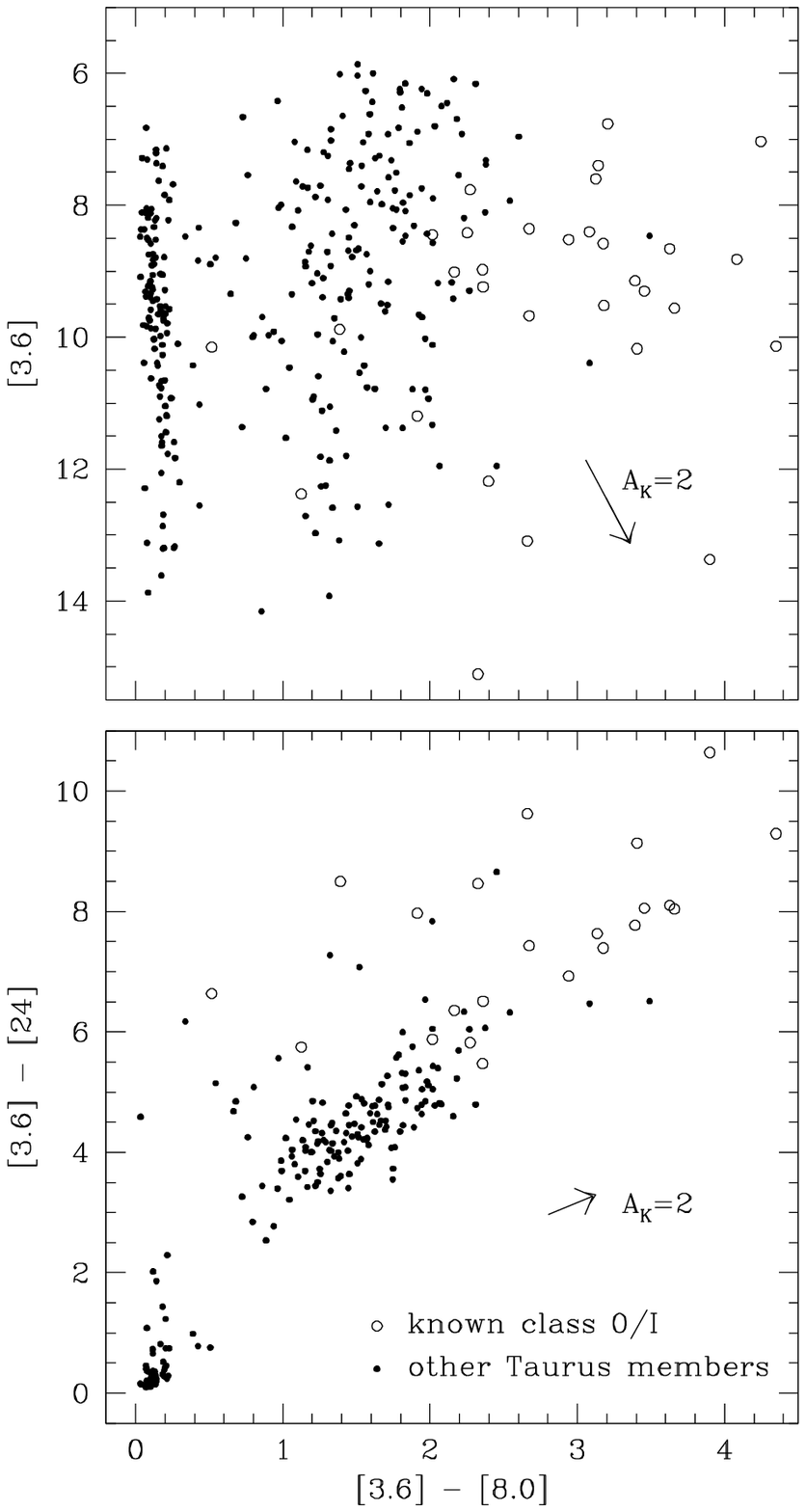}
\caption{
{\it Spitzer} color-magnitude and color-color diagrams for members of the 
Taurus star-forming region. Stars that have been previously identified as 
class~0 and class~I sources through
mid-IR spectroscopy and other diagnostics are indicated ({\it open circles},
\S~\ref{sec:proto}).
The reddening vectors are based on the extinction law from \citet{fla07}.
}
\label{fig:14}
\end{figure}

In Figure~\ref{fig:14}, we plot the members of Taurus in 
a color-color diagram that combines IRAC and MIPS data and
an IRAC color-magnitude diagram.
As in the IRAC color-color diagrams, two distinct populations of 
diskless and disk-bearing stars are apparent in both diagrams. 
The IRAC/MIPS color-color diagram is useful for identifying 
stars that exhibit excess emission at 24~\micron\ but not in the IRAC
bands, which is a signature of a disk with an inner hole. 
Disks of this kind are discussed in more detail in \S~\ref{sec:trans}.
Meanwhile, the color-magnitude diagram in Figure~\ref{fig:14} illustrates
the dependence of mid-IR colors on magnitude, which acts as a proxy for
luminosity, spectral type, and stellar mass. 
At fainter magnitudes, the sequence of diskless stars becomes slightly
redder, reflecting a variation of the photospheric colors with spectral type,
while the average excess at 8~\micron\ of the sources with disks
becomes smaller. The same trends have been found in previous
IRAC measurements for Taurus \citep{luh06tau2}.

\subsection{Classification with Spectral Slopes}
\label{sec:slopes}

Most members of Taurus are easily classified as either class~III or 
class~0/I/II with {\it Spitzer} color-color diagrams.
However, the color-color diagrams do not offer a straightforward means
of dividing disk-bearing stars among classes~0, I, and II.
In addition, some stars have ambiguous colors that are neither neutral nor 
very red, which can arise from diskless stars that are highly reddened or
very cool, or stars that harbor disks in advanced stages of evolution.
We can perform more refined and quantitative classifications by characterizing 
the SED of each star in terms of spectral slopes that are defined as
$\alpha= d$~log$(\lambda F_\lambda)/d$~log$(\lambda)$ 
\citep{adams87,lada87,gre94}, correcting the slopes for reddening, 
and comparing the resulting values to the typical slopes of 
stellar photospheres near the spectral type in question.

As done in Chamaeleon~I \citep{luh08cha1}, we have computed slopes in Taurus
between four pairs of bands, $K_s$/[8.0], $K_s$/[24], [3.6]/[8.0], and 
[3.6]/[24]. For most stars, we use measurements of $K_s$ (2.2~\micron) from the 
Point Source Catalog of the Two-Micron All-Sky Survey \citep[2MASS,][]{skr06}. 
We adopt photometry from multiplicity surveys for a few systems that are
marginally resolved by 2MASS.
As in the color-color diagrams, we have used the weighted average 
in a given band if an object has been observed at multiple epochs.
For binaries that are resolved in $K_s$ and the IRAC bands but not at 
24~\micron, we computed the 24~\micron\ slopes with the total system
fluxes at shorter wavelengths. These systems are noted in Table~\ref{tab:alpha}.
We have dereddened the observed slopes using the extinctions described by
\citet{luh09tau1} and the reddening law from \citet{fla07}, except for
members that lack reliable estimates of extinctions.
The observed and dereddened values of $\alpha_{2-8}$, $\alpha_{2-24}$, 
$\alpha_{3.6-8}$, and $\alpha_{3.6-24}$ are presented in Table~\ref{tab:alpha}.
All known members of Taurus are included in Table~\ref{tab:alpha}, even
those for which none of these slopes could be computed. The classifications
for the latter sources are determined from other data, as discussed at the
end of this section.

\begin{figure}
\epsscale{1}
\plotone{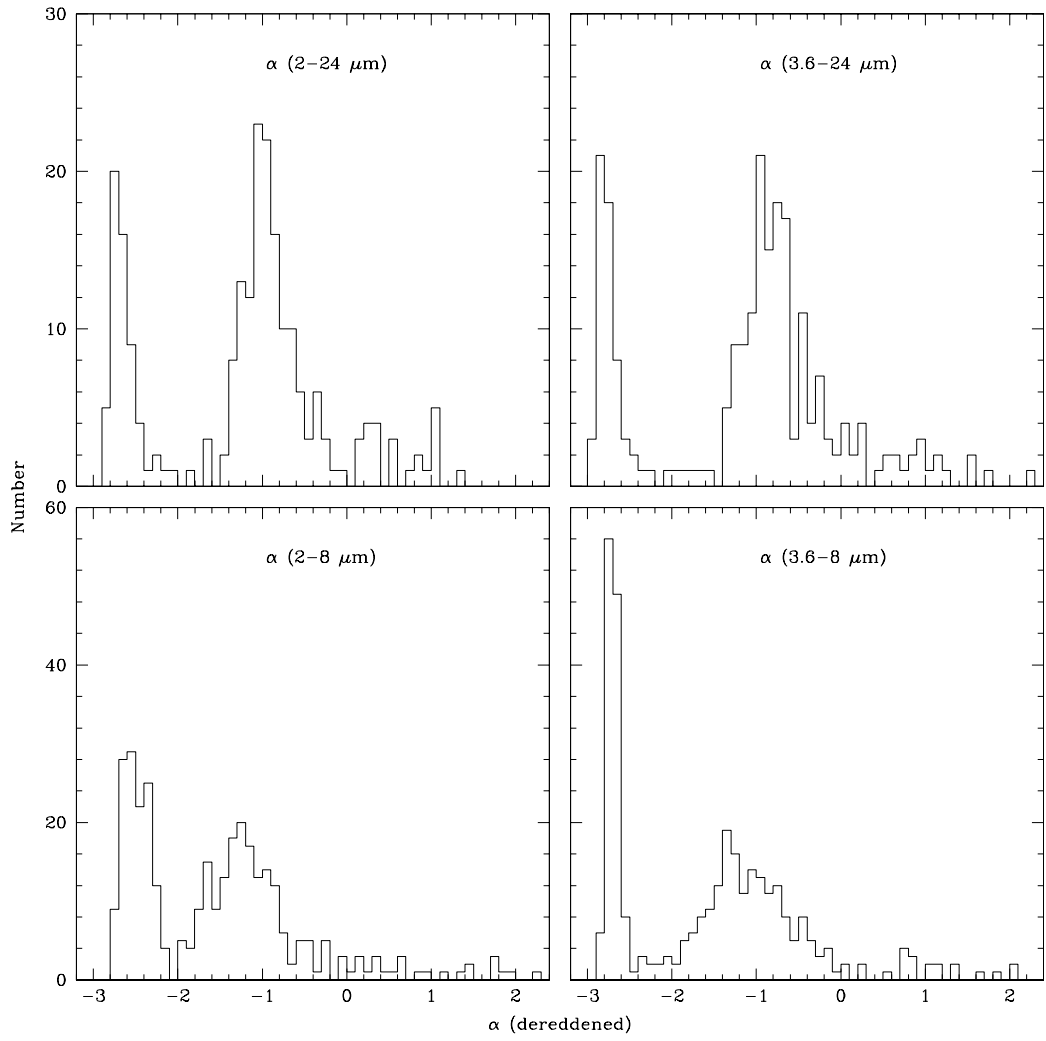}
\caption{
Distributions of spectral slopes for members of Taurus (Table~\ref{tab:alpha}).
}
\label{fig:alpha2}
\end{figure}

\begin{figure}
\epsscale{1}
\plotone{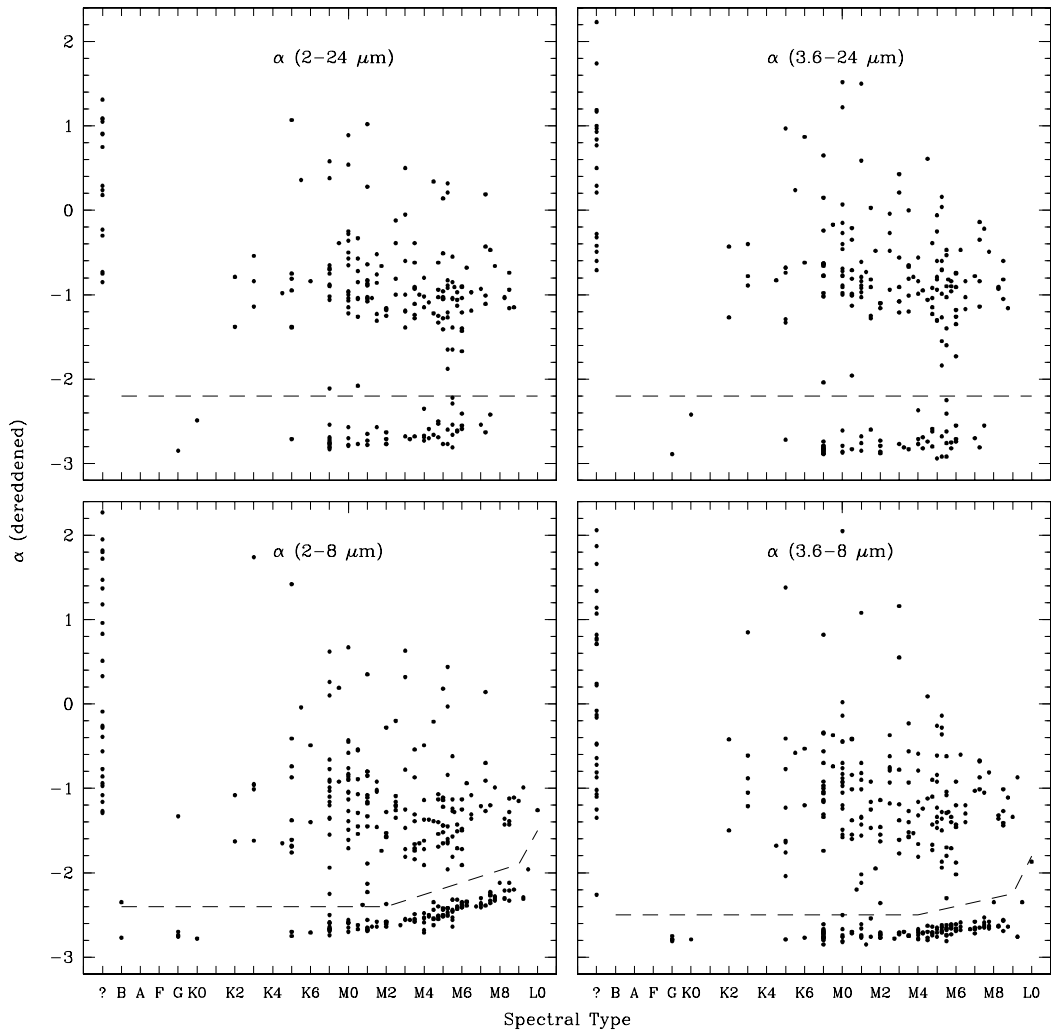}
\caption{
Spectral slopes as function of spectral type for members of Taurus
(Table~\ref{tab:alpha}). Our adopted boundaries for separating 
classes II and III are indicated ({\it dashed lines}).
}
\label{fig:alpha1}
\end{figure}

Histograms of the dereddened spectral slopes from Table~\ref{tab:alpha} are
shown in Figure~\ref{fig:alpha2}. The observed slopes are used for stars 
that lack extinction estimates. Like the data in the color-color diagrams,
the slopes form a narrow group of blue sources and a broader distribution
of redder objects, corresponding to stellar photospheres and stars with disks,
respectively. To better distinguish between these two populations, we plot
the slopes as a function of spectral type in Figure~\ref{fig:alpha1}.
Our adopted spectral types are listed with the spectral slopes in 
Table~\ref{tab:alpha} 
\citep{cow72,har94,ss94,kh95,tor95,wic96,bri98,bri02,ken98,lr98,mal98,duc99,
whi99,mar00,mar01,ste01,wg01,hk03,muz03,wal03,wb03,luh04tau,luh06tau1,
luh03tau,luh06tau2,luh09fu,luh09tau1,gui06,sle06,bec07,pra09}.
The sequences of class~III stars
are narrow and well-separated from the redder population of class~0/I/II 
sources. The class~III sequence is tighter at 8~\micron\ while the separations 
from the disk-bearing stars are larger at 24~\micron. The photospheric slopes
also vary with spectral type, 
particularly for $\alpha_{2-8}$. All of these characteristics apply
to the data for Chamaeleon~I from \citet{luh08cha1}. 
Indeed, we find that the photospheric sequences for Taurus and Chamaeleon~I
are very similar. As a result, we adopt the same thresholds for 
distinguishing between classes~II and III that were defined by 
\citet{luh08cha1}, except that we have used the colors of young field 
L dwarfs from \citet{luh09tau1} to revise the boundaries at L0.

We have used the spectral slopes in Figure~\ref{fig:alpha1} to classify the 
Taurus members that were not already assigned to classes~0 and I in 
\S~\ref{sec:proto} based on previous detections of envelopes.
Stars below the thresholds in Figure~\ref{fig:alpha1} are classified
as class~III while redder stars with $\alpha\leq0$ are designated as class~II 
\citep{lada87}.  Sources with $\alpha>0$ are labeled as ``class I?'' if
no other data are available to verify the presence of an envelope.
Unlike in our study of Chamaeleon~I, we do not employ the ``flat-spectrum"
class \citep[$-0.3\leq\alpha\leq0.3$,][]{gre94} because most Taurus members
within this range of slopes have been observed with mid-IR spectroscopy,
which is sensitive to the presence of envelopes and thus discriminates 
fairly reliably between classes~I and II \citep{fur06,fur08}.
If a star exhibits excess emission at 24~\micron\ but not at 8~\micron,
we classify it as class~II. 
A few objects that are slightly bluer than the II/III thresholds at 
24~\micron\ and are labeled as class~III may have small amounts of excess
emission. The SEDs of these stars are discussed in \S~\ref{sec:trans}.

We discuss in more detail the classifications of several sources,
particularly ones for which the four spectral slopes disagree with each 
other or with classifications derived from previous mid-IR spectroscopy.
Although IRAS~04385+2550, IRAS~04301+2608, IRAS~04260+2642,
CZ~Tau, CoKu~Tau/4, and 2MASS J04210795+2702204 have $\alpha_{3.6-24}>0$, 
we classify them as class~II rather than class~I based on the absence of
signatures of envelopes 
(e.g., strong silicate and ice absorption features) in IRS spectra from 
\citet{dal05}, \citet{fur06}, and unpublished observations in {\it Spitzer} 
programs PID=2, 3303, and 30765 (D. Watson, private communication).
The 8 and 24~\micron\ slopes for 2MASS J04381486+2611399 imply class~I and
class~II, respectively. It is probably a class~II object since 
high-resolution images and mid-IR spectroscopy do not show any evidence 
of an envelope, indicating that it is not a class~I source \citep{luh07edgeon}.
Those data do reveal an edge-on disk, which is the cause of its red SED.
Although the dereddened 24~\micron\ slopes of 2MASS J04194657+2712552 
are slightly below the I/II threshold of $\alpha=0$, we list it as 
``class I?" since it has the reddest SED of any known member of Taurus 
with a measured spectral type later than M6 (excluding the edge-on 
system 2MASS J04381486+2611399). 
For IRAS~04187+1927, $\alpha_{2-8}$ and $\alpha_{3.6-8}$ produce different
classifications (II vs.\ I). Slopes extended to 24~\micron\ are unavailable 
since this star was not observed by MIPS. We have assigned it
to class~II based on IRS data \citep{fur06}. 
J1-4872~A and B are class~II based on $\alpha_{2-8}$ but are class~III
according to the other slopes. We adopt the latter
classification for these stars because of the possibility of variability
between the dates of the $K_s$ and {\it Spitzer} observations.
For the same reason, we classify CIDA-9 using $\alpha_{3.6-8}$ (class~II) 
rather than with $\alpha_{2-8}$ (class~I).
Although 2MASS J04373705+2331080 is slightly redder in $\alpha_{2-8}$ than 
the II/III threshold, this color excess is not reliable given the
large uncertainty in the measurement of $K_s$. 
When compared to 3.6~\micron, the 5.8 and 8.0~\micron\ measurements for
this object do not exhibit significant excess emission relative to stellar 
photospheres \citep{luh09tau1}, indicating that it is a class~III source.
Similarly, V710~Tau~B has a small excess at 8~\micron, but the significance 
of this excess is marginal since the uncertainty in the 8~\micron\ measurement
is large. We classify this star as ``class II?".
IRAS~04173+2812 is class~I and class~II according to the 8 and 
24~\micron\ slopes, respectively. We adopt the 24~\micron\ classification
since the presence of an envelope is more easily detected at longer wavelengths.

Finally, we discuss the classifications of the Taurus members for which 
none of the four spectral slopes in Table~\ref{tab:alpha} could be computed 
because the necessary {\it Spitzer} photometry is not available.
We could not measure photometry at 8 and 24~\micron\ for 
V892~Tau, RY~Tau, T~Tau~N, and AB~Aur because they are either saturated
or outside of the images.  We assign these stars to class~II based
on the mid-IR spectroscopy from \citet{fur06}.
2MASS J04390637+2334179 was observed at 3.6 and 5.8~\micron\ but not
at 4.5 and 8.0~\micron. The photometry in the former two bands and
a non-detection at 24~\micron\ indicate that this star is class~III. 
Six known members of Taurus were not observed by IRAC in any of its bands.
Two of these stars, 2MASS J04381630+2326402 and 2MASS J04385871+2323595,
were within the field of view of MIPS images, but were not detected.
The detection limits at 24~\micron\ are sufficiently faint to demonstrate
that these stars are class~III. 
One star, MWC 480, is known to be a class II source from previous observations
\citep{man97}. The remaining three sources have weak H$\alpha$ emission 
\citep{luh04tau,luh06tau1,luh09tau1,sle06}, suggesting that they are not
undergoing significant accretion.  We classify these stars as ``class~III?".

The SED classifications for all known members of Taurus are presented in 
Table~\ref{tab:alpha}. For multiple systems that are unresolved by IRAC,
these classifications apply to the component that dominates at mid-IR 
wavelengths. However, it is possible for different components to dominate
at different IR wavelength ranges, as in the case of T~Tau~N and S 
\citep{fur06}. For that system, the designation of class~II in 
Table~\ref{tab:alpha} refers to T~Tau~N, whereas the southern component
may be a class~I source \citep{fur06}.

\section{Disk Fraction}
\label{sec:frac}

We would like to use our tabulation of SED classifications in Taurus to 
measure the fraction of members that have disks as a function of stellar mass. 
However, some of the known members of Taurus were originally discovered
because they exhibited evidence of disks. 
As a result, the census of Taurus could be biased in favor of stars with disks,
in which case the disk fraction computed from all known members would not 
be representative of the stellar population. 
To address this possibility, we have measured a disk fraction from
a sample of members that is likely to be unbiased in terms of disks. 
For this sample, we have selected all known members within the 
fields observed during the {\it XMM-Newton} Extended Survey of the Taurus 
Molecular Cloud \citep[XEST,][]{gud07}.
These fields have been searched extensively for new members with a variety of 
methods and offer relatively well-defined completeness limits \citep{luh09tau1}.
As done in Chamaeleon~I \citep{luh08cha1}, we compute the disk fraction 
as a function of spectral type, which acts as a proxy for stellar mass.
For both the XEST fields and all of Taurus, we list in Table~\ref{tab:sed}
the number of members as a function of SED class and spectral type. 
For members that lack measured spectral types, most of which are class~I 
sources, we estimated the spectral type bins in which they belong
by combining their luminosities \citep[e.g.,][]{fur08} with the
values predicted by the evolutionary models for an age of 1~Myr.
The two class~0 sources are excluded from Table~\ref{tab:sed}.

We now consider the question of how to define the disk fraction in a
young population.
The ultimate objective of studies of disk fractions is the measurement of
the average lifetime of disks by comparing disk fractions
among clusters with different ages \citep{hai01}. 
In these comparisons, the ages that are typically adopted for clusters
apply only to the class~II and III members since the class~I sources
usually lack the temperature and luminosity measurements that are needed
for isochronal ages. If class~I sources are younger than stars in
classes II and III (e.g., \S~\ref{sec:spatial}), then it would be 
inappropriate to include them in disk fractions when estimating disk lifetimes. 
Therefore, we define the disk fraction as N(II)/N(II+III) in our analysis
of Taurus. 
In comparison, many previous studies have included both classes I and II
in their disk fractions. Because the frequency of class~I sources decreases
rapidly with cluster age, disk fractions for clusters older than Taurus 
($\tau>1$~Myr) do not depend significantly on the treatment of these objects.
However, the adopted definition does have a noticeable effect on the
disk fractions in the youngest regions.

\begin{figure}
\epsscale{1}
\plotone{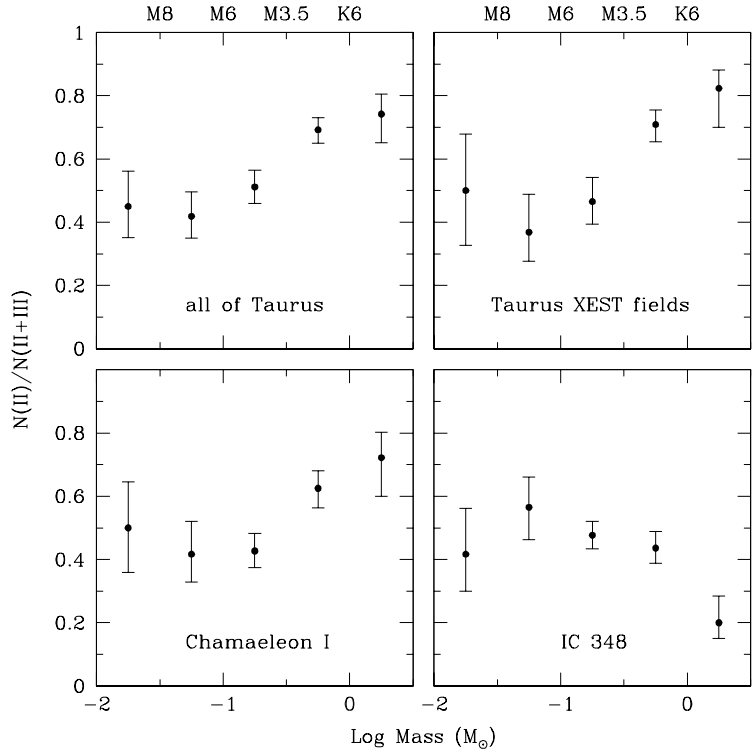}
\caption{
Disk fraction as a function of stellar mass and spectral type among all
known members of Taurus and the members within the XEST fields.
Similar measurements for Chamaeleon~I and IC~348 
are included for comparison \citep{lada06,mue07,luh05frac,luh08cha1}.
}
\label{fig:diskfraction}
\end{figure}

The values of N(II)/N(II+III) for Taurus from Table~\ref{tab:sed} are
plotted versus spectral type in Figure~\ref{fig:diskfraction}.
The boundaries of the spectral type bins 
in Figure~\ref{fig:diskfraction} have been converted to stellar masses
by combining the temperature scale of \citet{luh03ic} and the evolutionary 
models of \citet{bar98} and \citet{cha00}.
As shown in Figure~\ref{fig:diskfraction}, the disk fractions for
the known members in the XEST fields and across all of Taurus are similar. 
If the true disk fraction is the same within and outside of the XEST fields,
then the agreement in these disk fractions suggests that the total census 
of Taurus is not strongly biased for or against stars with disks.
These disk fractions in Taurus exhibit a dependence on spectral type
and stellar mass, declining steadily from $\sim75$\% for solar-mass stars to 
$\sim45$\% for low-mass stars and brown brown dwarfs. 
For comparison, we have included in Figure~\ref{fig:diskfraction} the disk 
fractions measured from {\it Spitzer} observations of Chamaeleon~I and 
IC~348 \citep{lada06,mue07,luh05frac,luh08cha1}. 
The fractions for Taurus and Chamaeleon~I vary with spectral type in a 
similar manner 
while an opposite dependence on spectral type is present in the data
for IC~348. Like Taurus and Chamaeleon~I, roughly half of its low-mass
members have disks, but the disk fraction is much lower among its solar-mass
stars. The $\sigma$~Ori cluster closely resembles IC~348 in this
respect \citep{her07a,luh08sig}.
Given that Chamaeleon~I, IC~348, and $\sigma$~Ori have similar ages
\citep[2--3~Myr,][]{luh07cha,luh08sig}, these differences in disk fractions are
presumably tied to the star-forming conditions of these regions. 
For instance, \citet{luh08cha1} suggested that the low disk fraction at
higher masses in IC~348 relative to Chamaeleon~I might be 
related to the higher stellar density of IC~348 combined with the segregation
of the more massive stars toward the center of the cluster.
The high and low disk fractions for solar-mass stars in Taurus and $\sigma$~Ori,
respectively, provide further support for this hypothesis since the former 
is even more sparse than Chamaeleon~I while the latter is comparable 
in density to IC~348. 
Thus, it appears that the disk lifetimes for stars more massive than 
0.5~$M_\odot$ may be shorter in denser clusters.

Finally, we note that the ratio of the number of class~II sources to the
number of class~I sources in Taurus ($\sim4.4$) is similar to the values
measured in previous studies of Taurus \citep{kh95} and in {\it Spitzer}
surveys of other nearby regions of ongoing star formation
\citep[$\tau<5$~Myr,][]{eva09,gut09}. This ratio is typically interpreted
as the ratio of the lifetimes of these stages. For instance, if 
an average lifetime of 3~Myr is adopted for class~II sources, then the
measurement of N(I)/N(II) in Taurus implies a lifetime of 0.7~Myr for class~I
objects.

\section{Spatial Distributions of SED Classes}
\label{sec:spatial}

In addition to measuring the disk fraction as a function of stellar
mass in Taurus, we can also use our SED classifications to compare the spatial 
distributions of the SED classes. To arrive at meaningful results
in this comparison, the completeness of the stellar census should not 
vary significantly with SED class.  The images of Taurus from {\it Spitzer} 
have demonstrated that the current census has a high level of 
completeness for classes~I and II across most of the cloud 
complex \citep{luh06tau2,luh09tau1}.
The completeness for class~III stars is not well-defined outside 
of the denser stellar aggregates and may be lower than that of the class~I/II
sources in those outer areas, but correcting for such incompleteness
would reinforce the primary result regarding the class~III sources that
we are about to describe, which is that they are more widely distributed
than the less evolved stars.

\begin{figure}
\epsscale{1.2}
\plotone{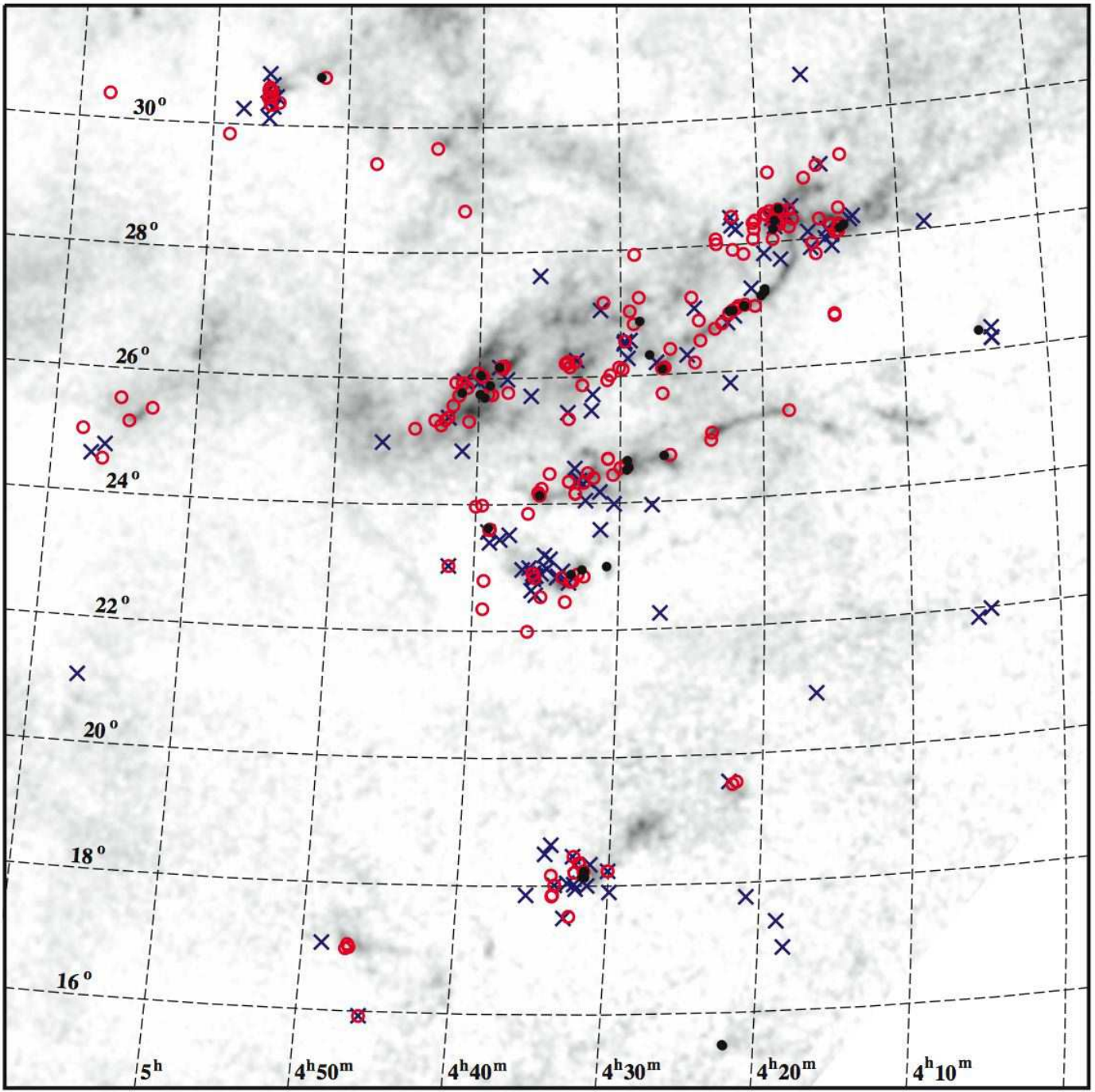}
\caption{
Spatial distributions of classes 0/I ({\it black filled circles}),
II ({\it red open circles}), and III ({\it blue crosses}) in the
Taurus star-forming region.
}
\label{fig:mapclass1}
\end{figure}

\begin{figure}
\epsscale{1.2}
\plotone{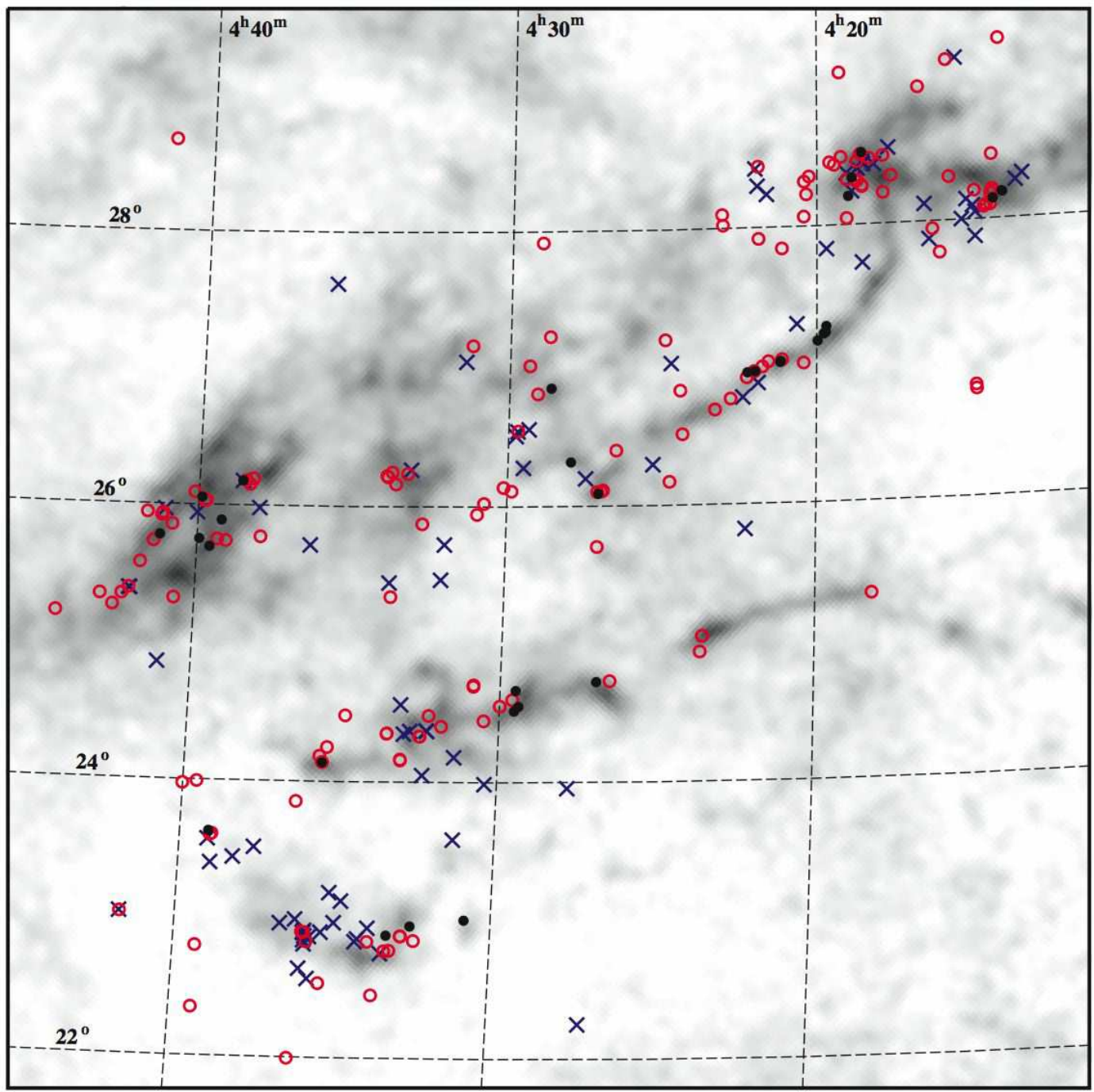}
\caption{
Same as Figure~\ref{fig:mapclass1}, but for the central clouds in Taurus.
}
\label{fig:mapclass2}
\end{figure}

\begin{figure}
\epsscale{1}
\plotone{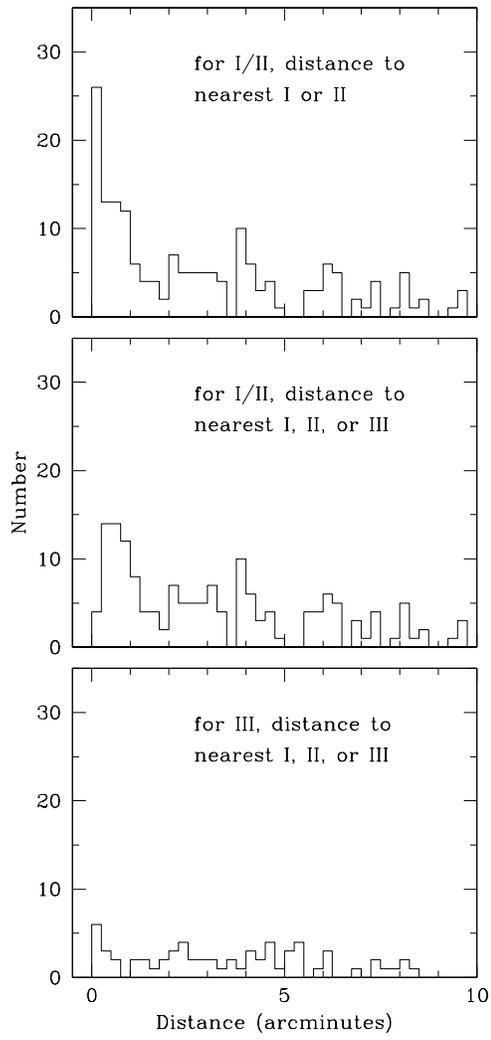}
\caption{
Distributions of projected angular distances to the nearest neighbor for 
different SED classes in Taurus. The bin size is $15\arcsec$, corresponding to 
$\sim0.01$~pc at the distance of Taurus. 
}
\label{fig:dist}
\end{figure}

The locations of Taurus members in classes~I, II, and III are
shown on maps of the cloud complex in Figures~\ref{fig:mapclass1} 
and \ref{fig:mapclass2}.
The class~I/II sources tend to appear near the dense gas, as found 
previously in Taurus and other star-forming regions 
\citep[][references therein]{har02,gut09}, while the class~III stars
are more frequently scattered far from the dark clouds.
To perform a quantitative comparison of these spatial distributions, 
we have computed for each star the distance to the nearest known member
from any SED class. The median of these distances is 2.2, 3.3, and $5\farcm5$
for classes~I, II, and III, respectively, which 
corresponds to 0.09, 0.13, and 0.22~pc at the distance of Taurus.
Histograms of the nearest neighbor distances for classes~I+II and III
are presented in Figure~\ref{fig:dist}. 
According to a two-sided Kolmogorov-Smirnov test, the probability that those 
two samples are drawn from the same parent distribution is 0.3\%.
The probability that classes~I and II are drawn from the same distribution is
10\%, which does not represent a significant difference.

{\it Spitzer} has been used to measure the spatial distributions of young
stars in dozens of other star-forming regions \citep{gut09}. 
Because most of those clusters have not been searched extensively for members
prior to the {\it Spitzer} imaging, and only the disk-bearing members can
be identified with mid-IR photometry, the {\it Spitzer} surveys have primarily
measured the spatial distributions of classes~I and II. 
For instance, the nearest neighbor distances from \citet{gut09} were
computed among class~I/II sources without the inclusion of class~III stars.
To enable a comparison of our results to those of \citet{gut09}, 
we have computed nearest neighbor distances in Taurus in the same manner.
We find that the median of the distance to the nearest member in classes I or
II is 2.2, 4.1, and $3\farcm6$ for classes~I, II, and I+II,
respectively, or 0.09, 0.17, and 0.15~pc at the distance of Taurus.
The class~II sources again appear to be more widely distributed than
the protostars. The probability that the nearest neighbor distances for
classes~I and II are drawn from the same distribution is 2\%. 
Thus, the exclusion of the class~III stars has enhanced the 
difference between the class~I and II distributions.
A marginally significant difference of this kind also was found
by \citet{gut09} in their analysis of 36 young clusters.

The median of the nearest neighbor distances for the class~I+II sample in
Taurus, 0.15~pc, is twice as large as the average median for the clusters
from \citet{gut09}, which is consistent with the measurement of larger
protostellar envelopes in Taurus than in denser clusters like Ophiuchus
\citep{mot98b}. 
For several clusters, \citet{gut09} found that the distributions of nearest 
neighbor distances among classes~I and II peaked near 0.02--0.05~pc, while
the distribution in Taurus reaches a sharp maximum in the lowest bin below
$15\arcsec$, or 0.01~pc, as shown in the top panel of Figure~\ref{fig:dist}. 
This difference is likely a reflection of the close proximity of Taurus
relative to the clusters from \citet[$d=140$--1700~pc]{gut09}, which
allows higher spatial resolution.
To give the Taurus sample a resolution that is comparable to that from
\citet{gut09}, we can count neighboring members in Taurus that are within
$10\arcsec$ of each other as single sources. By doing so, the maximum
in the Taurus distribution moves out to 0.02--0.03~pc and the median
increases by $\sim20$\%.

\citet{gui07} measured {\it Spitzer} photometry for 23 late-type (M5.5--M9.5) 
members of Taurus and used these data to check for spatial variations in
the disk fraction of brown dwarfs.
They found that the disk fraction for low-mass objects in their sample was
lower in the southern filaments than in northern ones.
In comparison, the more massive members did not appear to exhibit the 
same trend. 
\citet{gui07} suggested that the spatial dependence of the disk fraction
for low-mass sources could result from an older age for southern filaments 
combined with shorter disk lifetimes for brown dwarfs.
We have investigated this issue with our SED classifications.
The northern and southern filaments compared by \citet{gui07} can be
separated by a declination of $\delta=25\arcdeg10\arcmin$ (J2000).
We have computed disk fractions on each side of this
boundary for low- and high-mass members, arriving at 
$128/175=0.73\pm0.03$ for $\leq$M6 and $15/30=0.50\pm0.09$ for $>$M6
in the north and $75/121=0.62\pm0.04$ for $\leq$M6 and
$5/22=0.23^{+0.11}_{-0.06}$ for $>$M6 in the 
south\footnote{The statistical errors in the disk fractions are computed in the
manner described by \citet{bur03}.}.
The ratio of southern to northern disk fractions is 
$0.85\pm0.07$ and $0.46^{+0.23}_{-0.14}$ for $\leq$M6 and $>$M6, respectively.
Since these ratios differ by only slightly more than 1~$\sigma$, 
we find that there is no significant evidence to indicate that the disk
fraction of brown dwarfs varies with position in a different manner than
the disk fraction of stars.  The suggestion by \citet{gui07} that disk 
lifetimes are shorter for brown dwarfs also is
not supported by the disk fractions measured in older clusters like 
Chamaeleon~I, IC~348, and $\sigma$~Ori that were described in the previous 
section.

\section{Disk Variability}
\label{sec:var}

A majority of the members of Taurus have been observed more than once with the
cameras on board {\it Spitzer}.  We can use these multi-epoch measurements to 
measure the mid-IR variability of disks in Taurus. 

\begin{figure}
\epsscale{1}
\plotone{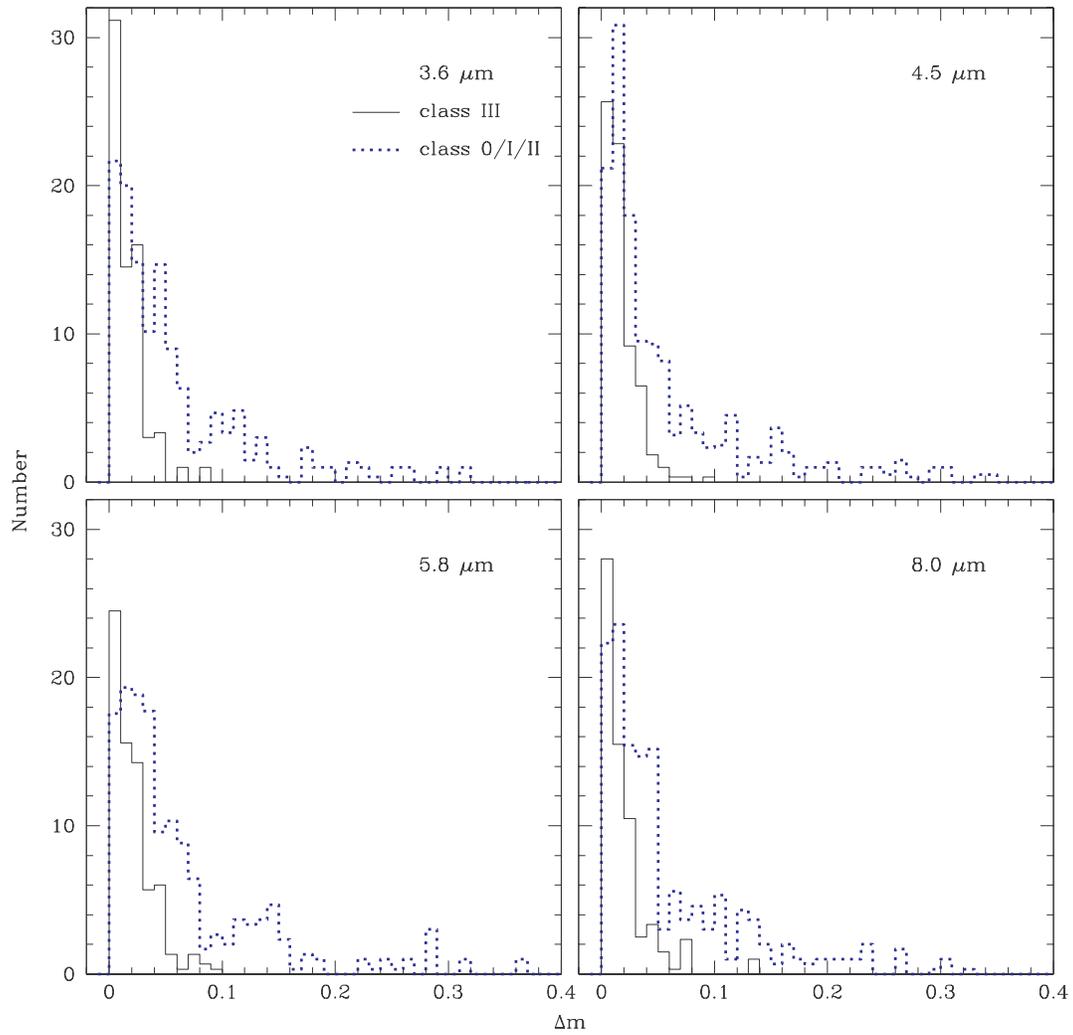}
\caption{
Variability of IRAC magnitudes for members of Taurus that have
disks (classes~0, I, and II, {\it dotted histograms}) and
members without disks (class~III, {\it solid histograms}).
The histograms represent distributions of differences between a magnitude
and the average magnitude for a given source and band. Each magnitude
difference is weighted by the inverse of the number of measurements so that
all members contribute equally.
}
\label{fig:var}
\end{figure}

We have measured photometry at multiple epochs for 201, 206, 214, and 207
members of Taurus at 3.6, 4.5, 5.8, and 8.0~\micron, respectively. 
As done in Chamaeleon~I by \citet{luh08cha1}, we have quantified
the variability of the data in Taurus by computing the difference between 
each magnitude and the average magnitude for a given object and wavelength
($\Delta m$). In Figure~\ref{fig:var}, we present histograms of $\Delta m$
for class~III stars and for sources that are class~0, I, or II.  
Each measurement of $\Delta m$ is weighted by the inverse of the number
of measurements so that all members contribute equally to the distributions.
The data in Figure~\ref{fig:var} demonstrate that stars with disks 
exhibit significantly greater levels of mid-IR variability than diskless stars,
which agrees with variability measurements in Chamaeleon~I \citep{luh08cha1}.
Among stars that have been observed at multiple epochs in at least two bands,
92 of 144 class 0/I/II sources and 16 of 75 class III stars
have $\Delta m>0.05$ in one or more bands. If we require $\Delta m>0.05$ in at 
least two bands, then these fractions become 64/144 and 2/75 respectively. 
Thus, the difference in the variability fractions is more pronounced when 
variability in two bands is required, which is a reflection of the fact
that the different IRAC bands usually vary simultaneously for stars with 
disks but not for class~III stars.

It is useful to compare our variability statistics for Taurus to those
in Chamaeleon~I. We have updated the variability measurements from 
\citet{luh08cha1} by combining the IRAC photometry from that study 
with additional data from \citet{luh08cha2}.
Multiple epochs of photometry are available for 49, 52, 88, and 76 
members of Chamaeleon~I at 3.6, 4.5, 5.8, and 8.0~\micron, respectively.
For sources that have been observed more than once in at least two bands,
15 of 57 class 0/I/II sources and 2 of 48 class III stars 
exhibit $\Delta m>0.05$ in two or more bands.
Thus, the variability fraction for stars with disks is higher in Taurus than 
in Chamaeleon~I ($64/144=0.44\pm0.04$ vs.\ $15/57=0.26^{+0.07}_{-0.05}$).
Since Chamaeleon~I is older than Taurus, this comparison indicates that
the variability of disks decreases with age.

\section{Advanced Stages of Disk Evolution}
\label{sec:trans}

Taurus offers a unique opportunity for studying the advanced stages of disk
evolution. Because Taurus is nearby and does not suffer from crowding or bright 
nebular emission, mid-IR photometry can be measured more accurately for
its stellar population than for most other star-forming regions.
Furthermore, since most members of Taurus have been classified 
spectroscopically, we can reliably estimate the SEDs of 
their underlying stellar photospheres. 
The high quality of both the observed and photospheric SEDs makes it possible 
to detect the small levels of IR excess emission that arise from 
the most evolved disks, even those surrounding low-mass stars and brown dwarfs.
Because the stellar population in Taurus is large and contains significant
numbers of both protostars and diskless stars, the intervening stages of
disk evolution should be sampled with the best available uniformity and
number statistics. 
For these reasons, Taurus was the site of much of the earliest work
on the advanced stages of disk evolution \citep{str89,skr90} and is
the one region to which most other young clusters are compared.

In this section, we use our {\it Spitzer} data in Taurus to 
refine the classification of the evolutionary phases of disks and
to identify the Taurus members that harbor the most evolved disks.
We then apply our classification criteria to other young populations that
have been observed with {\it Spitzer} so that we can constrain the timescale
of disk clearing.

\subsection{Terminology}
\label{sec:term}

A variety of names, definitions, and observational criteria for the stages
of disk evolution have been employed in previous studies.
We describe our adopted terms in this section and 
develop observational criteria for them in \S~\ref{sec:evolve}.

We define {\it primordial disks} as disks that are optically thick at IR
wavelengths and have not experienced inner clearing\footnote{
Alternatively, primordial disks can be defined to include the
transitional and evolved stages, i.e., any disks that are less evolved than
debris disks.}.
Observationally, primordial disks in Taurus form a large, continuous population
in terms of mid-IR color excesses, as demonstrated in \S~\ref{sec:evolve}.
There are several possible mechanisms that may drive the evolution of 
primordial disks, including giant planet formation and the growth and 
settling of dust grains \citep[][references therein]{naj07}.

{\it Pre-transitional disks} are primordial disks that have formed gaps
in which dust is optically thin or absent. 
These disks exhibit less excess emission at
$\lambda\lesssim10$~\micron\ than the average primordial disk but still
retain strong disk emission at longer wavelengths.
Because they are not distinct from the main population of disks
in terms of their mid-IR excess emission (\S~\ref{sec:evolve}), 
we treat pre-transitional disks as a subset of primordial disks.
Examples of pre-transitional disks in Taurus include UX~Tau~A and LkCa~15
\citep{fur06,esp07b,esp08a}.

The IR excesses from members of Taurus do not decrease monotonically
from primordial disks to stellar photospheres.
Instead, stars in Taurus are found predominantly in two groups that are
well-separated from each other in terms of their IR colors \citep{skr90}.
The gap between these populations does contain a few stars, including
those with {\it transitional disks} and {\it evolved disks} 
\citep{skr90,her07a}. Transitional disks have large inner holes, and thus 
produce little or no emission at shorter IR wavelengths followed by a sudden
onset of strong emission at longer wavelengths.
Taurus members that are known to harbor transitional disks include 
CoKu~Tau/4, GM~Aur, and DM~Tau \citep{ric03,for04,qui04,cal05,dal05},
although the inner hole in the disk of CoKu~Tau/4 was probably produced
by a stellar companion rather than processes associated with disk evolution
and planet formation \citep{ire08}.
Meanwhile, evolved disks are full disks without inner holes that are in
the process of becoming optically thin. As a result, they exhibit
weak emission at all mid-IR wavelengths and their spectral slopes
do not show the abrupt increase that is observed in transitional disks.
Other terms used for these disks include anemic, thin, weak, and 
homologously depleted \citep{lada06,bar07,dahm07,cur09a}.

An optically thin disk that contains an inner hole could arise when
an evolved disk is inwardly truncated or a transitional disk becomes
optically thin. We refer to these disks as {\it evolved transitional disks}.
They produce little or no excess emission at $\lambda<10$~\micron\ and
weak emission at longer wavelengths. The SEDs of evolved transitional disks
closely resemble those of {\it debris disks}, which are disks of
second-generation dust that is generated by collisions among planetesimals
\citep{ken05,rie05,her06,cur08,car09}.

\subsection{Classification of Evolutionary Stages}
\label{sec:evolve}

\begin{figure}
\epsscale{1}
\plotone{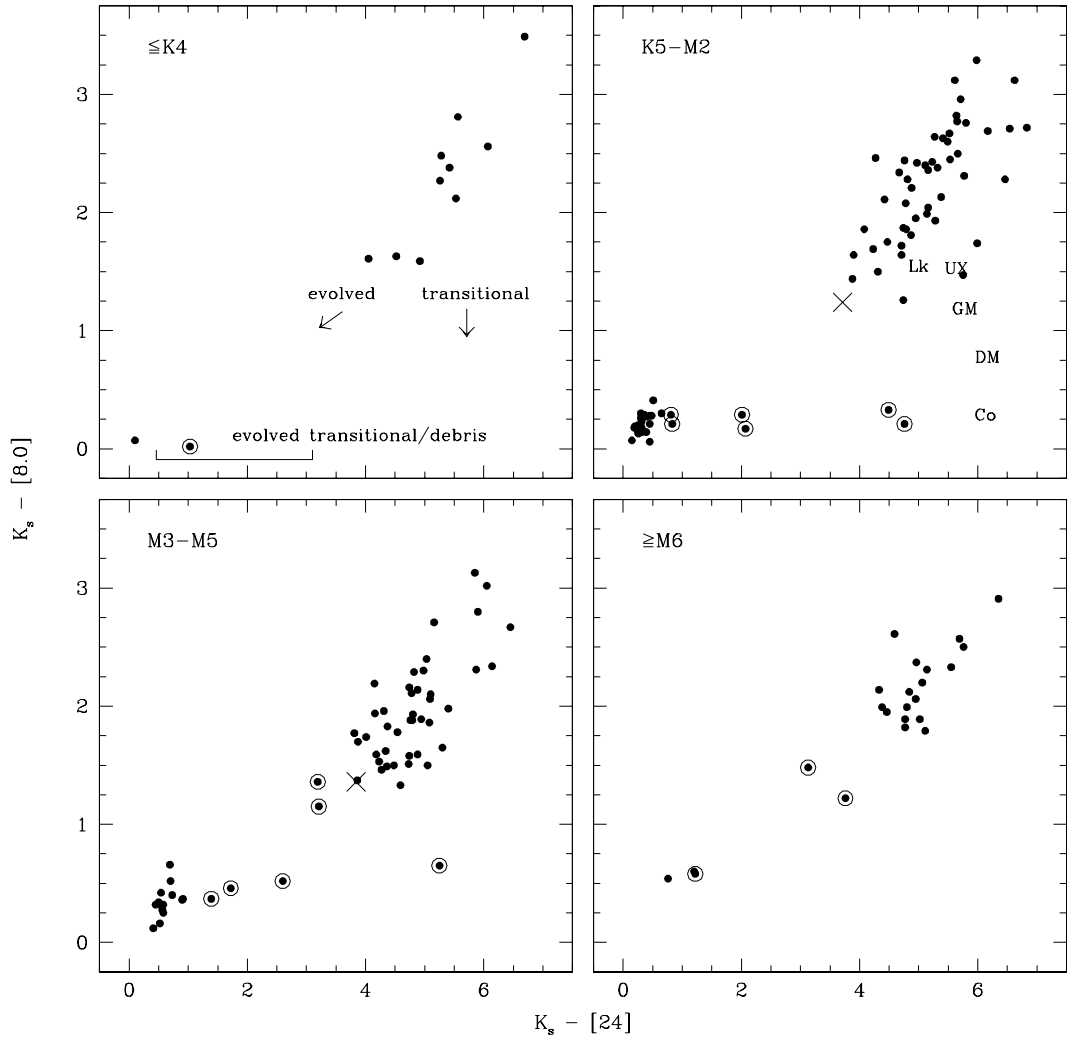}
\caption{
$K_s-[8.0]$ versus $K_s-[24]$ for class~II and class~III 
members of Taurus in four ranges of spectral types.  
The colors predicted by a model of an optically thick disk with a low 
accretion rate and a high degree of dust settling are shown for stellar 
masses of 0.2 and 0.7~$M_\odot$ (M3--M5 and K5--M2, {\it crosses}).
These models demonstrate that the large, continuous population of red colors
can be explained in terms of optically thick disks, which we refer to as
primordial disks. 
In the upper left panel, we indicate the colors expected for optically thick
disks that have developed inner holes (transitional), disks that are becoming
optically thin throughout their inner regions (evolved), optically thin
disks with inner holes (evolved transitional), and disks of
second-generation dust (debris).  
The SEDs of systems that appear to be in these stages 
according to this diagram ({\it circles}) are shown in Figs.~\ref{fig:sed1}, 
\ref{fig:sed2}, and \ref{fig:sed3}.
For reference, abbreviated names are used as the symbols for the 
pre-transitional and transitional systems UX~Tau~A, LkCa~15, GM~Aur, DM~Tau,
and CoKu~Tau/4.
}
\label{fig:k424}
\end{figure}

\begin{figure}
\epsscale{1}
\plotone{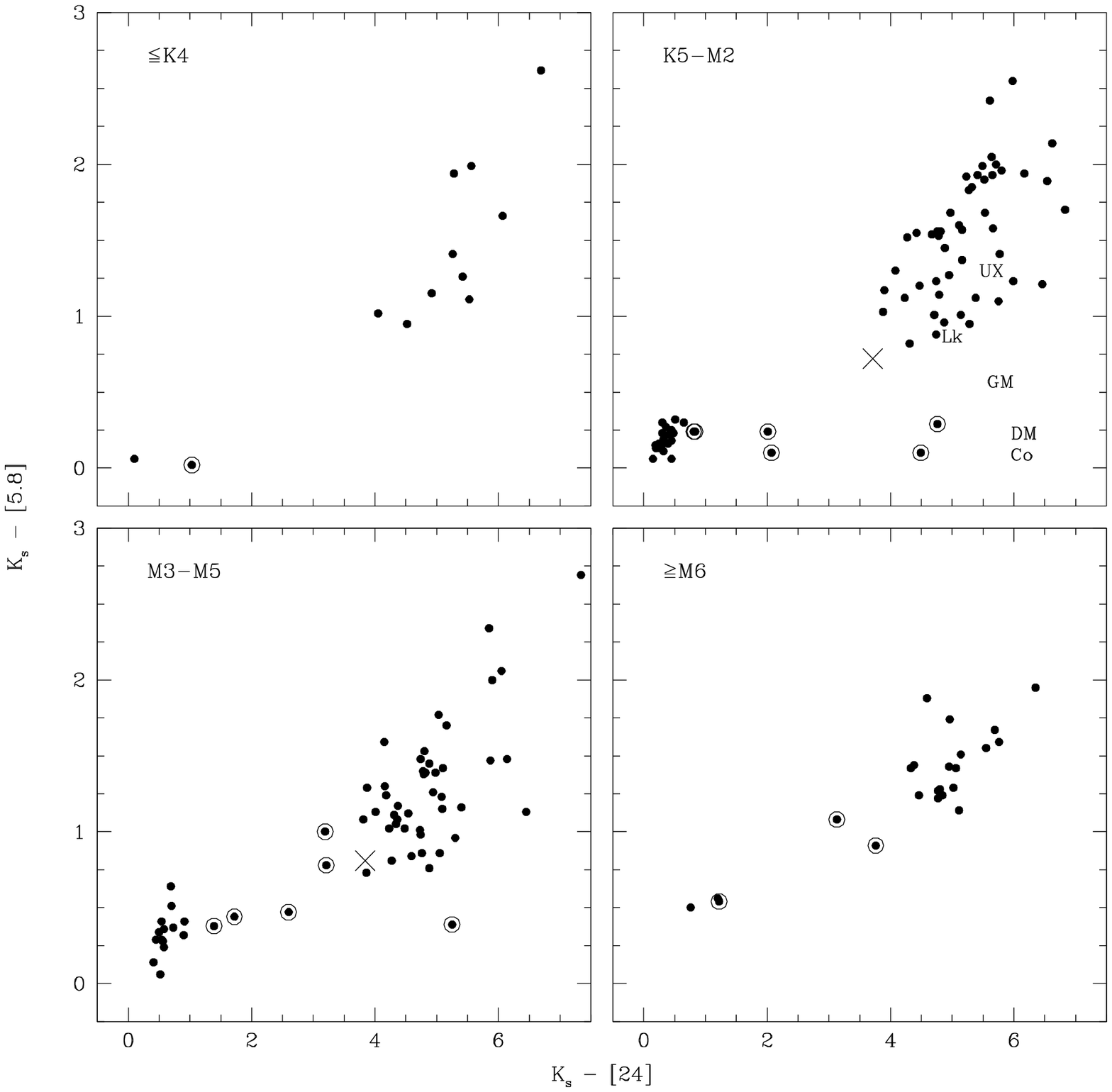}
\caption{
Same as Figure~\ref{fig:k424}, but for $K_s-[5.8]$ versus $K_s-[24]$.
}
\label{fig:k324}
\end{figure}

\begin{figure}
\epsscale{1}
\plotone{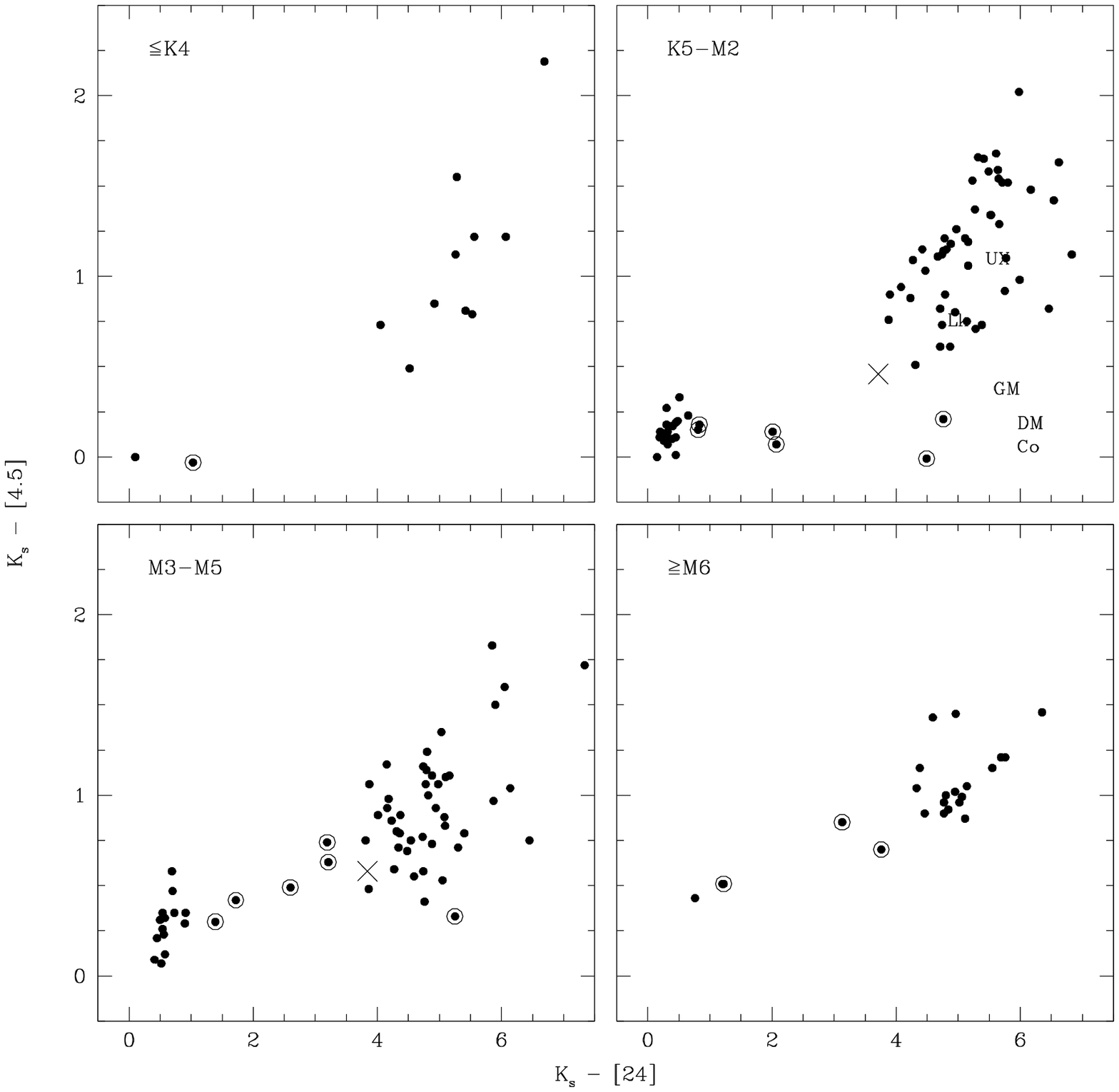}
\caption{
Same as Figure~\ref{fig:k424}, but for $K_s-[4.5]$ versus $K_s-[24]$.
}
\label{fig:k224}
\end{figure}

The stages of disk evolution described in the previous section differ
from each other in terms of the strength of mid-IR excess emission as
a function of wavelength. 
We can characterize these differences by examining excesses at multiple 
wavelengths for all members of Taurus that have been observed by {\it Spitzer}.
In a transitional disk, the wavelength beyond which excess emission appears 
depends on the size of the inner hole. Therefore, we consider a range of 
wavelengths covered by {\it Spitzer}, namely 4.5, 5.8, 8.0, and 24~\micron.
To measure the excess emission in each band, we subtract the {\it Spitzer}
data from photometry at $K_s$ (2.2~\micron), which is short enough in
wavelength that the stellar photosphere usually dominates the total flux while
red enough that extinction is low.  These colors were corrected for extinction
in the manner described for the spectral slopes in \S~\ref{sec:slopes}. 
We plot $K_s-[8.0]$, $K_s-[5.8]$, and $K_s-[4.5]$ versus
$K_s-[24]$ for class~II and III members of Taurus 
in Figures~\ref{fig:k424}, \ref{fig:k324}, and \ref{fig:k224}, respectively.
Since both the average excess emission and the photospheric
colors vary with spectral type \citep[Figure~\ref{fig:alpha1},][]{luh08cha1},
we divide the data into four ranges of spectral types ($sp$), consisting of
$sp\leq$K4.5, K4.5$<sp\leq$M2.5, M2.5$<sp\leq$M5.75, and $sp\geq$M6.
We exclude members that lack measured spectral types and that have
photometric uncertainties greater than 0.1~mag.
The positions of the known pre-transitional and transitional systems
UX~Tau~A, LkCa~15, GM~Aur, DM~Tau, and CoKu~Tau/4 are indicated in the
color-color diagrams. Since DM~Tau was not observed by MIPS, we have
substituted the 25~\micron\ measurement from IRAS. 

As found in previous studies of Taurus \citep{skr90}, the IR colors
in Figures~\ref{fig:k424}, \ref{fig:k324}, and \ref{fig:k224} exhibit two
distinct populations. The gap between them is larger for bands
at longer wavelengths and for earlier spectral types, and is
nearly absent in $K_s-[4.5]$ for late-type stars.
We define the large, continuous population of red sources as primordial
disks and objects within the gaps as evolved disks, transitional disks,
and evolved transitional or debris disks.
The approximate locations of these disks within a gap are indicated in
Figure~\ref{fig:k424}. The colors in which a transitional disk can be
distinguished from the primordial population depends on
the size of its inner hole. For instance, a disk that is primordial
according to its 4.5--24~\micron\ excesses could have a small hole
that produces a deficit of excess emission only at shorter wavelengths.

\subsection{Models of Settled Disks}
\label{sec:model}

We have performed modeling of accretion disks with large degrees of dust
settling to assess whether the sources near the bottom of the primordial
populations in Figures~\ref{fig:k424}--\ref{fig:k224}
can be explained in terms of optically thick disks.
For these calculations, we have employed models of irradiated accretion
disks from \citet{dal05,dal06}. For comparison to the K5--M2 and M3--M5
populations in Taurus, we consider models for two stars that have
spectral types of M0 and M4. We adopt stellar masses of 0.7 and 0.2~$M_\odot$,
effective temperatures of 3850 and 3240~K, and stellar radii of 2.3
and 1.4~$R_\odot$, respectively. 
The model disks are composed of amorphous silicates and graphite grains with
opacities and dust-to-gas mass ratios from \citet{dl84}.
The grain size distribution follows the form $a^{-3.5}$ where $a$ varies
between $a_{min}$ and $a_{max}$.  The upper disk layers contain grains that
are similar in size to those in the interstellar medium
(i.e., $a_{min}$=0.005~{\micron} and $a_{max}$=0.25{\micron}).
In the midplane, the maximum grain size is 1~mm.
We adopt an outer disk radius of 300 AU and the inner disk edge or
``wall'' is located at the radius where the dust sublimation temperature
(1400~K) is reached. The disks are given accretion rates of
10$^{-10}$~$M_\odot$~yr$^{-1}$, a viscosity parameter ($\alpha$)
of 0.01, and a settling parameter ($\epsilon$) of 0.001, where
$\epsilon$ is the dust-to-gas mass ratio in the upper disk layers
relative to the standard dust-to-gas mass ratio.
We selected values for these parameters, particularly $\epsilon$, that
are likely to minimize the IR excess emission for comparison to the
bluest primordial disks in Taurus. 

The IR colors produced by our two models are shown with the data for
K5--M2 and M3--M5 members of Taurus in Figures~\ref{fig:k424}--\ref{fig:k224}.
For both ranges of spectral types, the model colors appear near the
upper boundary of the gaps, indicating that the objects in the main
population of disks above the gap can be interpreted as optically thick disks.
Thus, the observational criteria defined in the previous section (i.e., gap 
boundaries) appear to be physically meaningful and consistent with the 
theoretical distinction between primordial disks and the later stages of
disk evolution (\S~\ref{sec:term}).
In addition, our comparison of the observed and model colors demonstrates
that Taurus contains a substantial number of optically thick disks that 
are highly settled, which is not surprising given that $\sim1/3$ of the
members of Taurus have fully cleared their inner disks (Table~\ref{tab:sed}).

\subsection{New Transitional and Evolved Disks in Taurus}

\begin{figure}
\epsscale{1}
\plotone{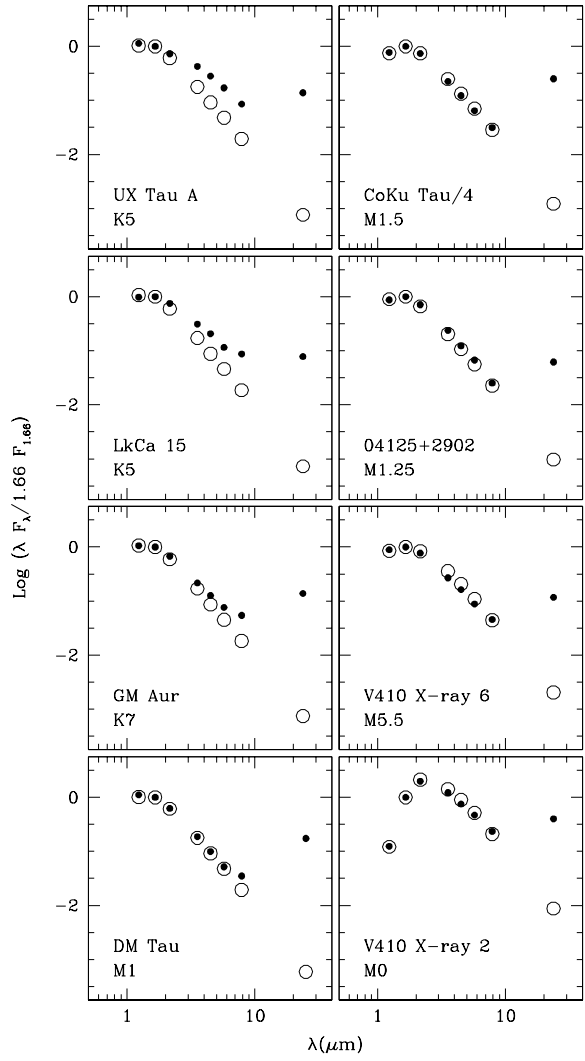}
\caption{
SEDs in Taurus that exhibit large 24~\micron\ excesses but
less excess emission at shorter wavelengths, which indicate the presence of 
disks with gaps and inner holes (pre-transitional and transitional disks).
Each SED is compared to the SED of a stellar photosphere at the same
spectral type ({\it circles}), which has been reddened by the
extinction of the Taurus source and scaled to its $H$-band flux.
}
\label{fig:sed1}
\end{figure}

\begin{figure}
\epsscale{1}
\plotone{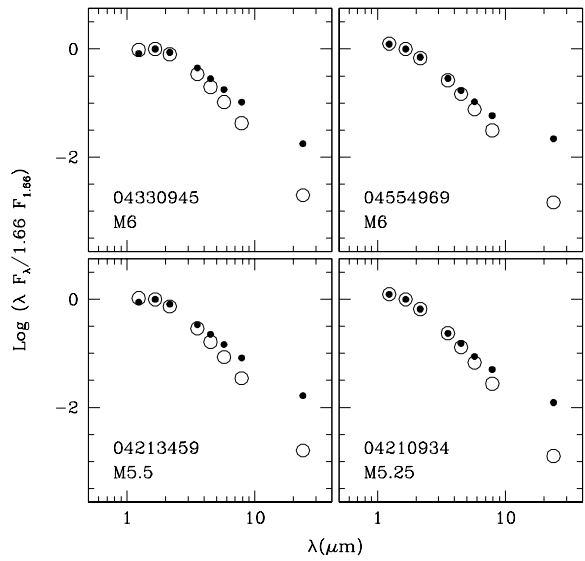}
\caption{
SEDs in Taurus that exhibit small excesses at 8 and 24~\micron,
which indicate the presence of disks that are becoming
optically thin at these wavelengths (evolved disks). 
Each SED is compared to the SED of a stellar photosphere at the same
spectral type ({\it circles}), which has been reddened by the
extinction of the Taurus source and scaled to its $H$-band flux.
}
\label{fig:sed2}
\end{figure}

\begin{figure}
\epsscale{1}
\plotone{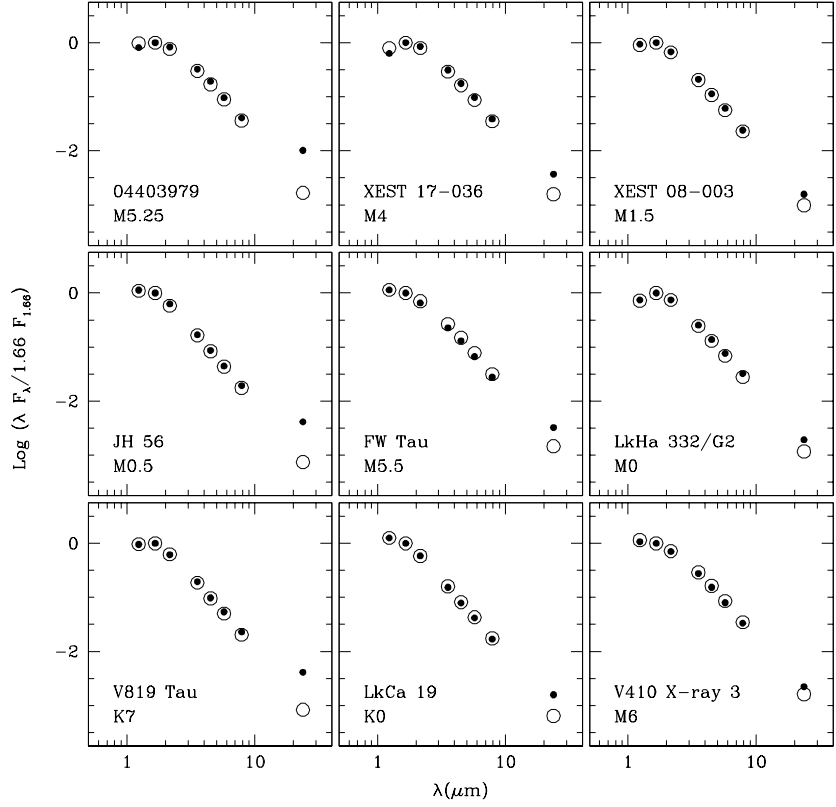}
\caption{
SEDs in Taurus that exhibit no excess emission at $\lambda<10$~\micron\ and
small excesses at 24~\micron, which may indicate the presence of
disks that have inner holes and are becoming optically thin at 
24~\micron\ (evolved transitional disks) or disks that are composed of 
second-generation dust from collisions among planetesimals (debris disks). 
Each SED is compared to the SED of a stellar photosphere at the same
spectral type ({\it circles}), which has been reddened by the
extinction of the Taurus source and scaled to its $H$-band flux.
}
\label{fig:sed3}
\end{figure}

The SEDs of Taurus members that have transitional, evolved, and evolved 
transitional/debris disks according to their 4.5--24~\micron\ colors are 
presented in Figures~\ref{fig:sed1}, \ref{fig:sed2}, and \ref{fig:sed3}.
For reference, we include the pre-transitional systems
UX~Tau~A and LkCa~15 in Figure~\ref{fig:sed1}.
These SEDs consist of our 3.6--24~\micron\ photometry from {\it Spitzer} and
$JHK_s$ data (1.2--2.2~\micron) from the 2MASS Point Source Catalog. 
Since DM~Tau was not observed at 24~\micron\ by MIPS, we have used the
25~\micron\ measurement from IRAS in its SED. 
The sizes of the solid points that comprise the SEDs are equivalent to 
$\pm0.12$~mag. All of the measurements in these SEDs have uncertainties
less than this value. To characterize the excess emission in each SED,
we compare it to an estimate of the SED of the stellar photosphere, 
which is based on the average colors of class~III stars 
near the spectral type in question (\S~\ref{sec:app}).
We have reddened the photospheric SEDs by combining the extinction estimates 
described by \citet{luh09tau1} with the reddening laws from \citet{rl85} 
and \citet{fla07}.

Evidence of evolved and transitional disks has been previously presented for
several of the stars in Figures~\ref{fig:sed1}--\ref{fig:sed3}.
CoKu~Tau/4, GM~Aur, and DM~Tau were observed with IRS spectroscopy
and the resulting data were successfully modeled in terms of disks with 
inner holes \citep{dal05,cal05,fur06}.
Similar analysis was applied to UX~Tau~A and LkCa~15, which
show evidence of disk gaps \citep{fur06,esp07b}.
\citet{fur06,fur09} detected excess emission longward of 12~\micron\ in 
an IRS spectrum of V819~Tau.
\citet{fur06} also reported a tentative detection of excess emission beyond
20~\micron\ in an IRS spectrum of V410~X-ray~3, but that excess is no
longer present in a new reduction of those observations \citep{fur09}.
Nevertheless, we do find that V410~X-ray~3 exhibits a small excess of
$\sim0.4$~mag at 24~\micron, which is below the detection limit of the
IRS data \citep{fur09}.
\citet{and05} suggested that FW~Tau could be a transitional disk based
on a detection at submillimeter wavelengths. The spatial resolution of
those observations is insufficient to determine whether the source
of the submillimeter emission is the $0\farcs08$ binary FW~Tau~A+B or 
their $2\farcs3$ companion FW~Tau~C, which is known to have a disk 
\citep{wg01}. Although MIPS cannot fully resolve a pair with
this separation, the 24~\micron\ emission does appear to be centered on
the former rather than the latter, indicating that FW~Tau~A+B indeed may
be responsible for the submillimeter emission. If so, the disk around
FW~Tau~A+B is an evolved transitional disk or debris disk rather than a
transitional disk based on the small size of the 24~\micron\ excess.
The SED of IRAS~04125+2902 in Figure~\ref{fig:sed1} was also shown
in the study that confirmed its membership in Taurus \citep{luh09tau1}.
The remaining stars in Figures~\ref{fig:sed1}--\ref{fig:sed3} have
not been previously identified as having disks in advanced stages of evolution.
We note that one star in Figure~\ref{fig:sed3} (XEST~17-036) and the stars 
in Figure~\ref{fig:sed3} have 24~\micron\ excesses that are too small to 
be classified as class~II with the II/III thresholds that were 
adopted in Figure~\ref{fig:alpha1}. 

Additional data are needed to definitively assess the nature of 
our new candidates for transitional and evolved disks.
For instance, the SEDs in Figure~\ref{fig:sed3} could arise from both
evolved transitional disks and debris disks.
When SEDs of this kind are observed for early-type stars, they are cited as 
convincing evidence of debris disks since the dust in an (optically thin)
evolved transitional disk would be removed very quickly. However, a
source of dust replenishment may not be necessary to explain the SEDs
in Figure~\ref{fig:sed3} since the dust removal timescale is roughly 
comparable to the age of Taurus for low-mass stars \citep{bac93,cur09b}.
Because debris disks are depleted of gas, measurements that trace gas
can help distinguish between debris disks and evolved transitional disks.
The strengths of the H$\alpha$ emission lines from the candidate evolved
transitional/debris systems do not indicate the presence of active accretion,
and thus allow for the possibility of gas depletion
\citep{str89,ken98,wg01,muz03,luh04tau,luh09tau1}.

Binarity also is relevant to the interpretation of our candidates.
Excess emission that appears only at longer wavelengths can
arise from the disk of an unresolved low-mass companion rather than from
a disk with an inner hole.
Even if the latter is present, the inner truncation of the disk may be caused
by interactions with a stellar companion rather than processes associated
with disk evolution \citep{gue07,ire08}.
Multiplicity measurements are required to address these possibilities.
Finally, millimeter and submillimeter observations would be valuable for 
verifying the presence of disks for the stars with the smallest
24~\micron\ excesses \citep{and05} and for resolving inner holes 
\citep{bro08,dut08,hug07,hug09} while mid-IR spectroscopy would provide
the high-resolution SEDs that are necessary for detailed modeling of the
disks \citep{cal05,dal05,esp07b}.

As with V819~Tau and V410~X-ray~3, \citet{fur06} found excess emission 
at long wavelengths in their IRS spectrum of HBC~427. However,
they suggested that a star at a distance of $15\arcsec$ from HBC~427 may 
have contributed light into the IRS slit during the observations of
HBC~427, resulting in a spurious excess. 
To assess this possibility, we have inspected the MIPS 24~\micron\ images 
of HBC~427. We find that the nearby star cited by \citet{fur06} is
sufficiently faint and well-resolved that it probably did not 
contaminate the IRS spectrum of HBC~427. However, another object that is
$9\arcsec$ from HBC~427, corresponding to HBC~427/1 from \citet{mas05}, is 
$\sim1.3$~mag brighter than HBC~427 at 24~\micron\ and very likely is
the source of the apparent excess emission in the IRS data. 
This conclusion is supported by the fact that our 24~\micron\ photometry
of HBC~427 does not exhibit excess emission. The contaminating source 
is probably a galaxy based on its near- and mid-IR colors.
\citet{fur09} arrived at the same conclusion through analysis of the MIPS
data as well as peak-up images from IRS.

\subsection{Comparison to Clusters at 2--10 Myr}
\label{sec:compare}

\begin{figure}
\epsscale{1}
\plotone{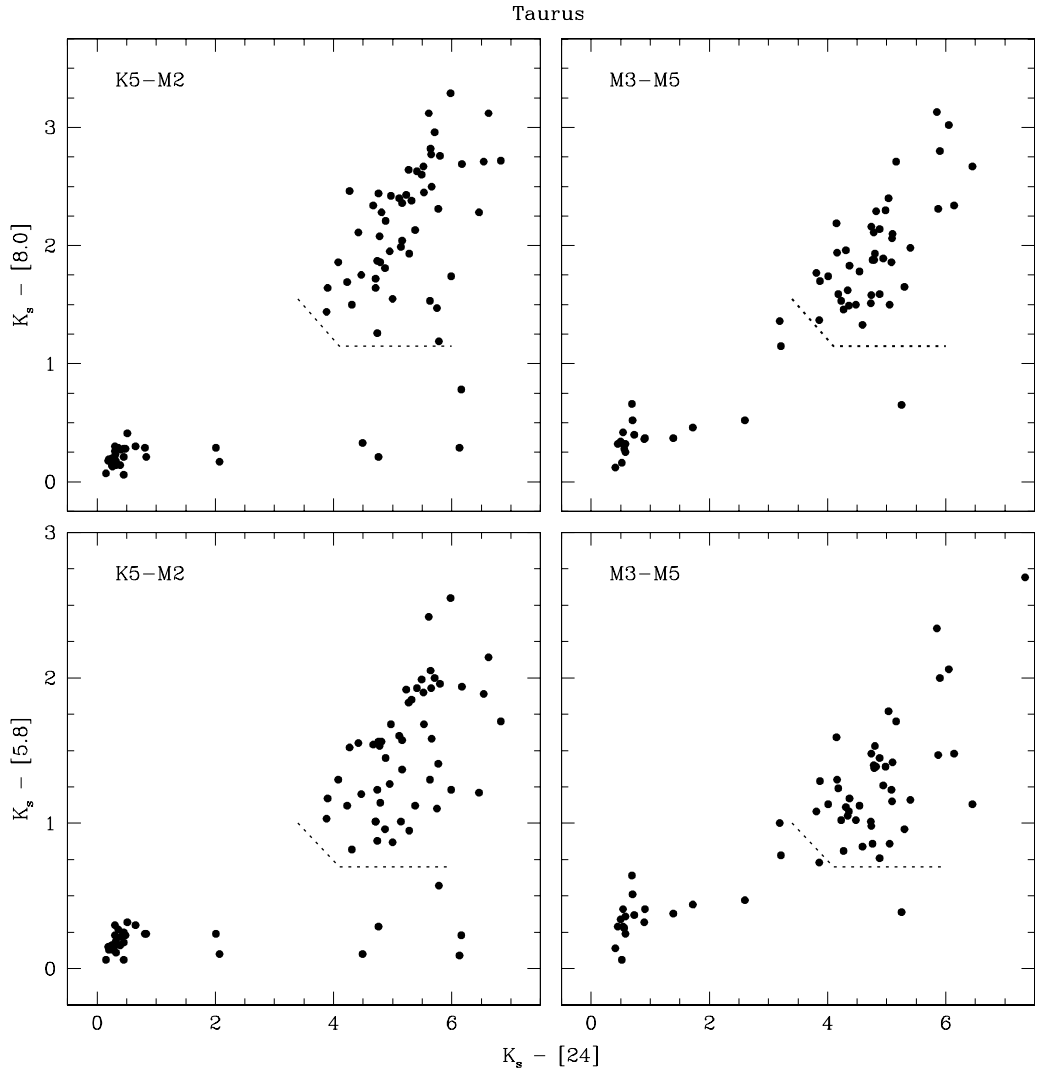}
\caption{
$K_s-[8.0]$ and $K_s-[5.8]$ versus $K_s-[24]$ for K5--M2 and
M3--M5 members of Taurus ($\tau\sim1$~Myr). We have marked the lower boundary
of the primordial disks ({\it dotted lines}), which is plotted
with data for other clusters in Figs.~\ref{fig:colcha}--\ref{fig:colus}.
}
\label{fig:coltau}
\end{figure}

Imaging with the {\it Spitzer Space Telescope} has been widely used to 
classify the evolutionary stages of disks in nearby clusters and associations.
We can apply the observational criteria developed for Taurus to 
these populations to provide the uniformly-derived classifications that are
necessary for meaningful comparisons among them.
We have selected clusters and associations that were observed with both IRAC
and MIPS down to spectral types of $\sim$M2 ($M\sim0.5$~$M_\odot$) and that are 
older than Taurus but within the era of primordial disks ($\tau\sim2$--10~Myr).
For each region, we will plot $K_s-[5.8]$ and $K_s-[8.0]$ versus $K_s-[24]$ 
for spectral types of K5--M2 and M3--M5. The {\it Spitzer} data for
a few clusters do not reach the latter range of types.
We begin by showing the colors for the class~II and class~III sources
in Taurus in the same format in Figure~\ref{fig:coltau}.
We have defined boundaries in Figure~\ref{fig:coltau} that follow the
upper edges of the color gaps in Taurus, which will be used for
identifying transitional and evolved disks in the other clusters.
We find that the same boundaries are adequate for the two ranges of spectral 
types. These boundaries are defined as lines connecting ($K_s-[5.8]$, 
$K_s-[8.0]$, $K_s-[24])=(1,1.55,3.4)$,(0.7,1.15,4.1), and (0.7,1.15,6).

\subsubsection{Chamaeleon I}

\begin{figure}
\epsscale{1}
\plotone{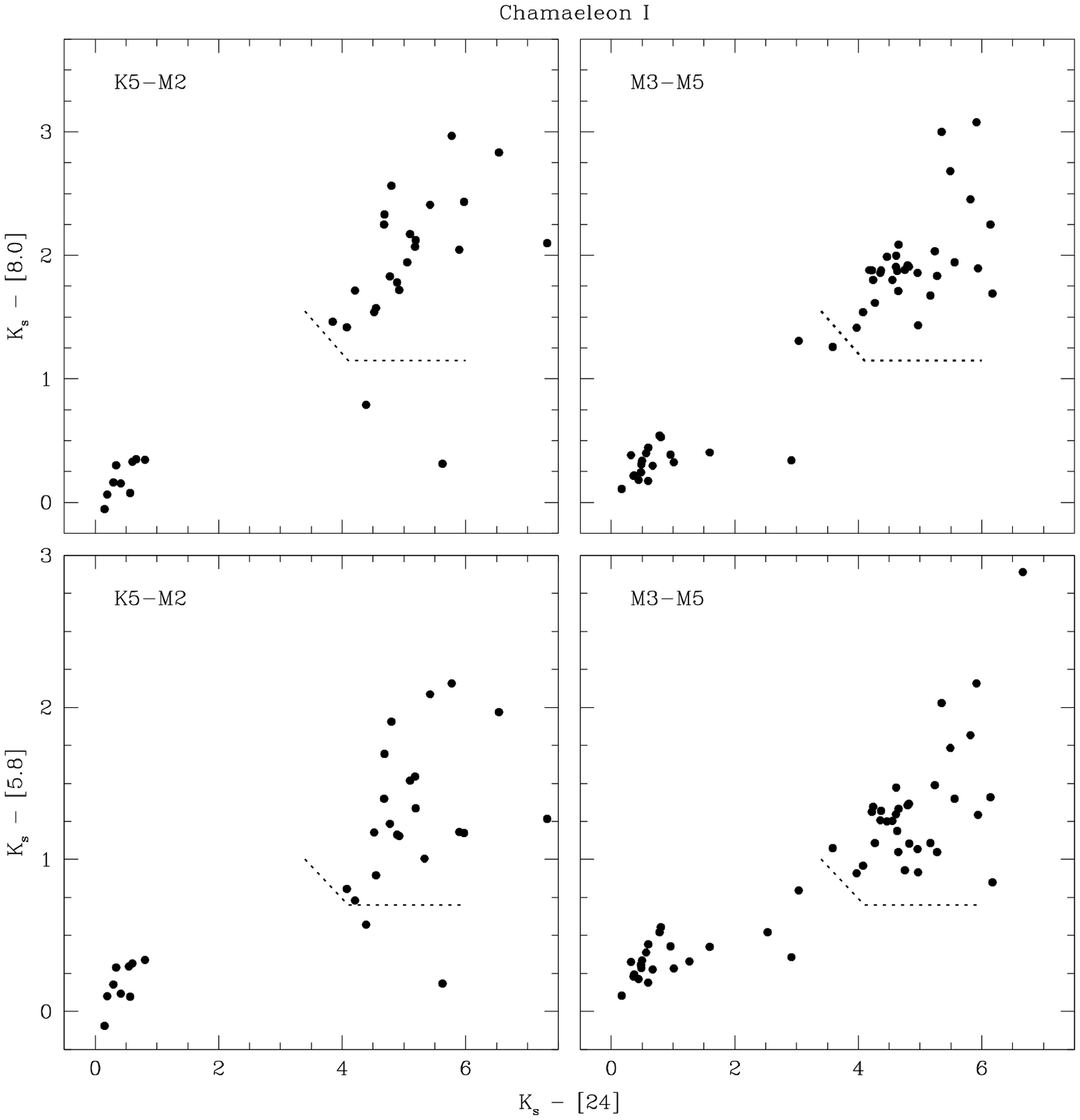}
\caption{
$K_s-[8.0]$ and $K_s-[5.8]$ versus $K_s-[24]$ for K5--M2 and M3--M5 
members of Chamaeleon~I \citep[$\tau\sim2-3$~Myr,][]{luh08cha1,luh08cha2}.
The lower boundary of the primordial disks in Taurus 
is indicated ({\it dotted lines}).
}
\label{fig:colcha}
\end{figure}

The dereddened values of $K_s-[5.8]$, $K_s-[8.0]$, and $K_s-[24]$ for K5--M2 
and M3--M5 members of Chamaeleon~I \citep[$\tau\sim2$--3~Myr,][]{luh07cha}
are plotted in Figure~\ref{fig:colcha}. The {\it Spitzer} data are from 
\citet{luh08cha1} and \citet{luh08cha2} and the extinction estimates that were 
used for dereddening the colors are from \citet{luh07cha} and \citet{luh08cha2}.
We adopt the $K_s$ data from 2MASS and \citet{luh07cha}.
As shown in Figure~\ref{fig:colcha}, the distribution of colors in Chamaeleon~I
closely resembles that in Taurus.  The same gap in colors is present in both 
regions.
Several stars below the boundaries in Figure~\ref{fig:colcha}
have been previously recognized as possible transitional systems, consisting of
CS~Cha, T35, C7-1, CHXR~22E, CHXR~71, and CHXR~76 
\citep{tak03,dam07,esp07a,luh08cha1,kim09}.
Using our terminology (\S~\ref{sec:term}, Figure~\ref{fig:k424}) and the colors
in Figure~\ref{fig:colcha}, we classify CS~Cha, T35, and CHXR~22E
as transitional disks, C7-1 and CHXR~71 as evolved disks, and 
CHXR~76 as an evolved transitional disk or a debris disk.
In addition, we identify CHSM~9484 as a new candidate for an evolved disk.
The ratio of the number of transitional and evolved disks to the number
of primordial disks in Figure~\ref{fig:colcha} is 7/56, which is similar 
to the value of 15/98 in Taurus for K5--M5.
SZ Cha, T21, T25, T54, and T56 also exhibit evidence of transitional disks
\citep{kim09}, but are absent from Figure~\ref{fig:colcha} because measurements 
with IRAC or MIPS are unavailable, or they have spectral types earlier than K5.

\subsubsection{IC~348}

\begin{figure}
\epsscale{1}
\plotone{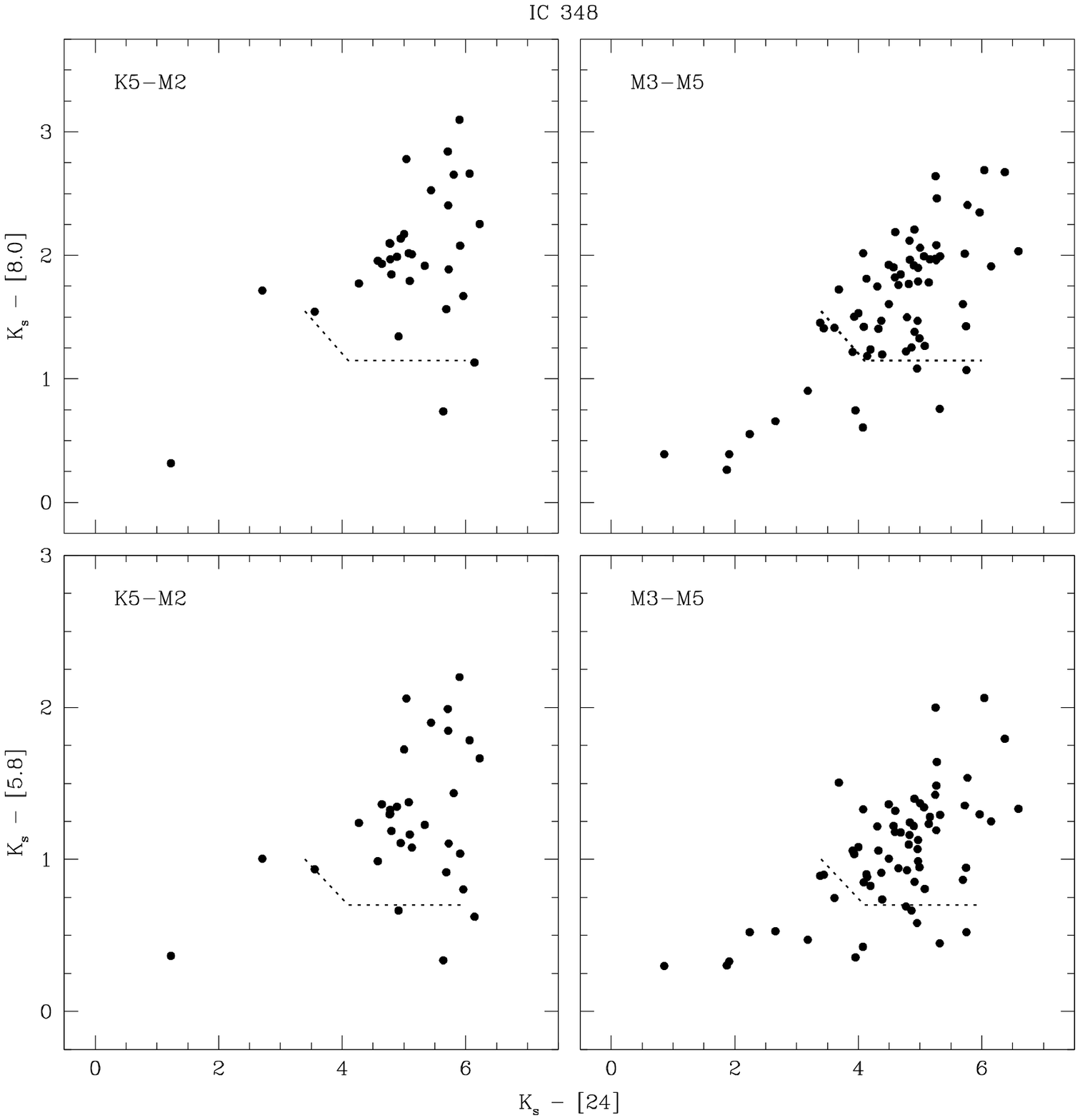}
\caption{
$K_s-[8.0]$ and $K_s-[5.8]$ versus $K_s-[24]$ for K5--M2 and
M3--M5 members of IC~348 \citep[$\tau\sim2-3$~Myr,][]{lada06,mue07}.
The lower boundary of the primordial disks in Taurus 
is indicated ({\it dotted lines}).
}
\label{fig:colic348}
\end{figure}

The dereddened colors for K5--M2 and M3--M5 members of IC~348
\citep[$\tau\sim2$--3~Myr,][]{luh03ic} are shown in 
Figure~\ref{fig:colic348}. The {\it Spitzer} measurements and extinction
estimates are from \citet{lada06}, \citet{mue07}, and \citet{cur09b}
and the $K_s$ data are from 2MASS.
The sensitivity of the 24~\micron\ images of IC~348 is lower than
that in Taurus and Chamaeleon~I because of brighter background emission.
As a result, those images do not detect most of the stellar photospheres
for spectral types later than K5, making it more difficult to characterize
the distribution of colors.  However, the available data that are shown
in Figure~\ref{fig:colic348} are consistent with the presence of a gap
in colors that is similar to the ones observed in Taurus and Chamaeleon~I.
According to Figure~\ref{fig:colic348}, sources 67, 72, 133, and 190 
from \citet{luh03ic} have transitional disks and source 176 from that
study has an evolved disk. Source 72 was previously noted as a possible
transitional disk by \citet{lada06}. Several additional stars appear below the
boundaries in Figure~\ref{fig:colic348}, but their candidacy as
evolved disks or transitional disks is questionable since the uncertainties
in their 24~\micron\ data are large \citep[0.1--0.5~mag,][]{cur09b}.
\citet{cur09b} suggested that the frequency of transitional and evolved disks
is higher in IC~348 than in Taurus. However, we find that the frequencies
are consistent with each other when we apply the same observational 
criteria to both regions. For instance, even if the IC~348 members with
uncertain 24~\micron\ photometry are included, the ratio of the number of 
transitional and evolved disks to the number of primordial disks is 15/98 in
Taurus and 20/83 in IC~348 for spectral types of K5--M5 based on the 
thresholds adopted in Figure~\ref{fig:colic348}.

\subsubsection{$\sigma$ Ori}

\begin{figure}
\epsscale{1}
\plotone{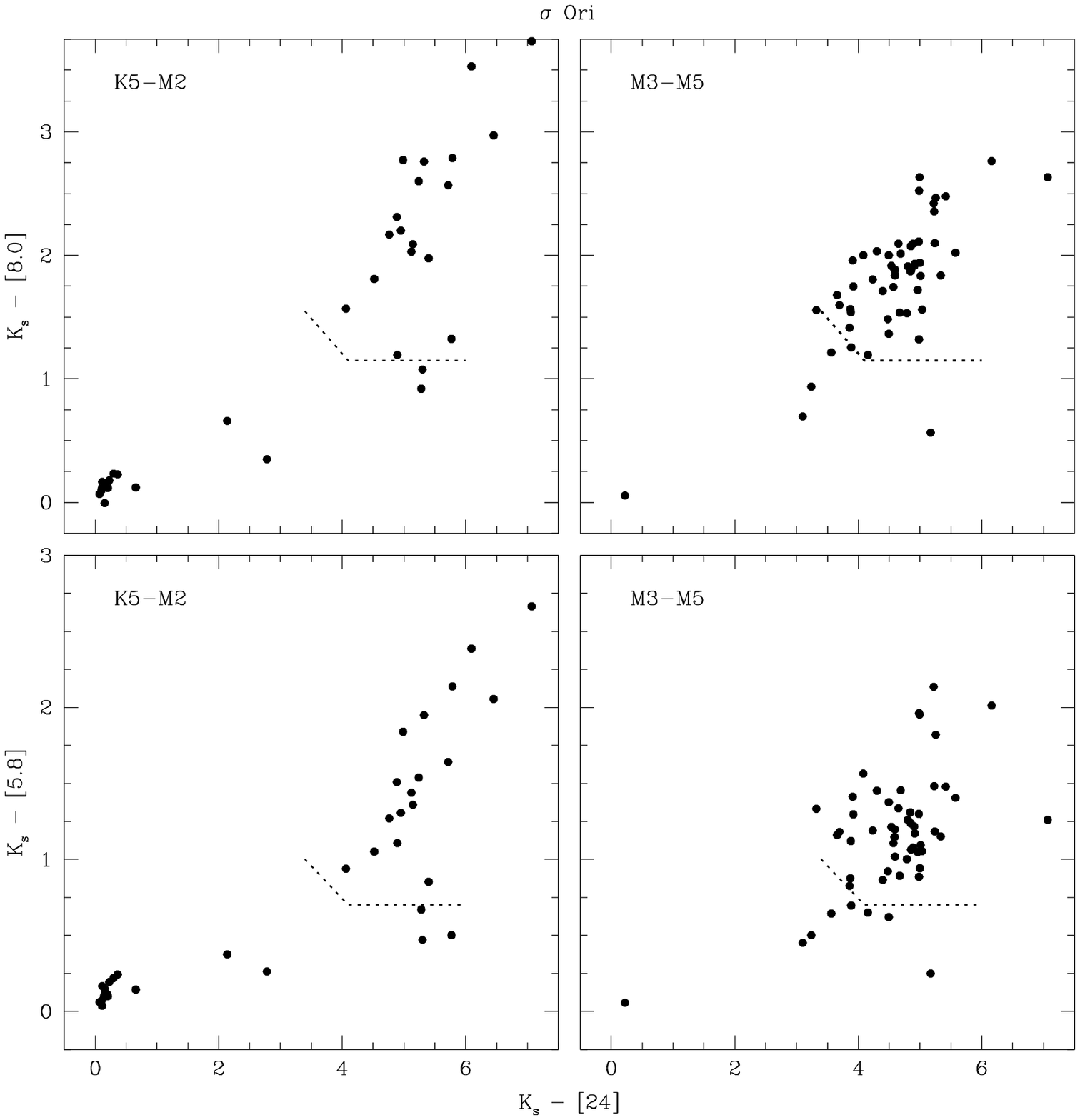}
\caption{
$K_s-[8.0]$ and $K_s-[5.8]$ versus $K_s-[24]$ for K5--M2 and
M3--M5 members of $\sigma$~Ori \citep[$\tau\sim2-3$~Myr,][]{her07a,luh08sig}.
The lower boundary of the primordial disks in Taurus 
is indicated ({\it dotted lines}).
}
\label{fig:colsig}
\end{figure}

For the $\sigma$~Ori cluster 
\citep[$\tau\sim3$~Myr,][references therein]{her07a}, we have adopted 
the membership list that was compiled by \citet{luh08sig}. 
The colors measured in that study and by 2MASS for K5--M2 and M3--M5 stars
are plotted in Figure~\ref{fig:colsig}. We have not corrected the colors
for extinction since the cluster exhibits little reddening.
Since many members of $\sigma$~Ori have not been spectroscopically 
classified, we assume that stars with $2.1<V-J<3.5$ and $3.5\leq V-J<5.7$ 
have types of K5--M2 and M3--M5, respectively. For stars that lack $V$-band 
measurements, we apply criteria of $J-K_s>0.7$ and $J<11.7$ for K5--M2 and 
$J-K_s>0.7$ and $11.7\leq J<14.5$ for M3--M5. We adopt measurements of 
$V$ and $J$ from \citet{her07a} and 2MASS, respectively.

The distribution of colors in Figure~\ref{fig:colsig} for $\sigma$~Ori is
similar to that of Taurus, showing a prominent gap between stellar photospheres
and the main population of disks.  Our classifications of transitional and 
evolved disks differ from those of \citet{her07a} in only a few cases.
\citet{her07a} classified sources 540 and 1156 from their study as class~II
while we identify them as possible transitional disks based on 
Figure~\ref{fig:colsig}. Inspection of the SEDs of these stars from
\citet{her07a} tends to confirm our designations.
Sources 615, 818, and 1267 appear above our adopted thresholds in
Figure~\ref{fig:colsig} but were listed as evolved or transitional systems by 
\citet{her07a}. The latter two stars exhibit color excesses in all of the
IRAC bands relative to the $K_s$ but not in $[3.6]-[4.5]$ and $[4.5]-[5.8]$,
which can be explained by variability between the 2MASS and {\it Spitzer} 
observations. Thus, these two stars probably do have transitional disks
as found by \citet{her07a}.

\subsubsection{Tr 37}

For Trumpler 37 \citep[Tr 37, $\tau\sim4$~Myr,][]{sic05}, we have adopted the 
membership list, extinction estimates, and spectral types that were presented
by \citet{sic05}, the {\it Spitzer} photometry measured by \citet{sic06a},
and the $K_s$ data from 2MASS.
The dereddened colors for K5--M2 members of Tr~37 are shown in
Figure~\ref{fig:coltr}. We have omitted stars for which extinctions were
not estimated by \citet{sic05}. 

The data in Figure~\ref{fig:coltr} are incomplete at bluer colors because
of the detection limit of the 24~\micron\ image, which does not reach stellar
photospheres between K5 and M2. Nevertheless, the available colors
are consistent with a gap between primordial disks and photospheres
that is similar to the one in Taurus.
\citet{sic06a} suggested that the median SED for stars with disks
in Tr~37 exhibits less excess emission than the median SED in Taurus.
However, we find that the median SEDs do not differ significantly when
we utilize our photometry in Taurus.
For instance, the dereddened colors for class~II K5--M2 stars have
median values of $K_s-[5.8]=1.38$ and 1.50, $K_s-[8.0]=2.20$ and 2.06, 
and $K_s-[24]=5.16$ and 5.11 for Taurus and Tr~37, respectively.

\citet{sic06a} defined transitional objects as stars that have photospheric
colors at $\lambda\leq4.5$~\micron\ and excess emission at longer wavelengths.
They identified as many as 14 sources of this kind in Tr~37. 
Eleven of these stars were not detected at 24~\micron\ and one additional
star is outside of the range of spectral types that we are considering.
As a result, these 12 sources are absent from Figure~\ref{fig:coltr}.
Among the remaining two candidate transitional disks from \citet{sic06a},
we classify one as a primordial disk (14-11) and the other as
a transitional disk (13-52) based on our adopted criteria.

\subsubsection{NGC 2362}
\label{sec:2362}

\begin{figure}
\epsscale{1}
\plotone{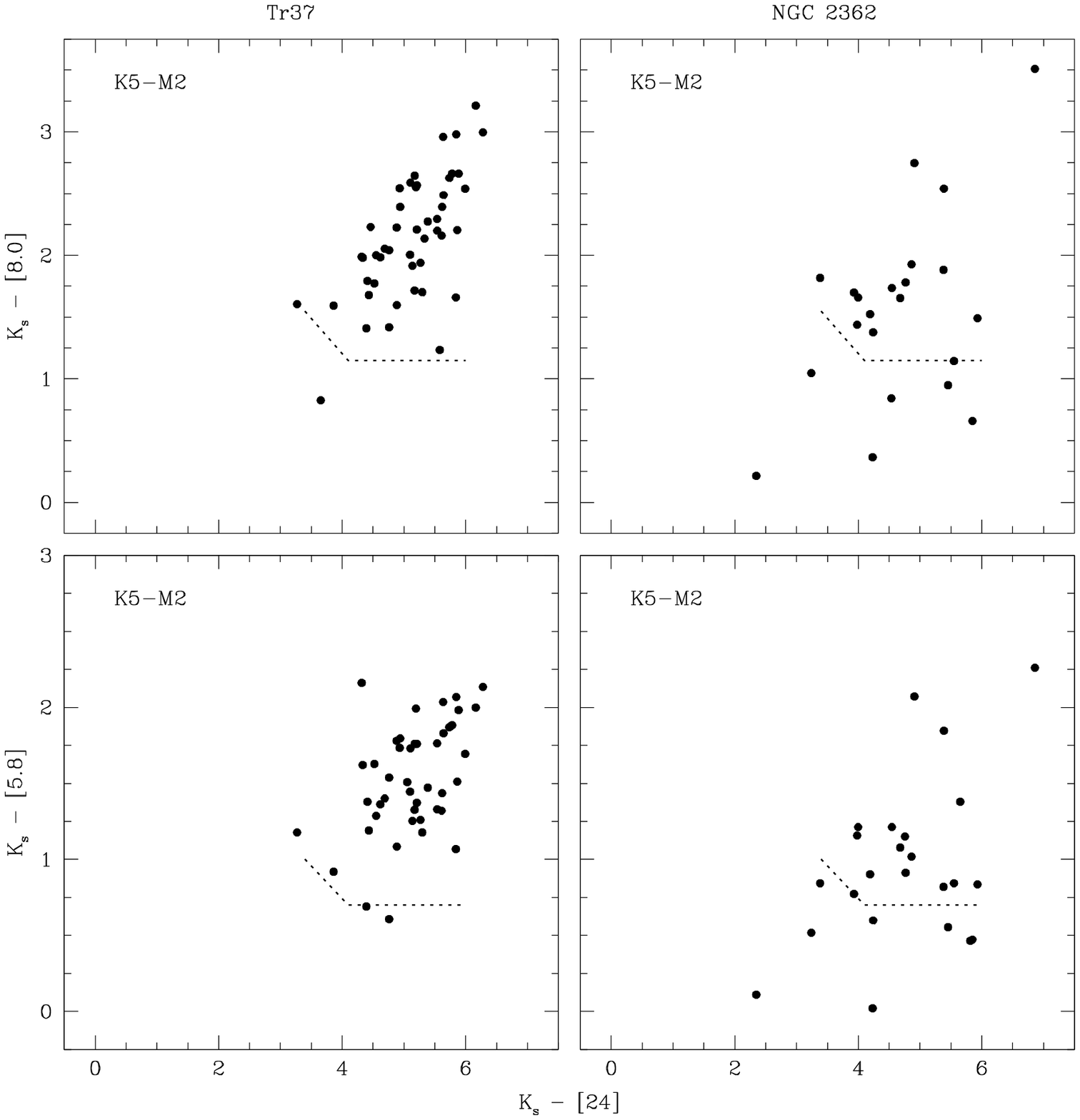}
\caption{
$K_s-[8.0]$ and $K_s-[5.8]$ versus $K_s-[24]$ for K5--M2 members 
of Tr~37 and NGC~2362 \citep[$\tau\sim4$ and 5~Myr,][]{sic06a,dahm07,cur09a}.
The lower boundary of the primordial disks in Taurus 
is indicated ({\it dotted lines}).
}
\label{fig:coltr}
\end{figure}

For NGC~2362 \citep[$\tau\sim5$~Myr,][]{bal96,moi01,dahm05,may08}, 
we have used the spectral types measured by \citet{dahm05} and estimated 
from colors by \citet{cur09a}, the IRAC and MIPS photometry from
\citet{cur09a}, and the $K_s$ photometry from 2MASS.
The IRAC data from \citet{cur09a} are similar to those
measured from the same images by \citet{dahm07}.
Before using these data for our color-color diagrams, we 
inspected the 24~\micron\ images from MIPS for the faintest detections
that were presented by \citet{cur09a}.
In contrast to \citet{cur09a}, we find that source 1091 from \citet{irw08}
was not detected in the MIPS images. The nearest intensity peak in the MIPS 
data is $\sim3\farcs5$ from the optical and IRAC coordinates of 1091, which
is too far to correspond to a 24~\micron\ counterpart. We also find
that source 663 from \citet{irw08} was not detected by MIPS.
Sources 809 and 931 from \citet{irw08} and sources 41 and 63 from 
\citet{dahm07} were detected with signal-to-noise ratios of $\lesssim2$--3, 
which is too low for useful photometry. We exclude from our analysis the 
24~\micron\ measurements for these six stars that were reported by
\citet{cur09a}. After doing so, there remain 29 stars that were found to have
disks by \citet{dahm07} and \citet{cur09a} and
that have 24~\micron\ photometry. The colors of the 25 stars between
K5 and M2 are plotted in Figure~\ref{fig:coltr}.
We have not corrected the colors for reddening since little extinction
is present in NGC~2362 \citep[$A_V<1$,][]{moi01}.
The four disk-bearing stars outside of the range K5--M2 consist of 
one K3 star and three M3 stars. We classify the former as a transitional
disk and the latter stars as primordial disks with our criteria developed
for Taurus. We include the classifications of these four stars with those
of the K5--M2 stars in the statistics that we are about to describe.

Sixteen of the 47 disk-bearing stars that were identified by \citet{dahm07}
have 24~\micron\ photometry. These MIPS-detected sources 
consist of 11 primordial disks and five weak (evolved) disks according to the 
criteria adopted by \citet{dahm07}.  \citet{cur09a} classified only four
of the 16 disks as primordial, suggesting that \citet{dahm07} 
overestimated the number of primordial disks.
In comparison, we classify 15 of these disks as primordial since they
appear above the boundaries indicated in Figure~\ref{fig:coltr} and thus
fall within the region inhabited by the main population of disks in Taurus.

\citet{cur09a} presented a sample of 35 stars with MIPS photometry and
detections of disks, which they classified as six primordial disks, 17
homologously-depleted (evolved) disks, and 12 transitional disks. 
Among the latter two categories, we find that six stars have unreliable
MIPS photometry (see above) and nine stars appear above both of our thresholds 
for primordial disks in Figure~\ref{fig:coltr}.
\citet{cur09a} underestimated the number of primordial disks because
they defined these disks to be close to the median SED of disks in Taurus
without accounting for the spread in the colors of the Taurus disks.

As seen in a comparison of Figures~\ref{fig:coltau} and \ref{fig:coltr},
the primordial disk population in NGC~2362 is weighted more heavily 
toward bluer colors, or flatter disks, than the primordial disks in Taurus.
This feature suggests the youngest members of NGC~2362 are old enough that
most of them have already evolved from flared disks to settled disks.
In other words, if the reddest (and youngest) primordial disks in Taurus
were removed from Figure~\ref{fig:coltau}, the resulting distribution of 
colors would resemble the distribution in NGC~2362. The similarity in those
distributions of colors extends to the evolved and transitional regime.
For instance, the ratio of the number of evolved and transitional disks
to the number of flat primordial disks ($K_s-[8.0]<2$, $K_s-[5.8]<1.4$)
in NGC~2362 (7/16) does not differ significantly from the value in Taurus (7/22)
for the range of spectral types in the NGC~2362 sample of disks (K4.5--M3).

\citet{cur09a} concluded that the ratio of evolved and transitional
disks to all primordial disks is much higher in NGC~2362 than in Taurus, 
which they interpreted as evidence for a long timescale for the 
evolved/transitional phases (\S~\ref{sec:time}). 
However, that ratio is only meaningful if the rate of star formation
has been roughly constant from the earliest point at which stars enter
the class II stage (reddest and most flared disks) through the 
evolved and transitional stages.
In other words, without ongoing star formation in a cluster, the ratio of
evolved/transitional disks to primordial disks will naturally increase
over time as the supply of primordial disks is exhausted.
It is clear that star formation is not continuing at a constant rate in 
NGC~2362 based on the paucity of red primordial disks in Figure~\ref{fig:coltr} 
and the
positions of cluster members on the color-magnitude diagram from \citet{moi01}.
The appropriate metric for comparing two clusters
with different star formation histories is the number
ratio of evolved/transitional disks to stars in the stage immediately
preceding stage, i.e., primordial disks that have experienced large degrees of 
dust settling, which is the ratio that we employed in the previous paragraph.
In addition, \citet{cur09a} counted as evolved/transitional disks the
six stars that we excluded because of questionable MIPS detections,
and they did not apply the same observational criteria for evolved, 
transitional, and primordial disks to both NGC~2362 and Taurus, making their 
comparison of the two populations less reliable.

\subsubsection{$\gamma$~Velorum, 25~Ori, and Orion OB1b}

\begin{figure}
\epsscale{1}
\plotone{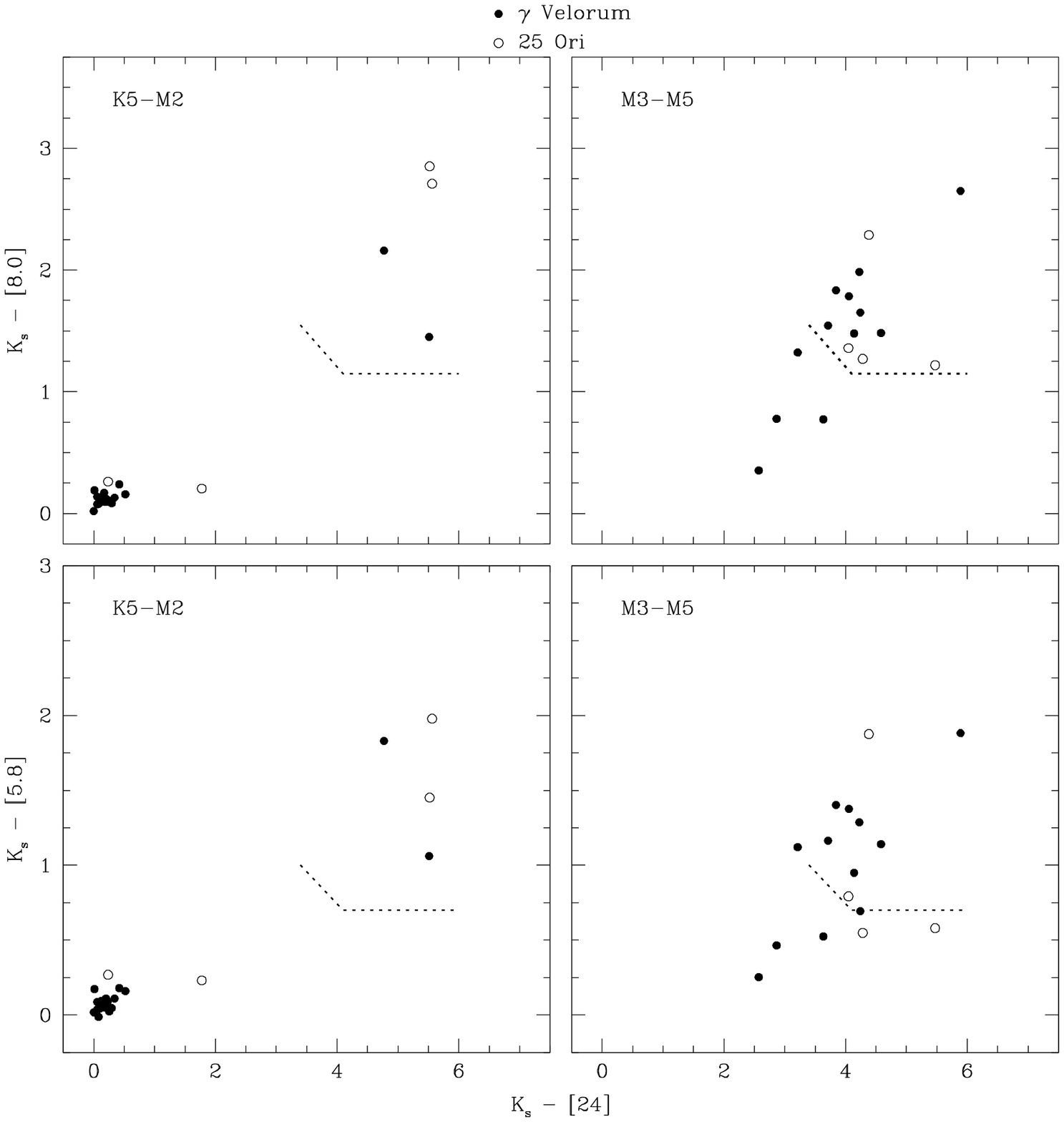}
\caption{
$K_s-[8.0]$ and $K_s-[5.8]$ versus $K_s-[24]$ for K5--M2 and
M3--M5 members of $\gamma$ Velorum and 25~Ori 
\citep[$\tau\sim5$ and 10~Myr,][]{her07b,her08}.
The lower boundary of the primordial disks in Taurus 
is indicated ({\it dotted lines}).
}
\label{fig:colgam}
\end{figure}

For the $\gamma$~Velorum, 25~Ori, and Orion OB1b stellar populations
\citep[$\tau\sim5$, 7--10, and 5~Myr,][]{bri07,her08},
we have adopted the {\it Spitzer} photometry from \citet{her07b,her08}
and the $K_s$ data from 2MASS.
We have used the spectral types measured by \citet{bri05,bri07} and 
\citet{dow08} when possible, and otherwise estimated them from $V-J$ 
using $V$ data from \citet{her07b,her08} and $J$ photometry from 2MASS.
The colors for K5--M2 and M3--M5 members of the three populations
are presented in Figures~\ref{fig:colgam} and \ref{fig:colob}.
We have not dereddened the colors since these stars exhibit little extinction
\cite[$A_V<1$,][]{poz00,bri05,her06}.

Like NGC~2362, the primordial disks in $\gamma$~Velorum are dominated by bluer, 
flatter disks. The number ratio of transitional and evolved disks to those 
flatter disks (4/7) is consistent with the one described in \S~\ref{sec:2362}
for Taurus. The classifications based on our adopted criteria agree with those
of \citet{her08} with the exception of sources 115 and 414 from that study.
These stars appear on or above the lower boundaries for primordial disks
in Figure~\ref{fig:colgam} while \citet{her08} classified them as evolved disks.

The samples of disks in 25~Ori and OB1b are too small for quantitative
comparisons to Taurus. In both samples, a majority of the disks are primordial.
We classify source 905 from \citet{her07b} as a primordial disk since it 
is slightly above the boundaries in Figure~\ref{fig:colgam} while 
\citet{her07b} classified it as an evolved disk. 
Similarly, source 585 was listed as a transitional disk by \citet{her07b},
but its 5.8~\micron\ excess is slightly too large to satisfy our adopted
criteria for this category. Our remaining classifications in 25~Ori and OB1b 
agree with those of \citet{her07b}. One of the transitional disks in 25~Ori,
source 1200 from \citet{her07b}, has been studied in detail through IRS 
and H$\alpha$ spectroscopy and a comparison of its SED to the predictions
of disk models \citep{esp08b}.

\subsubsection{$\eta$~Cha}
\label{sec:eta}

\begin{figure}
\epsscale{1}
\plotone{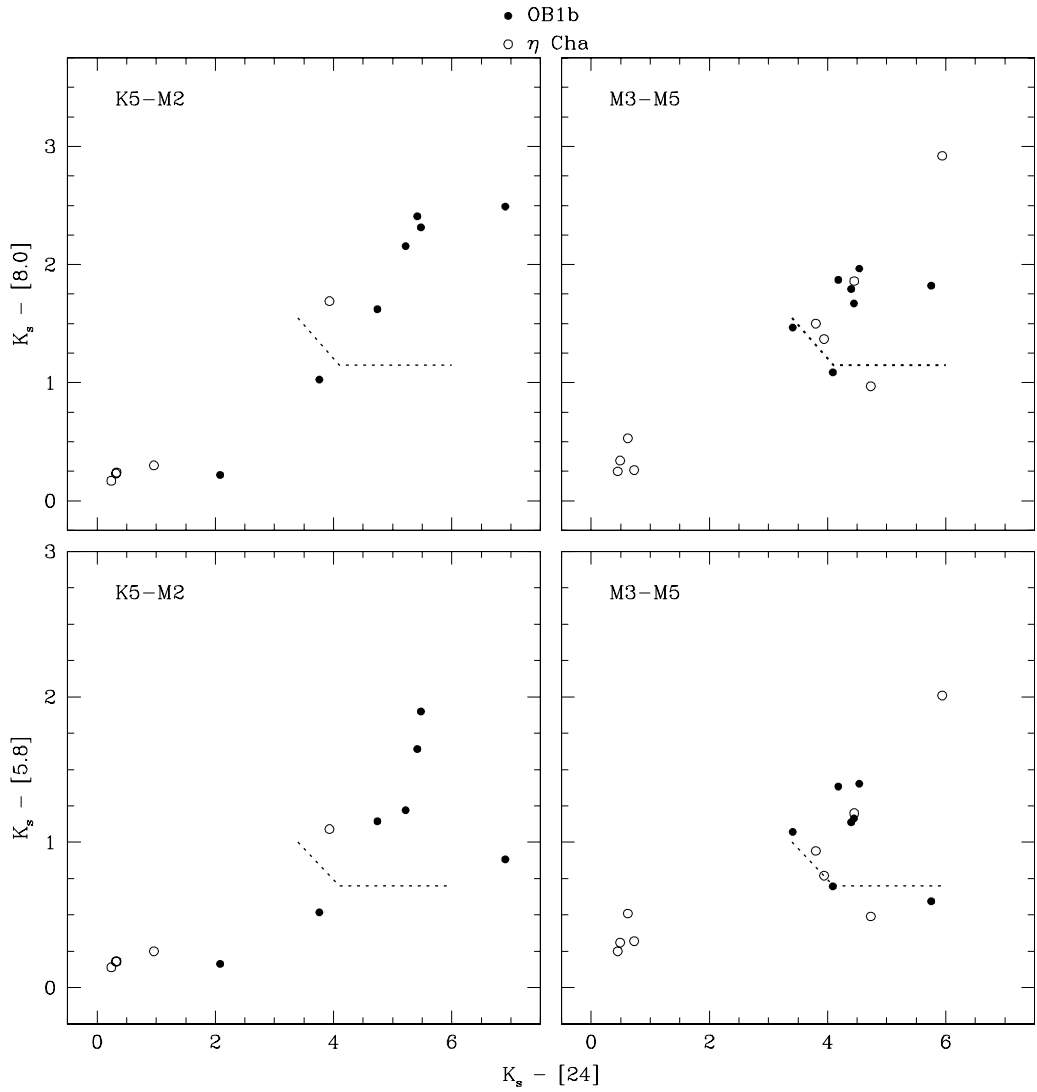}
\caption{
$K_s-[8.0]$ and $K_s-[5.8]$ versus $K_s-[24]$ for K5--M2 and
M3--M5 members of Orion OB1b and $\eta$~Cha 
\citep[$\tau\sim5$ and 6~Myr,][]{meg05,her07b}.
The lower boundary of the primordial disks in Taurus 
is indicated ({\it dotted lines}).
}
\label{fig:colob}
\end{figure}

For the $\eta$~Cha association \citep[$\tau\sim6$~Myr,][]{mam99,law01,ls04},
we have adopted the spectral types from \citet{ls04}, the IRAC 
photometry from \citet{meg05}, and the $K_s$ photometry from 2MASS.
\citet{gau08} and \citet{sic09} presented 
24~\micron\ photometry based on the same MIPS images of $\eta$~Cha.  
\citet{sic09} noted a systematic difference between the data from the two
studies. 
After reducing those images with the methods that were employed for Taurus, we 
arrive at photometry that is an average of 0.1~mag brighter and 0.05~mag
fainter than the measurements from \citet{gau08} and \citet{sic09},
respectively. In addition, we have measured
photometry from all other MIPS images of $\eta$~Cha, which were obtained
through {\it Spitzer} programs PID=40, 84, 173, and 50316.
We present our 24~\micron\ measurements in \S~\ref{sec:app} and 
adopt them for our analysis of $\eta$~Cha,
using the weighed average for each star that was observed more than once.
The colors for K5--M2 and M3--M5 members of the association are 
shown in Figure~\ref{fig:colob}. No correction for reddening has been applied
to these data \citep[$A_V<1$,][]{ls04}.
Because of their close proximity to the Sun \citep[$d\sim100$~pc,][]{mam99},
all of the members of $\eta$~Cha that were within the 24~\micron\ images were
detected in those data. As a result, the distributions of colors in
Figure~\ref{fig:colob} do not suffer from incompleteness among the
stellar photospheres, although they are poorly sampled because of the small
size of the association.

Through analysis of their IRAC data, \citet{meg05} found that six members
of $\eta$~Cha have excess emission at 8~\micron.
Four of these stars (M4--M5.75) exhibit little or no excess at 
$\lambda<6$~\micron, which \citet{meg05} interpreted as evidence of inner holes.
The resulting number ratio of transitional disks to all disks with 
8~\micron\ excesses (4/6) appeared to differ significantly from that in Taurus. 
However, through our full census of disks in Taurus, we find that
three of the four stars cited as transitional disks in $\eta$~Cha have
colors that fall within the main population of disks in Taurus, as demonstrated
by a comparison of Figures~\ref{fig:coltau} and \ref{fig:colob}, and that
can be explained in terms of optically thick disks with large degrees
of dust settling \citep[\S~\ref{sec:model},][]{erc09}. Only one of these
stars, RECX-5, qualifies as a transitional disk based on our adopted criteria.

\citet{sic09} studied the disk population in $\eta$~Cha by combining the
IRAC data with MIPS photometry and IRS spectroscopy.
They identified four stars (M1.75--M4.5) that have photospheric colors at 
$\lambda<6$~\micron\ and excess emission at longer wavelengths.
Two of these disks were classified as transitional by \citet{meg05} while
the remaining two stars lack excesses in any of the IRAC bands.
For low-mass stars, \citet{sic09} suggested that the absence of excess 
emission at $\lambda<6$~\micron\ indicates that the disk has an inner hole, 
or perhaps is relatively flat \citep{erc09}.
In either case, \citet{sic09} advocated for the classification of these disks 
as transitional. The resulting number ratio of transitional disks
to all disks with 24~\micron\ excesses (4/8) appeared to be much larger
than the value in Taurus. As a result, \citet{sic09} concluded that the
transitional phase is not rapid relative to the lifetimes of primordial disks.

As discussed in \S~\ref{sec:term}, transitional objects were originally
defined as stars whose colors appear within a gap between stellar 
photospheres and most stars with disks in Taurus. These sources were
interpreted as disks in which the dust in the inner regions is optically
thin or completely cleared. The criterion for transitional disks proposed by 
\citet{sic09} for low-mass stars --- an absence of excess emission at
$\lambda<6$~\micron\ --- does not satisfy either of these
observational and theoretical definitions since it encompasses stars that are 
within the main population of disks in Taurus in terms of their IR colors,
and since those colors can be explained with optically disks, as shown in
\S~\ref{sec:model} and Figures~\ref{fig:k424}--\ref{fig:k224}. 
Furthermore, because the definition of transitional disks from \citet{sic09} 
included flat optically thick disks, it does not provide a measurement of
the timescale of inner disk clearing.

When we apply our classification criteria to the data for $\eta$~Cha in
Figure~\ref{fig:colob}, we find that the association contains five primordial
disks, one transitional disk, and two stars that have weak excess emission at
24~\micron, indicating the presence of either evolved transitional disks or
debris disks. As done for NGC~2362 (\S~\ref{sec:2362}), we compare $\eta$~Cha
to Taurus in terms of their number ratios of evolved and transitional disks
to the bluest and flattest of the primordial disks. For spectral types of 
K5 to M5, this ratio 
is (1--3)/4 for $\eta$~Cha and (11--15)/43 for Taurus, where the range of values
for each numerator corresponds to the uncertainty in the nature of the
disks with weak 24~\micron\ emission. These ratios are consistent with each
other within the statistical uncertainties.

\subsubsection{Upper Sco}
\label{sec:us}

\begin{figure}
\epsscale{1}
\plotone{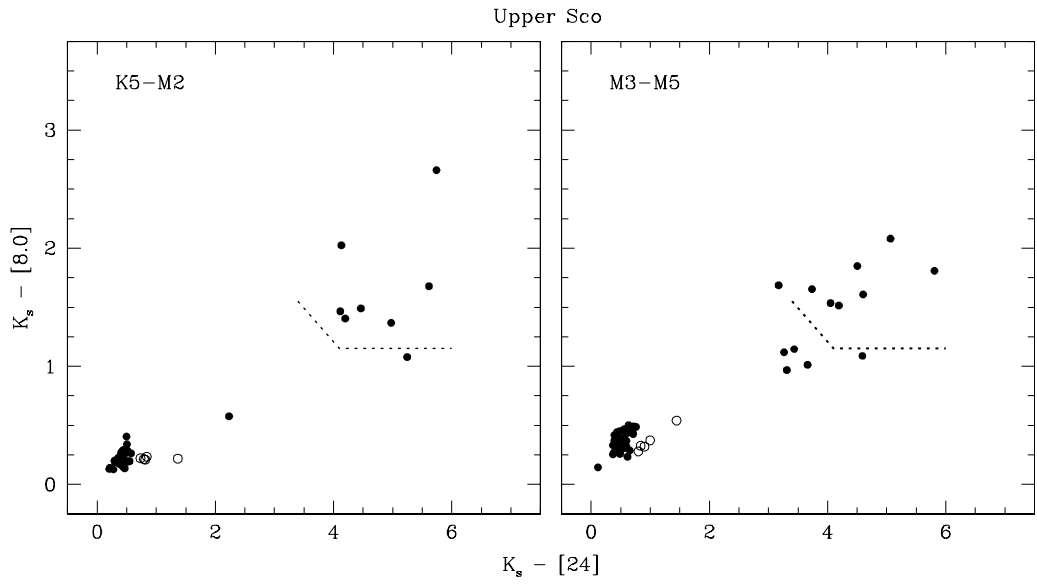}
\caption{
$K_s-[8.0]$ versus $K_s-[24]$ for K5--M2 and
M3--M5 members of Upper Sco \citep[$\tau\sim5$~Myr,][]{car06,car09}.
Candidate debris disks that were identified by \citet{car09} are shown as open
circles. The lower boundary of the primordial disks in Taurus 
is indicated ({\it dotted lines}).
}
\label{fig:colus}
\end{figure}

For the Upper Sco association \citep[$\tau\sim5$~Myr,][]{pm08},
we have adopted the {\it Spitzer} photometry from \citet{car06,car09},
the $K_s$ data from 2MASS, and the compilation of spectral types and
extinctions from \citet{car09}. The {\it Spitzer} observations were performed
at 4.5, 8.0, and 24~\micron\ and did not include 3.6 and 4.5~\micron.
The dereddened values of $K_s-[8.0]$ and $K_s-[24]$ for
K5--M2 and M3--M5 members of Upper Sco are plotted in Figure~\ref{fig:colus}.
We have omitted sources that have questionable 24~\micron\ data according
to \citet{car09}.


Because of the close proximity of Upper Sco \citep[$d\sim145$~pc,][]{pm08},
the 24~\micron\ images were able to detect the stellar photospheres of
a large number of low-mass stars.
Like Taurus and the other young populations that we have examined, Upper Sco
exhibits a distinct gap between stellar photospheres and redder sources.
Our classification criteria suggest that the following seven stars between
K5 and M5 have disks that are evolved or transitional:
[PBB2002] USco J155729.9$-$225843,
[PBB2002] USco J160525.5$-$203539,
[PBB2002] USco J160600.6$-$195711,
[PBB2002] USco J160622.8$-$201124,
[PBB2002] USco J160643.8$-$190805,
[PBB2002] USco J160827.5$-$194904,
and ScoPMS 31.
\citet{car09} identified 10 additional K5--M5 stars with smaller
24~\micron\ excesses that are candidates for debris disks
(see Figure~\ref{fig:colus}). Given that the excesses in these latter sources
could also arise from evolved transitional disks, 
the ratio of the number of evolved and transitional disks
to the number of relatively flat primordial disks ($K_s-[8.0]<2$,
\S~\ref{sec:2362}) is 7--17/12. This ratio is higher than the value of
11--15/43 for Taurus, although they are consistent within the statistical
errors.
A higher ratio in Upper Sco could be explained by a star formation history that
is not uniform across the phases represented in the numerator and denominator
(\S~\ref{sec:2362}). In other words, if the youngest members of Upper Sco
are old enough to have begun clearing their disks, then one would expect
a ratio of transitional/evolved disks to primordial disks that is larger
than that in Taurus.

\subsection{Timescale of Disk Clearing}
\label{sec:time}

The frequency of evolved and transitional disks provides a constraint on
the timescale of the clearing of optically thick disks.
Based on the small number of transitional disks in Taurus, \citet{skr90}
concluded that this process occurs rapidly. By combining a value of 3/30
for the number ratio of transitional disks to primordial disks with an average 
lifetime of $\sim3$~Myr for primordial 
disks, they estimated a timescale of $\sim0.3$~Myr for the transitional phase. 
Other studies of Taurus have arrived at even shorter timescales 
\citep[$\tau\lesssim0.1$~Myr,][]{sim95,wol96}.
Using our classifications of the {\it Spitzer} data in Taurus, 
we find that the number ratio of evolved and transitional disks to primordial 
disks is 15/98 for spectral types of K5--M5. 
Thus, we derive a somewhat longer timescale of 
$\sim0.45$~Myr, assuming a primordial disk lifetime of 3~Myr.
We have overestimated this timescale if some of the stars with weak 
24~\micron\ emission have debris disks rather than evolved transitional disks, 
or if stellar companions are responsible for the inner holes in some of the 
transitional disks \citep[CoKu Tau/4,][]{ire08}.

The timescale of disk clearing has been investigated further through the 
{\it Spitzer} observations of older clusters that were described in 
\S~\ref{sec:compare}. 
Some of those studies have concluded that the frequency of evolved and
transitional disks is much higher in older clusters than in Taurus, implying 
a timescale for this phase that is longer than previous estimates 
\citep{cur09a,sic09}. However, in our analysis of the two clusters in 
question, NGC~2362 and $\eta$~Cha (\S~\ref{sec:2362}, \S~\ref{sec:eta}),
we found that \citet{cur09a} and \citet{sic09} 
overestimated the number of evolved and transitional disks because they adopted
criteria that encompassed optically thick disks that are relatively flat.
In addition, when computing the frequencies of evolved and transitional disks
in NGC~2362 and $\eta$~Cha for comparison to Taurus, \citet{cur09a} and 
\citet{sic09} divided by the number of all primordial disks, whereas only the 
flatter primordial disks should have been counted given that star formation
is no longer occurring in these clusters.
In \S~\ref{sec:2362} and \S~\ref{sec:eta}, we addressed these issues by
applying classification criteria that exclude flat optically
thick disks from the transitional stage, and by comparing the populations
in terms of the ratio of evolved and transitional disks to flatter primordial 
disks. By doing so, we found that the frequencies of evolved and transitional
disks in NGC~2362 and $\eta$~Cha do not differ significantly from
the value in Taurus.  The other clusters that were examined in 
\S~\ref{sec:compare} also exhibit distributions of color excesses that
are consistent with the same timescale of disk clearing that is implied by the
data in Taurus.

Based on the large frequency of evolved and transitional disks that
they measured in NGC~2362, \citet{cur09a} concluded that the timescale of 
these phases may be comparable to the lifetime of primordial disks.
\citet{cur09a} attempted to explain how the low number of evolved and 
transitional disks in Taurus (i.e., the gap in IR colors) is consistent 
with such a long lifetime for those disks.
They suggested that Taurus contains few evolved and
transitional disks because most disks in Taurus are too young 
to have evolved beyond the primordial stage.
However, that hypothesis is inconsistent with the fact that a large fraction of 
the stellar population in Taurus (40\%) has already fully cleared their 
disks and become class~III stars (Table~\ref{tab:sed}). 
If Taurus is old enough to have a significant number of diskless stars, then 
the evolved and transitional phases should be heavily populated as well if 
they have a long timescale.

To account for the existence of class~III stars in Taurus, \citet{cur09a} 
contended that the bulk of these stars may have lost their disks through 
interactions with stellar companions rather than mechanisms associated with 
the dispersal of disks during the evolved and transitional phases.
However, if disruption by companions has been the dominant mechanism for disk 
removal in Taurus, then the disk fraction as a function of stellar mass should 
be anti-correlated with binary frequency, but this is not observed in Taurus
\citep[Figure~\ref{fig:diskfraction};][]{kra06,luh07ppv}. 
In addition, according to the scenario proposed by \citet{cur09a}, 
class~III stars should have the same ages as class~II sources in Taurus.
However, class~III stars are more widely distributed than members with disks 
(\S~\ref{sec:spatial}), which indicates that they are older on average.
In other words, if the fundamental difference between class~II and III
sources is the absence or presence of tight binaries, then they should share
the same spatial distribution.
Overall, the suggestion that the lifetime of evolved and transitional disks is
comparable to that of primordial disks is incompatible with the observed
properties of stars and disks in Taurus. The same conclusion applies to 
other clusters that also exhibit clear evidence of a paucity of evolved and 
transitional disks (e.g., Chamaeleon~I).

\section{Conclusions}

We have performed a census of the circumstellar disk population of the
Taurus star-forming region ($\tau\sim1$~Myr) 
using mid-IR images obtained with the {\it Spitzer 
Space Telescope}. The results of this study are summarized as follows:

\begin{enumerate}

\item
We have analyzed nearly all images of the Taurus cloud complex
at 3.6, 4.5, 5.8, 8.0, and 24~\micron\ that were collected
by {\it Spitzer} during its cryogen mission.
The IRAC and MIPS cameras on board {\it Spitzer} each covered a total area 
46~deg$^{2}$ and encompassed 346 and 299 members of Taurus, respectively,
corresponding to 99\% of the known stellar population.
We have presented photometry for all members that were
detected in these images. 

\item
We have used our mid-IR photometry in conjunction with other available 
observations to determine the likely evolutionary stages of all members 
of Taurus (classes~0--III).
Stars that exhibit evidence of circumstellar envelopes in previous
measurements (e.g., IRS spectra) are assigned to classes~0 and I.
We have classified the remaining members of Taurus using their mid-IR 
SEDs, or H$\alpha$ emission for the few stars that lack {\it Spitzer} data.

\item
Based on our classifications, the disk fraction in Taurus, N(II)/N(II+III),
is $\sim75$\% for solar-mass stars and declines to $\sim45$\% for low-mass
stars and brown dwarfs (0.01--0.3~$M_\odot$). 
A similar dependence on stellar mass has been observed in Chamaeleon~I
\citep{luh08cha1}.
In contrast to Taurus and Chamaeleon~I, IC~348 exhibits a disk fraction of
only $\sim20$\% for solar-mass stars \citep{lada06,mue07,luh05frac}.
Given that IC~348 is roughly coeval with Chamaeleon~I ($\tau\sim2$--3~Myr), 
the comparison of these three regions suggests that
disk lifetimes for solar-mass stars are longer in star-forming regions
like Taurus and Chamaeleon~I that have lower stellar densities. 

\item
As previously observed in Taurus \citep{har02}, we find that 
the positions of the class~I and II members closely follow the distribution
of dense gas while the class~III stars are more widely distributed.
An analysis of the nearest neighbor distances also 
suggests that class~II sources may have a wider distribution than
class~I sources, although this difference is only marginally significant.
The median of the nearest neighbor distances for classes~I and II 
is 0.15~pc, which is twice the average value in nearby star-forming
clusters \citep{gut09}.

\item
Our study has produced multiple epochs of mid-IR photometry for $\sim200$
members of Taurus. In these data, the mid-IR variability
of class~I/II sources is much greater than that of class~III stars, which
agrees with similar measurements with {\it Spitzer} in Chamaeleon~I
\citep{luh08cha1}. The fraction of disk-bearing stars that are variable
is higher in Taurus than in Chamaeleon~I, indicating that the variability 
of disks decreases with age. 

\item
We have used our {\it Spitzer} photometry for the disk population in Taurus to
refine the observational criteria for the evolutionary phases of disks.
When plotted in terms of $K_s-[5.8]$, $K_s-[8.0]$, and $K_s-[24]$,
the members of Taurus appear predominantly within two distinct 
groups that are well-separated from each other, as found in
early mid-IR studies of Taurus \citep{skr90}. 
The colors of the stars in the bluer group are consistent with stellar
photospheres.
We have defined the large, continuous population of much redder sources as
primordial disks.
Using our models of accretion disks, we have demonstrated that the colors
of these primordial disks can be explained in terms of optically thick disks.
The sources that fall within the gap between primordial disks and stellar
photospheres are defined as evolved disks (weak excess in all bands),
transitional disks (weak or no excess at $\lambda<10$~\micron, large excess
at longer $\lambda$), and evolved transitional disks or debris disks
(only weak 24~\micron\ excess). We have identified 19 members of Taurus that
are candidates for disks in these later stages of disk evolution, 11 of 
which have not been previously recognized as such.

\item
We have applied our classification criteria for disks to nearby clusters and 
associations with ages of 2--10~Myr that have been observed with {\it Spitzer}.
We find that the number of evolved and transitional disks in those
regions is consistent with the paucity of such disks in Taurus.
Some of the sources that have been classified as evolved and transitional 
disks in previous {\it Spitzer} studies have colors that are similar to those
of the bluer primordial disks in Taurus (i.e., flatter optically thick disks).
The number ratio of evolved and transitional disks to primordial disks
in Taurus is 15/98 for spectral types of K5--M5, indicating a timescale of 
inner disk clearing that is $\sim15$\% of the lifetime of primordial disks
($\sim3$~Myr). 
The data in Taurus and the older populations that we have examined 
are inconsistent with notion that the lifetime of the evolved and transitional 
phases (i.e., disk clearing timescale) is comparable to that of primordial 
disks. 

\end{enumerate}

\acknowledgements

K.~L. was supported by grant AST-0544588 from the National Science Foundation.
C.~E. and N.~C. were supported by grant NNX08AH94G from NASA and grant 1344183
from the Jet Propulsion Laboratory.
We thank Dan Watson and Melissa McClure for their analysis of unpublished IRS 
spectra. We are grateful to Jesus Hern\'andez for helpful comments.
This work makes use of data from the {\it Spitzer Space Telescope} and 2MASS.
{\it Spitzer} is operated by the Jet Propulsion Laboratory, California 
Institute of Technology under a contract with NASA.
2MASS is a joint project of the University of Massachusetts and the Infrared
Processing and Analysis Center/California Institute of Technology, funded
by NASA and the NSF.
The Center for Exoplanets and Habitable Worlds is supported by the
Pennsylvania State University, the Eberly College of Science, and the
Pennsylvania Space Grant Consortium.

\appendix

\section{Infrared Colors of Young Stellar Photospheres}
\label{sec:app}

We have estimated the intrinsic IR colors of young stellar photospheres
as a function of spectral type from K4 to L0 ($M\sim0.01$--1~$M_\odot$).
For this analysis, we examined the colors of probable members of the
$\eta$~Cha, $\epsilon$~Cha, and TW~Hya associations 
\citep[TWA,][]{mam99,web99,law02,giz02,fei03,ls04,luh04eta,lyo04,sz04,zuc04,
mam05,sch05,loo07,luh08cha1,kas08} and young late-type dwarfs in the solar 
neighborhood \citep{kir06,kir08,cru07,cru09}.
These sources should have negligible extinction ($A_V<1$) since they
are relatively nearby ($d\lesssim100$~pc) and are not associated with
molecular clouds. As a result, their observed colors should not depart from
the intrinsic photospheric values unless emission from circumstellar disks
is present.
Although the members of embedded clusters are subject to both reddening and
excess emission from disks, the bluest sources provide useful constraints
on the intrinsic colors of photospheres.
Taurus and Chamaeleon~I are the best star-forming regions for this purpose
since accurate spectral types and {\it Spitzer} photometry have been
measured for most of their members and the census of each population is
large and reaches the end of the M spectral sequence
\citep[][references therein]{luh07cha,luh08cha2,luh09tau1}.

We have adopted measurements of $J$, $H$, and $K_s$ from the 2MASS Point
Source Catalog when they are available. For the faintest members of
Chamaeleon~I, we used photometry measured from deeper near-IR
images that were calibrated with 2MASS sources \citep{luh07cha}.
In the {\it Spitzer} bands between 3.6 and 24~\micron, we have
used our photometry in Taurus and our previous measurements for Chamaeleon~I
\citep{luh08cha1,luh08cha2}, $\epsilon$~Cha \citep{luh08cha1},
and young late-type dwarfs \citep{luh09tau1}.
We have analyzed all {\it Spitzer} images of $\eta$~Cha,
TWA, and additional members of $\epsilon$~Cha 
with the same methods that we have applied to Taurus.
The resulting photometry for the members of these associations is
presented in Tables~\ref{tab:eta1}--\ref{tab:twa}. 
Stars that were not observed by either IRAC or MIPS are excluded from
these tabulations.

\begin{figure}
\epsscale{1}
\plotone{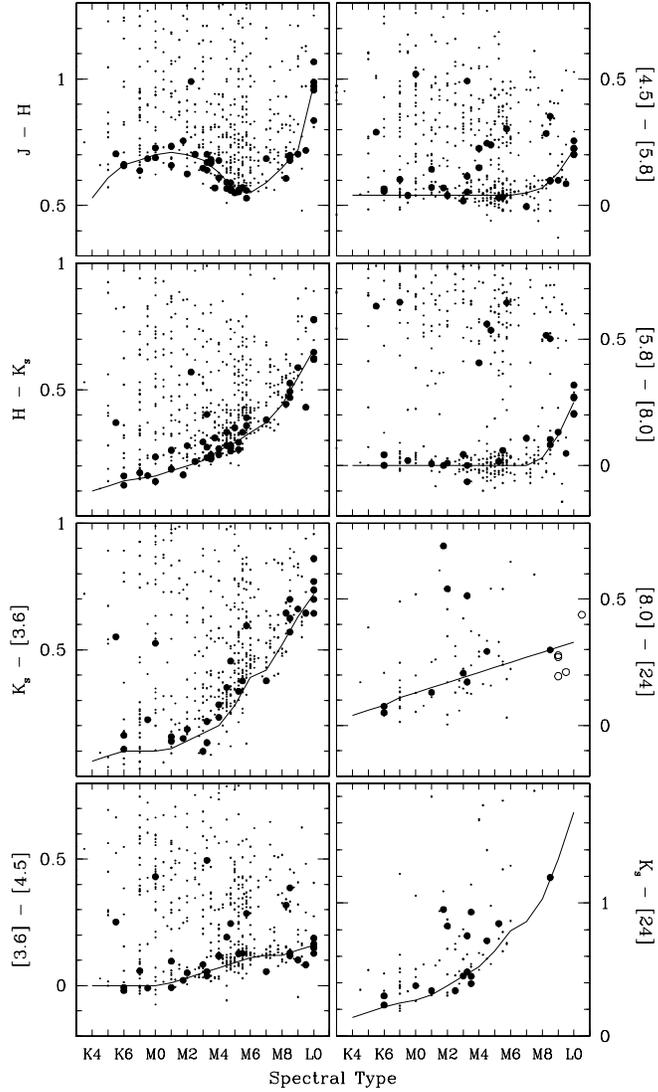}
\caption{
Infrared colors as a function of spectral type for members of the Taurus
and Chamaeleon~I star-forming regions ({\it small filled circles}) and
young sources in the $\eta$~Cha, $\epsilon$~Cha, and TW Hya associations and in 
the solar neighborhood ({\it large filled circles}). The latter samples 
should have negligible extinction ($A_V<1$). We have used these data to
estimate the intrinsic colors of young stellar photospheres ({\it solid lines},
Table~\ref{tab:colors}). Because few young late-type objects have been
measured at 24~\micron, we also show data for field dwarfs at M9--L0 in
the diagram of $[8.0]-[24]$ colors ({\it open circles}). 
}
\label{fig:colors}
\end{figure}

The IR colors for Taurus, Chamaeleon~I, $\eta$~Cha, $\epsilon$~Cha, TWA,
and young dwarfs are plotted as a function of spectral type in
Figure~\ref{fig:colors}. The limits of these diagrams encompass
only the bluest colors since we wish to examine the colors of the stellar
photospheres. As a result, many of the stars with disks are too red to
appear in Figure~\ref{fig:colors}, particularly in the colors at longer
wavelengths. 
The spectral types that we have measured in TWA from unpublished spectra
are a few subclasses later than some of the classifications from previous
studies. Therefore, we have plotted only the TWA members for which
we have measured types, or that have been classified with similar
methods \citep{loo07,her09}. 
We also have excluded measurements that have photometric
uncertainties greater than 0.1~mag.
Few of the class~III sources at the latest types have 
accurate 24~\micron\ data. To help constrain the photospheric colors
at those types, we have measured IRAC and MIPS photometry for field
dwarfs from M9--L0 and have included them in the diagram for $[8.0]-[24]$.
These stars consist of BRI~0021$-$0214, LHS~2065, LP944$-$20, LHS~2924,
and 2MASS J07464256+2000321.
The $[8.0]-[24]$ color of the TWA brown dwarf 
2MASSW J1139511$-$315921 (M8.5) is similar to colors of these fields dwarfs,
which illustrates the absence of 24~\micron\ excess emission from
this object that was discussed by \citet{mor08}.

For each IR color, we have estimated the photospheric values as a function
of spectral type by performing a fit to the sequence of class~III objects
from the young associations and the solar neighborhood. This fit was also
constrained to agree with the blue envelope of colors from Taurus and
Chamaeleon~I. These color relations are plotted in Figure~\ref{fig:colors}
and are presented in Table~\ref{tab:colors}.

\clearpage

\LongTables

\begin{deluxetable}{lrlcccrc}
\tabletypesize{\scriptsize}
\tablewidth{0pt}
\tablecaption{IRAC Observing Log\label{tab:iraclog}}
\tablehead{
\colhead{} &
\colhead{} &
\colhead{Date} &
\colhead{3.6/5.8 Center} &
\colhead{4.5/8.0 Center} &
\colhead{Dimensions} &
\colhead{Angle\tablenotemark{a}} &
\colhead{Exp Time\tablenotemark{b}} \\
\colhead{AOR} &
\colhead{PID} &
\colhead{(UT)} &
\colhead{(J2000)} &
\colhead{(J2000)} &
\colhead{(arcmin)} &
\colhead{(deg)} &
\colhead{(s)}}
\startdata
 3963904 &     37 & 2004 Feb 9 & 04 46 42.3 +24 58 58 & 04 46 42.3 +24 58 58 & 5.3$\times$5.3 &   85.0 & 31.2 HDR \\
         &        &            & 04 46 46.6 +24 52 21 & 04 46 38.5 +25 05 43 & 5.3$\times$5.3 &   85.0 & 31.2 HDR \\
         &        &            & 04 44 31.0 +25 12 21 & 04 44 31.0 +25 12 21 & 5.3$\times$5.3 &   85.0 & 31.2 HDR \\
         &        &            & 04 44 34.9 +25 05 50 & 04 44 27.2 +25 19 06 & 5.3$\times$5.3 &   85.0 & 31.2 HDR \\
         &        &            & 04 43 03.4 +25 20 10 & 04 43 03.4 +25 20 10 & 5.3$\times$5.3 &   85.0 & 31.2 HDR \\
         &        &            & 04 43 07.2 +25 13 44 & 04 43 00.0 +25 26 56 & 5.3$\times$5.3 &   85.0 & 31.2 HDR \\
         &        &            & 04 42 38.0 +25 15 29 & 04 42 38.0 +25 15 29 & 5.3$\times$5.3 &   85.0 & 31.2 HDR \\
         &        &            & 04 42 41.4 +25 08 54 & 04 42 34.1 +25 22 14 & 5.3$\times$5.3 &   85.0 & 31.2 HDR \\
\enddata
\tablenotetext{a}{Position angle of the long axis of the map.}
\tablenotetext{b}{Total exposure time for each position and filter.
``HDR" indicates that 0.4~s exposures were also obtained.}
\end{deluxetable}

\begin{deluxetable}{lrlccrr}
\tabletypesize{\scriptsize}
\tablewidth{0pt}
\tablecaption{MIPS 24~\micron\ Observing Log\label{tab:mipslog}}
\tablehead{
\colhead{} &
\colhead{} &
\colhead{Date} &
\colhead{Center} &
\colhead{Dimensions} &
\colhead{Angle\tablenotemark{a}} &
\colhead{Exp Time} \\
\colhead{AOR} &
\colhead{PID} &
\colhead{(UT)} &
\colhead{(J2000)} &
\colhead{(arcmin)} &
\colhead{(deg)} &
\colhead{(s)}}
\startdata
 3662080 &      6 & 2004 Feb 20 & 04 31 40.6 +18 08 45 & 30.5$\times$54 & 170.7 &    40 \\
 5675776 &    173 & 2004 Sep 18 & 04 20 24.3 +31 23 26 & 7.4$\times$8.2 & 167.9 &   277 \\
         &        &            & 04 20 07.3 +31 35 21 & 5.6$\times$8 & 167.9 &   357 \\
 5680384 &    173 & 2004 Sep 18 & 04 33 34.6 +24 21 09 & 7.4$\times$8.2 & 171.0 &    36 \\
         &        &            & 04 33 21.6 +24 33 15 & 5.6$\times$8 & 171.0 &    31 \\
 5675264 &    173 & 2004 Sep 19 & 04 17 39.0 +28 33 03 & 7.4$\times$8.2 & 168.2 &    73 \\
         &        &            & 04 17 22.8 +28 45 00 & 5.6$\times$8 & 168.2 &   357 \\
 5676032 &    173 & 2004 Sep 19 & 04 21 58.9 +28 18 07 & 7.4$\times$8.2 & 168.8 &    36 \\
\enddata
\tablenotetext{a}{Position angle of the long axis of the map.}
\end{deluxetable}

\begin{deluxetable}{lllll}
\tabletypesize{\scriptsize}
\tablewidth{0pt}
\tablecaption{IRAC Aperture Corrections\label{tab:apcor}}
\tablehead{
\colhead{Band} &
\colhead{2/14/15\tablenotemark{a}} &
\colhead{3/3/4\tablenotemark{a}} &
\colhead{4/4/5\tablenotemark{a}} &
\colhead{4/4/10\tablenotemark{a}} 
}
\startdata
$[3.6]$ & 0.450 & 0.270 & 0.180\tablenotemark{b} & \nodata \\
$[4.5]$ & 0.485 & 0.300 & 0.170\tablenotemark{b} & \nodata \\
$[5.8]$ & 0.645 & 0.515 & \nodata & 0.165\tablenotemark{b} \\
$[8.0]$ & 0.750 & 0.735 & \nodata & 0.260\tablenotemark{b} \\
\enddata
\tablenotetext{a}{Aperture radius and inner and outer radii of the sky annulus
in pixels (1 pixel = $0\farcs86$).}
\tablenotetext{b}{Default aperture and sky annulus.}
\end{deluxetable}

\begin{deluxetable}{llrrrrl}
\tabletypesize{\scriptsize}
\tablewidth{0pt}
\tablecaption{IRAC Photometry for Members of Taurus\label{tab:irac}}
\tablehead{
\colhead{2MASS\tablenotemark{a}} &
\colhead{Name} &
\colhead{[3.6]} &
\colhead{[4.5]} &
\colhead{[5.8]} &
\colhead{[8.0]} &
\colhead{Date} 
}
\startdata
J04034930+2610520 &         HBC 358 A+B+C   &  9.15$\pm$0.02 &  9.11$\pm$0.02 &  9.05$\pm$0.03 &  9.05$\pm$0.03 & 2005 Feb 20 \\
J04034997+2620382 &      XEST 06-006   &        out &        out &        out &     out & \nodata \\
J04035084+2610531 &          HBC 359   &  9.33$\pm$0.02 &  9.26$\pm$0.02 &  9.23$\pm$0.03 &  9.23$\pm$0.03 & 2005 Feb 20 \\
J04043936+2158186 &          HBC 360   &  9.71$\pm$0.02 &  9.69$\pm$0.02 &  9.62$\pm$0.03 &  9.62$\pm$0.03 & 2004 Feb 11 \\
J04043984+2158215 &          HBC 361   &  9.87$\pm$0.02 &  9.81$\pm$0.02 &  9.76$\pm$0.03 &  9.76$\pm$0.03 & 2004 Feb 11 \\
J04044307+2618563 &  IRAS 04016+2610   &  6.76$\pm$0.02 &  5.46$\pm$0.02 &  4.42$\pm$0.03 &  3.56$\pm$0.03 & 2005 Feb 20 \\
J04053087+2151106 &          HBC 362   &  9.83$\pm$0.02 &  9.82$\pm$0.02 &  9.76$\pm$0.03 &  9.75$\pm$0.03 & 2005 Feb 22 \\
J04080782+2807280 &    \nodata &        out &        out &        out &     out & \nodata \\
\enddata
\tablecomments{Entries of ``$\cdots$", ``sat", ``out", and ``bad"
indicate measurements that are absent because of non-detection, saturation,
a position outside the field of view of the camera, and 
contamination by a cosmic ray, respectively.}
\tablenotetext{a}{2MASS Point Source Catalog.}
\end{deluxetable}

\begin{deluxetable}{lllrrrrl}
\tabletypesize{\scriptsize}
\tablewidth{0pt}
\tablecaption{IRAC Photometry for Members of Taurus with Extended 
Emission\label{tab:ext}}
\tablehead{
\colhead{2MASS\tablenotemark{a}} &
\colhead{Name} &
\colhead{Aperture\tablenotemark{b}} &
\colhead{[3.6]} &
\colhead{[4.5]} &
\colhead{[5.8]} &
\colhead{[8.0]} &
\colhead{Date} 
}
\startdata
          \nodata &  IRAS 04111+2800G  &         15 & 12.60$\pm$0.05 & 11.39$\pm$0.05 & 10.79$\pm$0.05 & 10.18$\pm$0.05 & 2005 Feb 19 \\
                  &                 &         15 & 12.74$\pm$0.05 & 11.47$\pm$0.05 & 10.70$\pm$0.05 & 10.12$\pm$0.05 & 2007 Mar 29 \\
          \nodata &     IRAS 04166+2706   &         30 & 11.15$\pm$0.1 & 10.07$\pm$0.1 & 9.58$\pm$0.1 & 9.17$\pm$0.1 & 2005 Feb 20 \\
                  &                 &         30 & 11.22$\pm$0.1 & 10.07$\pm$0.1 & 9.76$\pm$0.1 & 9.13$\pm$0.1 & 2005 Feb 21 \\
                  &                 &         30 & 11.25$\pm$0.1 & 10.13$\pm$0.1 & 9.73$\pm$0.1 & 9.37$\pm$0.1 & 2007 Oct 17 \\
J04195844+2709570 &     IRAS 04169+2702   &         30 & 8.32$\pm$0.05 & 7.13$\pm$0.05 &    \nodata & \nodata & 2005 Feb 20 \\
                  &                 &         30 & 8.29$\pm$0.05 &        out &    \nodata & \nodata & 2005 Feb 21 \\
                  &                 &         30 & 8.19$\pm$0.05 & 7.03$\pm$0.05 &    \nodata & \nodata & 2007 Oct 17 \\
\enddata
\tablecomments{Entries of ``out" indicate measurements that are absent
because of a position outside the field of view of the camera.
We have not measured photometry with a large aperture if a source does
not exhibit significant extended emission in a given band ($\cdots$).
Meausrements with the default 4-pixel apertures are available in
Table~\ref{tab:irac}.
}
\tablenotetext{a}{2MASS Point Source Catalog.}
\tablenotetext{b}{Aperture radius in pixels (1 pixel = $0\farcs86$).}
\end{deluxetable}

\begin{deluxetable}{llrl}
\tabletypesize{\scriptsize}
\tablewidth{0pt}
\tablecaption{MIPS 24~\micron\ Photometry for Members of Taurus\label{tab:mips}}
\tablehead{
\colhead{2MASS\tablenotemark{a}} &
\colhead{Name} &
\colhead{[24]} &
\colhead{Date} 
}
\startdata
J04034930+2610520 &         HBC 358 A+B+C   &  8.98$\pm$0.08 & 2005 Feb 26 \\
                  &            &  8.67$\pm$0.13 & 2005 Mar 6 \\
J04034997+2620382 &      XEST 06-006   &     out & \nodata \\
J04035084+2610531 &          HBC 359   &  9.23$\pm$0.08 & 2005 Feb 26 \\
J04043936+2158186 &          HBC 360   &     out & \nodata \\
J04043984+2158215 &          HBC 361   &     out & \nodata \\
J04044307+2618563 &  IRAS 04016+2610   &     sat & \nodata \\
J04053087+2151106 &          HBC 362   &  9.40$\pm$0.06 & 2004 Sep 25 \\
\enddata
\tablecomments{Entries of ``$\cdots$", ``sat", and ``out"
indicate measurements that are absent because of non-detection, saturation, 
and a position outside the field of view of the camera, respectively.}
\tablenotetext{a}{2MASS Point Source Catalog.}
\tablenotetext{b}{2MASS J04142639+2805597.}
\end{deluxetable}

\begin{deluxetable}{lllrrrrc}
\tabletypesize{\scriptsize}
\tablewidth{0pt}
\tablecaption{Spectral Slopes for Members of Taurus\label{tab:alpha}}
\tablehead{
\colhead{} &
\colhead{} &
\colhead{Spectral} &
\colhead{} &
\colhead{} &
\colhead{} &
\colhead{} &
\colhead{SED} \\
\colhead{2MASS\tablenotemark{a}} &
\colhead{Name} &
\colhead{Type} &
\colhead{$\alpha$(2--8~\micron)} &
\colhead{$\alpha$(2--24~\micron)} &
\colhead{$\alpha$(3.6--8~\micron)} &
\colhead{$\alpha$(3.6--24~\micron)} &
\colhead{Class}
}
\startdata
J04034930+2610520 &         HBC 358 A+B+C   &  M3.5 & $-2.49$/$-2.49$ & $-2.68$/$-2.68$ & $-2.74$/$-2.74$ & $-2.83$/$-2.83$ & III \\
J04034997+2620382 &      XEST 06-006   & M5.25 & \nodata & \nodata & \nodata & \nodata & III? \\
J04035084+2610531 &          HBC 359   &    M2 & $-2.58$/$-2.58$ & $-2.77$/$-2.77$ & $-2.74$/$-2.74$ & $-2.88$/$-2.88$ & III \\
J04043936+2158186 &          HBC 360   &  M3.5 & $-2.54$/$-2.55$ & \nodata & $-2.75$/$-2.75$ & \nodata & III \\
J04043984+2158215 &          HBC 361   &    M3 & $-2.55$/$-2.55$ & \nodata & $-2.73$/$-2.73$ & \nodata & III \\
J04044307+2618563 &  IRAS 04016+2610   &    K3 & $1.74$ & \nodata & $0.85$ & \nodata &   I \\
J04053087+2151106 &          HBC 362   &    M2 & $-2.58$/$-2.58$ & $-2.63$/$-2.63$ & $-2.77$/$-2.77$ & $-2.73$/$-2.73$ & III \\
J04080782+2807280 &    \nodata & M3.75 & \nodata & \nodata & \nodata & \nodata & III? \\
\enddata
\tablenotetext{a}{2MASS Point Source Catalog.}
\tablenotetext{b}{Because this multiple system is unresolved at 24~\micron, 
this slope was computed with the total flux of the system in the band
at shorter wavelengths.}
\tablenotetext{c}{The spectral type bin in which this source was placed
in Table~\ref{tab:sed} based on its bolometric luminosity.}
\tablecomments{Entries consist of observed and dereddened values, 
respectively. Only the observed values are listed if extinction estimates
are unavailable.}
\end{deluxetable}

\begin{deluxetable}{lrrrrrrrr}
\tabletypesize{\scriptsize}
\tablewidth{0pt}
\tablecaption{Statistics of SED Classes\label{tab:sed}}
\tablehead{
\colhead{} &
\multicolumn{3}{c}{All of Taurus} &
\colhead{} &
\multicolumn{3}{c}{XEST Fields} \\
\cline{2-4} \cline{6-8}
\colhead{Spectral Type\tablenotemark{b}} &
\colhead{I} &
\colhead{II} &
\colhead{III} &
\colhead{} &
\colhead{I} &
\colhead{II} &
\colhead{III}}
\startdata
                    $<$K6 &  10 &  23 &   8 &  &   9 &  14 &   3  \\
  $\geq$K6 and $\leq$M3.5 &  22 &  90 &  40 &  &  19 &  56 &  23  \\
        $>$M3.5 and $<$M6 &   5 &  44 &  42 &  &   2 &  20 &  23  \\
    $\geq$M6 and $\leq$M8 &   3 &  18 &  25 &  &   1 &   7 &  12  \\
                    $>$M8 &   0 &   9 &  11 &  &   0 &   3 &   3  \\
                    Total &  42 & 184 & 126 &  &  31 & 100 &  64  \\
\enddata
\tablenotetext{a}{Spectral types have been estimated from bolometric 
luminosities for stars that lack spectral classifications, most of which
are class~I sources. The spectral type bins in which these sources appear are
indicated in Table~\ref{tab:alpha}.}
\end{deluxetable}

\begin{deluxetable}{lrrrrl}
\tabletypesize{\scriptsize}
\tablewidth{0pt}
\tablecaption{IRAC Photometry for Probable Members of 
$\eta$~Cha\label{tab:eta1}}
\tablehead{
\colhead{ID\tablenotemark{a}} &
\colhead{[3.6]} &
\colhead{[4.5]} &
\colhead{[5.8]} &
\colhead{[8.0]} &
\colhead{Date}} 
\startdata
   1 &  7.18$\pm$0.02 &  7.20$\pm$0.02 &  7.13$\pm$0.03 &  7.09$\pm$0.03 & 2004 Jun 10 \\
   2 &            out &            out &  5.63$\pm$0.03 &            out & 2004 Jun 10 \\
   3 &  9.28$\pm$0.02 &  9.23$\pm$0.02 &  9.11$\pm$0.03 &  9.18$\pm$0.03 & 2004 Jun 10 \\
   4 &  8.46$\pm$0.02 &  8.44$\pm$0.02 &  8.37$\pm$0.03 &  8.37$\pm$0.03 & 2004 Jun 10 \\
   5 &  9.62$\pm$0.02 &  9.51$\pm$0.02 &  9.36$\pm$0.03 &  8.95$\pm$0.03 & 2004 Jun 10 \\
   6 &  9.19$\pm$0.02 &  9.11$\pm$0.02 &  9.09$\pm$0.03 &  9.04$\pm$0.03 & 2004 Jun 10 \\
   7 &  7.53$\pm$0.02 &  7.54$\pm$0.02 &  7.48$\pm$0.03 &  7.48$\pm$0.03 & 2004 Jun 10 \\
   8 &            sat &            sat &  5.38$\pm$0.03 &  5.41$\pm$0.03 & 2004 Jun 10 \\
\enddata
\tablenotetext{a}{Identifications from \citet{mam99} and \citet{ls04}.}
\end{deluxetable}

\begin{deluxetable}{lrl}
\tabletypesize{\scriptsize}
\tablewidth{0pt}
\tablecaption{MIPS Photometry for Probable Members of
$\eta$~Cha\label{tab:eta2}}
\tablehead{
\colhead{ID\tablenotemark{a}} &
\colhead{[24]} &
\colhead{Date}}
\startdata
   1 &  7.07$\pm$0.05 & 2005 Mar 9 \\             
     &  7.01$\pm$0.04 & 2005 Apr 8 \\             
   2 &  4.37$\pm$0.04 & 2004 Feb 21 \\             
     &  4.40$\pm$0.04 & 2005 Apr 8 \\             
   3 &  8.68$\pm$0.05 & 2005 Mar 9 \\             
     &  8.65$\pm$0.07 & 2005 Apr 8 \\             
   4 &  7.62$\pm$0.05 & 2005 Mar 8 \\             
     &  7.73$\pm$0.04 & 2005 Apr 8 \\
\enddata
\tablenotetext{a}{Identifications from \citet{mam99} and \citet{ls04}.}
\end{deluxetable}

\begin{deluxetable}{lrl}
\tabletypesize{\scriptsize}
\tablewidth{0pt}
\tablecaption{MIPS Photometry for Probable Members of 
$\epsilon$~Cha\label{tab:eps}}
\tablehead{
\colhead{ID\tablenotemark{a}} &
\colhead{[24]} &
\colhead{Date}}
\startdata
1 & \nodata & \nodata \\
2 & 4.99$\pm$0.04  &  2006 Feb 22 \\
3+4+5+6 & sat & \nodata \\
7 & 2.1$\pm$0.2 & 2006 Feb 22 \\
8 & 6.69$\pm$0.04 & 2006 Feb 27 \\
9 & 9.9$\pm$0.2 & 2006 Feb 27 \\
10 & 6.68$\pm$0.04 & 2006 Feb 27 \\
11 & 5.30$\pm$0.04  & 2006 Feb 27 \\
12 & 9.8$\pm$0.2 & 2006 Feb 27 \\
\enddata
\tablenotetext{a}{Identifications from \citet{fei03} and \citet{luh04eta}.}
\end{deluxetable}

\begin{deluxetable}{lrrrrlrl}
\tabletypesize{\scriptsize}
\tablewidth{0pt}
\tablecaption{{\it Spitzer} Photometry for Probable Members of 
TWA\label{tab:twa}}
\tablehead{
\colhead{TWA} &
\colhead{[3.6]} &
\colhead{[4.5]} &
\colhead{[5.8]} &
\colhead{[8.0]} &
\colhead{Date} &
\colhead{[24]} &
\colhead{Date}}
\startdata
   1 &  7.14$\pm$0.02 &  7.04$\pm$0.02 &  6.90$\pm$0.03 &  5.98$\pm$0.03 & 2003 Dec 20 &  1.24$\pm$0.04 & 2004 Feb 2 \\
   2 &            out &            out &            out &            out &    \nodata &  6.37$\pm$0.04 & 2004 Jan 30 \\
   3 &  6.49$\pm$0.02 &  6.37$\pm$0.02 &  6.15$\pm$0.03 &  5.15$\pm$0.03 & 2004 May 27 &  1.62$\pm$0.04 & 2004 Feb 2 \\
   4 &  5.59$\pm$0.02 &  5.54$\pm$0.02 &  5.43$\pm$0.03 &  4.79$\pm$0.03 & 2003 Dec 20 &            sat &    \nodata \\
   5 &            out &            out &            out &            out &    \nodata &  6.35$\pm$0.04 & 2004 Feb 2 \\
   6 &            out &            out &            out &            out &    \nodata &  7.66$\pm$0.04 & 2004 May 9 \\
   7 &            out &            out &            out &            out &    \nodata &  5.97$\pm$0.04 & 2004 Jan 30 \\
  8A &            out &            out &            out &            out &    \nodata &  6.98$\pm$0.04 & 2004 Jan 31 \\
\enddata
\end{deluxetable}

\begin{deluxetable}{lccccccc}
\tabletypesize{\scriptsize}
\tablewidth{0pt}
\tablecaption{Colors of Young Stellar Photospheres\label{tab:colors}}
\tablehead{
\colhead{Spectral Type} &
\colhead{$J-H$} &
\colhead{$H-K_s$} &
\colhead{$K_s-[3.6]$} &
\colhead{$[3.6]-[4.5]$} &
\colhead{$[4.5]-[5.8]$} &
\colhead{$[5.8]-[8.0]$} &
\colhead{$[8.0]-[24]$}}
\startdata
 K4 & 0.53 & 0.10 & 0.06 & 0.00 & 0.04 & 0.00 & 0.04 \\
 K5 & 0.61 & 0.12 & 0.08 & 0.00 & 0.04 & 0.00 & 0.06 \\
 K6 & 0.66 & 0.14 & 0.10 & 0.00 & 0.04 & 0.00 & 0.08 \\
 K7 & 0.68 & 0.15 & 0.10 & 0.00 & 0.04 & 0.00 & 0.11 \\
 M0 & 0.70 & 0.16 & 0.10 & 0.00 & 0.04 & 0.00 & 0.13 \\
 M1 & 0.71 & 0.18 & 0.11 & 0.01 & 0.04 & 0.00 & 0.15 \\
 M2 & 0.70 & 0.20 & 0.14 & 0.03 & 0.04 & 0.00 & 0.17 \\
 M3 & 0.68 & 0.22 & 0.17 & 0.05 & 0.04 & 0.00 & 0.19 \\
 M4 & 0.63 & 0.26 & 0.20 & 0.07 & 0.04 & 0.00 & 0.21 \\
 M5 & 0.56 & 0.29 & 0.28 & 0.09 & 0.04 & 0.00 & 0.23 \\
 M6 & 0.55 & 0.33 & 0.39 & 0.11 & 0.04 & 0.00 & 0.25 \\
 M7 & 0.59 & 0.37 & 0.42 & 0.12 & 0.05 & 0.00 & 0.27 \\
 M8 & 0.65 & 0.44 & 0.52 & 0.12 & 0.07 & 0.03 & 0.29 \\
 M9 & 0.72 & 0.55 & 0.63 & 0.14 & 0.13 & 0.12 & 0.31 \\
 L0 & 0.98 & 0.66 & 0.72 & 0.16 & 0.22 & 0.25 & 0.33 \\
\enddata
\end{deluxetable}

\clearpage

\clearpage

\end{document}